\documentstyle[12pt]{article}
\input epsf.tex

\newcommand{\bmat}{\left(\begin{array}}
\newcommand{\emat}{\end{array}\right)}
\def\NPB#1#2#3{Nucl. Phys. B{#1} (19#2) #3}
\def\PLB#1#2#3{Phys. Lett. B{#1} (19#2) #3}

\def\PRD#1#2#3{Phys. Rev. D{#1} (19#2) #3}
\def\PRL#1#2#3{Phys. Rev. Lett. {#1} (19#2) #3}

\def\yzero{\smash{\hbox{$y\kern-4pt\raise1pt\hbox{${}^\circ$}$}}}

\def\beq{\begin{equation}}
\def\eeq{\end{equation}}
\def\beqa{\begin{eqnarray}}
\def\eeqa{\end{eqnarray}}

\def\-{\hphantom{-}}
\def\ov{\overline}
\def\s2{\frac{1}{2}}

\def\beq{\begin{equation}}
\def\eeq{\end{equation}}
\def\beqa{\begin{eqnarray}}
\def\eeqa{\end{eqnarray}}
\def\tr{{\rm tr \,}}
\def\Tr{{\rm Tr \,}}
\def\diag{{\rm diag \,}}
\def\Dtb{{\ov{{\rm D}3}}}
\def\Dsb{{\ov{{\rm D}7}}}
\def\Dfb{{\ov{{\rm D}5}}}
\def\IF{\relax{\rm I\kern-.18em F}}
\def\II{\relax{\rm I\kern-.18em I}}
\def\IP{\relax{\rm I\kern-.18em P}}
\def\IC{\relax{\rm I\kern-.48em C}}
\def\IR{\relax{\rm I\kern-.18em R}}
\def\BL{B{\rm -}L}

\def\cp{{\cal P}}

\def\id{{\bf 1}}

\def\NN{{\cal N}}
\def\Dsl{\,\raise.15ex\hbox{/}\mkern-13.5mu D} 
\def\IZ{Z\kern-.4em  Z}
\def\id{{\rm I}}

 \def\cp#1{\relax\ifmmode {\IP\kern-2pt{}_{#1}}\else $\IP\kern-2pt{}_{#1}$\=fi}
\newcommand{\drawsquare}[2]{\hbox{%
\rule{#2pt}{#1pt}\hskip-#2pt
\rule{#1pt}{#2pt}\hskip-#1pt
\rule[#1pt]{#1pt}{#2pt}}\rule[#1pt]{#2pt}{#2pt}\hskip-#2pt
\rule{#2pt}{#1pt}}

\newcommand{\fund}{\raisebox{-.5pt}{\drawsquare{6.5}{0.4}}}
\newcommand{\antifund}{\overline{\fund}}

\catcode`\@=11
\newdimen\@rotdimen
\newbox\@rotbox

\def\@vspec#1{\special{ps:#1}}
\def\@rotstart#1{\@vspec{gsave currentpoint currentpoint translate
   #1 neg exch neg exch translate}}
\def\@rotfinish{\@vspec{currentpoint grestore moveto}}
\def\@rotr#1{\@rotdimen=\ht#1\advance\@rotdimen by\dp#1%
   \hbox to\@rotdimen{\hskip\ht#1\vbox to\wd#1{\@rotstart{90 rotate}%
   \box#1\vss}\hss}\@rotfinish}
\def\@rotl#1{\@rotdimen=\ht#1\advance\@rotdimen by\dp#1%
   \hbox to\@rotdimen{\vbox to\wd#1{\vskip\wd#1\@rotstart{270 rotate}%
   \box#1\vss}\hss}\@rotfinish}%
\def\@rotu#1{\@rotdimen=\ht#1\advance\@rotdimen by\dp#1%
   \hbox to\wd#1{\hskip\wd#1\vbox to\@rotdimen{\vskip\@rotdimen
   \@rotstart{-1 dup scale}\box#1\vss}\hss}\@rotfinish}%
\def\@rotf#1{\hbox to\wd#1{\hskip\wd#1\@rotstart{-1 1 scale}%
   \box#1\hss}\@rotfinish}%
\def\rotate{\@ifnextchar[{\@rotate}{\@rotate[l]}}
\def\@rotate[#1]#2{\setbox\@rotbox=\hbox{#2}\@nameuse{@rot#1}\@rotbox}

\catcode`\@=12

\begin{document}

\makeatletter \@addtoreset{equation}{section} \makeatother
\renewcommand{\theequation}{\thesection.\arabic{equation}}
\pagestyle{empty}
\rightline{FTUAM-00/10; IFT-UAM/CSIC-00-18, DAMTP-2000-42}
\rightline{ CAB-IB
2405300, CERN-TH/2000-127, \tt hep-th/0005067}
\vspace{0.3cm}
\begin{center}
\LARGE{\bf  
D-Branes at Singularities : A  Bottom-Up Approach 
to the String Embedding of the Standard Model  \\[10mm]}
\smallskip
\large{G.~Aldazabal$^{1}$, L.~E.~Ib\'a\~nez$^2$, F. Quevedo$^3$ and
A.~M.~Uranga$^4$
\\[2mm]}
\small{$^1$ Instituto Balseiro, CNEA, Centro At\'omico Bariloche,\\[-0.3em]
8400 S.C. de Bariloche, and CONICET, Argentina.\\[1mm]
$^2$ Departamento de F\'{\i}sica Te\'orica C-XI
and Instituto de F\'{\i}sica Te\'orica  C-XVI,\\[-0.3em]
Universidad Aut\'onoma de Madrid,
Cantoblanco, 28049 Madrid, Spain.\\[1mm]
$^3$ DAMTP, Wilberforce Road, Cambridge, CB3 0WA, England.\\[1mm]
$^4$ Theory Division, CERN, CH-1211 Geneva 23, Switzerland.
\\[10mm]}

\end{center}

{\small
\begin{center}
\begin{minipage}[h]{15.5cm}
We propose a bottom-up approach to the building of particle physics models
from string theory. Our building blocks are Type II D-branes which we
combine appropriately to reproduce desirable features of a particle theory
model: 1) Chirality ; 2) Standard Model group ; 3) $\NN=1$ or $\NN=0$
supersymmetry ; 4) Three quark-lepton generations. We start such a  program
by studying configurations of $D=10$, Type IIB D3-branes located at
singularities. We study in detail the case of $\IZ_N$ $\NN=1,0$ orbifold
singularities leading to the SM group or some left-right symmetric
extension. In general, tadpole cancellation conditions require the
presence of additional branes, e.g. D7-branes. For the $\NN=1$ 
supersymmetric case the unique twist leading to three quark-lepton
generations is $\IZ_3$, predicting $\sin^2\theta_W=3/14=0.21$. The models 
obtained are the simplest semirealistic string models ever built. In the
non-supersymmetric case there is a three-generation model for each
$\IZ_N$, $N>4$, but the Weinberg angle is in general too small. One can
obtain a large class of $D=4$ compact models  by considering the above 
structure embedded into a Calabi Yau compactification. We explicitly construct
examples of such compact models using $\IZ_3$ toroidal orbifolds and 
orientifolds, and discuss their properties. In these examples, global
cancellation of RR charge may be achieved by adding anti-branes stuck 
at the fixed points, leading to models with hidden sector gravity-induced
supersymmetry breaking. More general  frameworks, like F-theory
compactifications, allow completely $\NN=1$ supersymmetric embeddings of
our local structures, as we show in an explicit example.
\end{minipage}
\end{center}
}
\newpage

\setcounter{page}{1} \pagestyle{plain}
\renewcommand{\thefootnote}{\arabic{footnote}}
\setcounter{footnote}{0}


\section{Introduction}

One of the important motivations in favour of string theory in the
mid-eighties was the fact that it seemed to include in principle all the
ingredients required to embed the observed standard model (SM) physics
inside a fully unified theory with gravity. The standard approach  when
trying to embed the standard model into string theory has  traditionally
been an {\it top-down approach}. One starts from a string theory like e.g.
the $E_8\times E_8$ heterotic and  reduces the number of dimensions, 
supersymmetries and the gauge group by an appropriate compactification   
leading to a massless spectrum as similar as possible to the SM. The 
paradigm of this approach \cite{philip} has been the compactification of
the $E_8\times E_8$ heterotic on a CY manifold with Euler characteristic
$\chi =\pm 6$, leading to a three-generation $E_6$ model. Further gauge
symmetry breaking may be achieved e.g. by the addition of Wilson
lines \cite{wl} and a final breakdown of $D=4$, $\NN=1$ supersymmetry is
assumed to take place due to some field-theoretical non-perturbative
effects \cite{gaugino}. Other constructions using compact orbifolds
or fermionic string models follow essentially the same philosophy
\cite{review}.

Although since 1995 our view of string theory has substantially changed,
the concrete attempts to embed the SM into string theory have
essentially followed the same traditional approach. This is the case for
instance in the construction of a M-theory compactifications on CY$\times
S^1/Z_2$ \cite{witten,dlow}, or of F-theory compactifications on Calabi-Yau
(CY)
four-folds \cite{ack,lsw}, leading to new non-perturbative heterotic 
compactifications. This is still a top-down approach in which matching of
the observed low-energy physics is expected to be achieved by searching
among the myriads of CY three- or four-folds till we find the correct
vacuum \footnote{For recent attempts at  semirealistic  
model building based on Type IIB orientifolds
see refs. \cite{type1,lpt,type1b,akt,au,aiq1,ads,aiq2}.}.

The traditional top-down approach is in principle a reasonable
possibility but it does not exploit  fully  some 
of the lessons we have learnt about string theory in recent years, most
prominently the fundamental role played by different classes of
p-branes
(e.g. D-branes) in the structure of the full theory, and the important
fact that they localize gauge interactions on  their worldvolume without
any need for compactification at this level. It also requires an exact 
knowledge of the complete geography of the compact extra dimensions
(e.g. the internal CY space) in order to obtain the final effective action
for massless modes.

We know that, for example, Type IIB D3-branes have gauge theories with
matter fields living in their worldvolume. These fields are localized
in the four-dimensional world-volume, and their nature and behaviour
depends only on the local structure of the string configuration in the
vicinity of that four-dimensional subspace. Thus, as far as gauge
interactions are concerned, it seems that the most sensible approach
should be to look for D-brane configurations with world volume field
theories resembling as much as possible the SM field theory, even before
any compactification of the six transverse dimensions. Following this
idea, in the present article we propose a  {\it bottom-up approach} to the
embedding of the SM physics into string theory. Instead of  looking for
particular CY compactifications of a $D=10,11,12$ dimensional structure
down to four-dimensions we propose to proceed in two steps:

{\bf i) }
Look for {\it local} configurations of D-branes with worldvolume theories
resembling the SM as much as possible. In particular we should search
for a gauge group $SU(3)\times SU(2)\times U(1)$ but also for the presence
of three chiral quark-lepton generations. Asking also for $D=4$ $\NN=1$
unbroken supersymmetry may be optional, depending on what our assumptions
about what solves the hierarchy problem are. At this level the theory
needs no compactification and the D-branes may be embedded in the full
10-dimensional Minkowski space. On the other hand, gravity still remains
ten-dimensional, and hence this cannot be the whole story.

{\bf ii) }
The above {\it local} D-brane configuration may in general be part of a
larger {\it global} model. In particular, if the six transverse dimensions
are now compactified, one can in general obtain appropriate four-dimensional 
gravity  with Planck mass proportional to the compactification radii.

\medskip

\begin{figure}
\begin{center}
\centering
\epsfysize=6cm
\leavevmode
\epsfbox{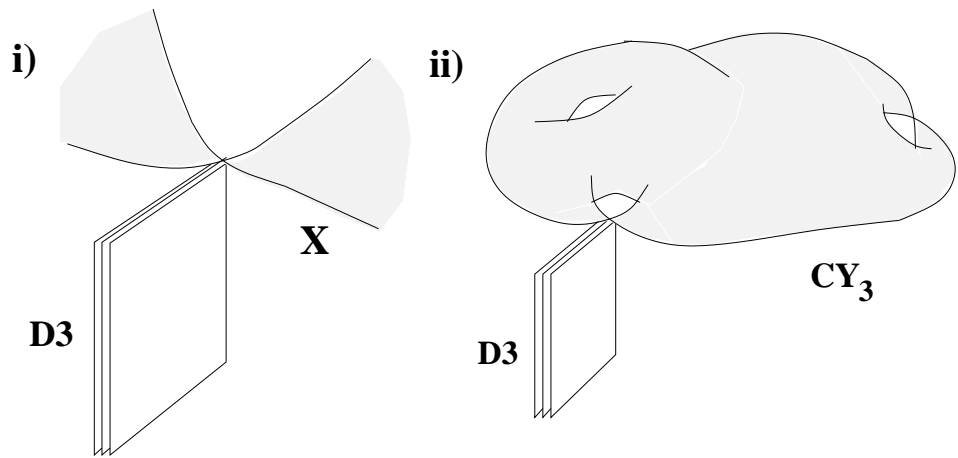}
\end{center}
\caption[]{\small Pictorial representation of the bottom-up approach to
the embedding of the standard model in string theory. In step {\bf
i)} the standard model is realized in the world-volume of D3-branes
sitting at a singular non-compact space $X$ (in the presence of D7-branes,
not depicted in the figure). In step {\bf ii)} this local
configuration is embedded in a global context, like a compact Calabi-Yau
threefold. Many interesting phenomenological issues depend essentially
only on the local structure of $X$ and are quite insensitive to the
details of the compactification in step {\bf ii)}. The global model may
contain additional structures (like other branes or antibranes) not shown
in the figure.}
\label{twostep}
\end{figure}

This two-step process is illustrated in Figure~\ref{twostep}.
An important point to realize is that, although taking the first step
i.e. finding a `SM brane configuration' may be relatively very
restricted, the second step may be done possibly in myriads of manners.
Some properties of the effective Lagrangian (e.g. the gauge group,
the number of generations, the normalization of coupling constants) will
depend only on the {\it local} structure of the D-brane configuration.
Hence many phenomenological questions can be addressed already at the
level of step {\bf i)}. On the other hand, other properties, like some
Yukawa couplings and K\"ahler metrics, will be dependent on the structure
of the full {\it global } model.

In this paper we present the first specific realizations of this {\it
bottom-up} approach. We believe that the results are very promising and
lead to new avenues for the understanding of particle theory applications
of string theory. In particular, and  concerning the two step bottom-up
approach described above we find that:

{\bf i) }
One can obtain simple configurations of Type IIB  D3, D7  branes with
world-volume theories remarkably close to the SM (or some left-right
symmetric generalizations). They correspond to collections of D3/D7 branes
located at orbifold singularities. The presence of additional branes
beyond D3-branes (i.e. D7-branes) is dictated by tadpole cancellation
conditions. Finding three quark-lepton generations and $\NN=1$ SUSY turns
out to be quite restrictive leading essentially to $\IC^3/\IZ_3$
singularities or some variations (including related models with discrete
torsion, $\IZ_3$ orbifolds of the conifold singularity, or a non-abelian
orbifold singularity based on the discrete group $\Delta _{27}$). In these 
models extra gauged $U(1)$'s with triangle anomalies (cured by a
generalized Green-Schwarz mechanism) are generically present but decouple
at low energies. The appearance of the weak hypercharge $U(1)_Y$ in these
models is particularly elegant, corresponding to a unique universal linear 
combination which is always anomaly-free and yields the SM  hypercharge
assignments automatically. In the case of non-supersymmetric $\IZ_N$
orbifold singularities, three quark-lepton generations may be obtained for
any $N>4$, but the resulting weak angle tends to be too small.

{\bf ii) }
These local `SM configurations'  may be embedded into compact models
yielding correct D=4 gravity. We construct specific compact Type IIB
orbifold and orientifold models which contain subsectors given by the
realistic D-brane
configurations discussed above. In order to cancel global tadpoles (i.e,
the total untwisted RR D-brane charges) one can add anti-D3 and/or anti-D7
branes. These antibranes are stuck at orbifold fixed  points (in order to
ensure stability against brane-antibrane annihilation) and lead to models
with hidden sector gravity-induced supersymmetry breaking. Other compact
models with unbroken $D=4$ $\NN=1$ supersymmetry may be easily obtained in
the more general framework of F-theory compactifications, and we construct
an specific example of this type. In this approach the embedding of SM
physics into F-theory is quite different from those followed previously:
the interesting physics resides on D3-branes, rather than D7-branes.

As we comment above, some properties of the low-energy physics of the
compact models will only depend on the local D3-brane configuration.
That is for example the case of hypercharge normalization. The D-brane
configurations leading to unbroken $\NN=1$ SUSY predict a tree-level value
for the weak angle $\sin^2\theta _W =3/14=0.215$, different from the
$SU(5)$ standard result 3/8. This is  compatible with standard
logarithmic gauge coupling unification if the string scale is of order
$M_s\propto 10^{11}$ GeV, as we discuss in the text. Other phenomenological 
aspects like Yukawa couplings depend not only on the singularity structure, 
but also on the particular form of the compactification. Notice in this
respect that although the physics of the D3-branes will be dominated by
the presence of the singularity, the D7-branes wrap subspaces in the
compact space and are therefore more sensitive to its global structure.

The structure of this paper is as follows. In the following chapter we
present some general results which will be needed in the remaining
sections. We describe the general massless spectrum and couplings of Type
IIB D3- and D7-branes on  Abelian orbifold singularities, both for the
$\NN=1$ and $\NN=0$  cases. We also discuss  the consistency
conditions of these configurations (cancellation of RR twisted tadpoles),
as well as the appearance of non-anomalous $U(1)$ gauge symmetries
in this class of theories. 
In chapter 3 we apply the formalism discussed in chapter
2 to the search of realistic three-generation D-brane configurations
sitting on $\IR^6/\IZ_N$ singularities. We present specific simple
$\NN=1$ models leading to the SM gauge group with three quark-lepton
generations. We also present an alternative $SU(3)\times SU(2)_L\times
SU(2)_R\times U(1)_{B-L}$ three generation model. We also discuss the
case of non-SUSY $\IZ_N$ singularities and present an specific three
generation non-SUSY  model based on a $\IZ_5$ singularity. 

Finally, we discuss different generalizations yielding also three
quark-lepton generations. In particular we discuss orbifold singularities
with discrete torsion, models based on non-abelian orbifolds, as well as
some models based on certain non-orbifold singularities. Although the
massless spectrum of these new possibilities is very similar to the models
based on the $\IZ_3$ singularity, some aspects like Yukawa couplings get
modified, which may be interesting phenomenologically. We argue that
locating the D3-branes on an orientifold (rather than orbifold) point
does not lead to standard model configurations and hence is not very
promising. Finally we discuss non-supersymmetric models constructed using
branes and antibranes.

In chapter 4 we proceed to the second step in our approach and embed the
realistic D3/D7 configurations found in the previous chapters into a
compact space. We present examples based on type IIB orbifolds and
orientifolds. As we mentioned above, in this models the global RR charges
may be canceled by the addition of anti-D-branes which are trapped at the
fixed points. Some of these models are T-dual to the models recently
studied in \cite{aiq1,aiq2}. We also discuss the construction of models
with unbroken $D=4$, $\NN=1$ SUSY by considering F-theory compactifications 
with the SM embedded on D3-branes, and construct an specific example. In
chapter 5 we briefly discuss some general phenomenological questions, like
gauge coupling unification and Yukawa couplings. We leave our final
comments and outlook for chapter 6. In order not to obstruct continuity in
the reading with many details, we have five appendices. The first four
give the details of each of the generalizations mentioned in section
3.6. The last appendix deals with the issue of $T$-duality on some of the
compact models in Section~4.

\section{Three-Branes and Seven-Branes at Abelian Orbifold Singularities}

In this section we introduce the basic formalism to compute the spectrum
and interactions on the world-volume of D3-branes at $\IR^6/\IZ_N$
singularities \footnote{Other cases, like $\IZ_N\times \IZ_M$ orbifolds
can be studied analogously, and we will skip their discussion.}. 
These have been discussed in \cite{dm,dgm,ks,lnv,hu}, but our treatment is
more general in that we allow the presence of D7-branes. We also
discuss several aspects of these field theories, to be used in the
remaining sections. 

Before entering the construction, we would like to explain our interest in
placing the D3-branes on top of singularities. The reason is that
D3-branes sitting at smooth points in the transverse dimensions lead to
$\NN=4$ supersymmetric field theories. The only known way to achieve
chirality in this framework is to locate the D3-branes at singularities,
the simplest examples being $\IR^6/\IZ_N$ orbifold singularities.
Another important point in our approach is that we embed all gauge
interactions in D3-branes. Our motivation for this is the fact that the
appearance of SM fermions in three copies is difficult to achieve if color
and weak interactions live in e.g. D3- and D7- branes, respectively (as
will be manifest from the general spectra below). 

\subsection{Brane Spectrum}

We start by considering  the case of a generic, not necessarily 
supersymmetric, singularity. Later on we discuss the specific $\NN=0$ non
supersymmetric and $\NN=1$ supersymmetric realizations. Consider a set of
$n$ D3-branes at a $\IR^6/\Gamma$ singularity with $\Gamma \subset SU(4)$
where, for simplicity,  we take $\Gamma=\IZ_N$. Before the projection, the
world-volume field theory on the D3-branes is a $\NN=4$ supersymmetric
$U(n)$ gauge theory. In $\NN=1$ language, it contains $U(n)$ vector
multiplets, and three adjoint chiral multiplets $\Phi^r$, $r=1,2,3$, with
interactions determined by the superpotential
\beqa
W & = & \sum_{r,s,t} \; \epsilon_{rst}\,\Tr (\Phi^r \Phi^s \Phi^t)
\label{supnfour}
\eeqa
In terms of component fields, the theory contains $U(n)$ gauge bosons,
four adjoint fermions transforming in the ${\bf 4}$  of the $SU(4)_R$
$\NN=4$ R-symmetry group, and six adjoint real scalar fields transforming
in the ${\bf 6}$. 

The $\IZ_N$ action on fermions is given by a matrix
\beq
{\bf R}_{\bf 4} = \diag (e^{2\pi i a_1/N}, e^{2\pi i a_2/N},
e^{2\pi i a_3/N}, e^{2\pi i a_4/N})
\eeq
with $a_1+a_2+a_3+a_4=0 \,\, {\rm mod} \,\, N$. 
The action of $\IZ_N$ on scalars  can be obtained from the definition of
the action on the ${\bf 4}$, and it is given by the matrix
\beq
{\bf R}_{\bf 6} = \diag (e^{2\pi i b_1/N}, e^{-2\pi i b_1/N},
e^{2\pi i b_2/N}, e^{-2\pi i b_2/N}, e^{2\pi i b_3/N},e^{-2\pi i b_3/N})
\label{rotesc}
\eeq
with $b_1=a_2+a_3$, $b_2=a_1+a_3$, $b_3=a_1+a_2$. Scalars can be
complexified, the action on them being then given by ${\bf R}_{esc}=
\diag(e^{2\pi ib_1/N}, e^{2\pi ib_2/N}, e^{2\pi i b_3/N})$. Notice that,
since scalars have the interpretation of brane coordinates in the
transverse space, eq. (\ref{rotesc}) defines the action of $\IZ_N$ on
$\IR^6$ required to form the quotient $\IR^6/\IZ_N$.

The action of the $\IZ_N$ generator $\theta$ must be embedded on the
Chan-Paton indices. In order to be more specific we consider the general
embedding given by the matrix
\beqa
\label{gamma3}
\gamma_{\theta,3} = \diag (\id_{n_0}, e^{2\pi i/N} \id_{n_1},\ldots,
e^{2\pi i(N-1)/N} \id_{n_{N-1}})
\eeqa
where $\id_{n_i}$ is the $n_i\times n_i$ unit matrix, and $\sum_i n_i=n$.
The theory on D3-branes at the $\IR^6/\IZ_N$ singularity is obtained by
keeping the states invariant under the combined (geometrical plus
Chan-Paton) $\IZ_N$ action \cite{dm,dgm}. World-volume gauge bosons
correspond to open string states in the NS sector, of the form 
$\lambda\psi^{\mu}_{-\frac 12}|0\rangle$, with $\mu$ along the D3-brane
world-volume, and $\lambda$ the Chan-Paton wavefunction. The projection
for gauge bosons is then given by
\beqa
\lambda = \gamma_{\theta,3}\ \lambda\ \gamma_{\theta,3}^{-1}
\eeqa
The projection for each of the three complex scalars,
$\lambda\Psi_{-\frac12}^r|0\rangle$ (with $r=1,2,3$ labeling a complex
plane transverse to the D3-brane) is
\beqa
\lambda = e^{-2\pi i b_r/N}\ \gamma_{\theta,3}\ \lambda\
\gamma_{\theta,3}^{-1}
\eeqa
The four fermions in the D3-brane world-volume, labeled by $\alpha=1,
\ldots,4$ are described by string states in the R sector, of the form
$\lambda |s_1,s_2,s_3,s_4 \rangle$, with $s_i=\pm \frac 12$ and $\sum_i
s_i={\rm odd}$. The projection for left-handed fermions, $s_4=-\frac 12$,
leads to 
\beq
\lambda = e^{2\pi i a_{\alpha}/N} \gamma_{\theta,3} \lambda
\gamma_{\theta,3}^{-1}
\eeq
The final spectrum in the $33$ sector is
\beqa
{\rm Vectors} & \prod_{i=0}^{N-1} U(n_i)  \nonumber\\
{\rm Complex} \;\; {\rm Scalars} & \sum_{r=1}^3 \sum_{i=0}^{N-1} 
(n_i,{\ov n}_{i-b_r}) \nonumber \\
{\rm Fermions} & \sum_{\alpha=1}^4 \sum_{i=0}^{N-1}  (n_i,{\ov
n}_{i+a_\alpha}) 
\label{specone}
\eeqa
where subindices will be understood modulo $N$ throughout the paper.
The interactions are obtained by keeping the surviving fields in the
interactions of the original $\NN=4$ theory. Notice that the spectrum is,
generically, non supersymmetric. Instead, when $b_1+b_2+b_3=0 $, we have
$a_4=0$ and the $\IZ_N$ action is in $SU(3)$. This case corresponds to a
supersymmetric singularity. The fermions with $\alpha=4$ transforming in
the adjoint representation of $U(n_i)$ become gauginos, while the other
fermions transform in the same bifundamental representations as the
complex scalars. The different fields fill out complete vector and chiral
multiplets of $\NN=1$ supersymmetry.

In general, we would also like to include D7-branes in the configuration.
Let us center on D7-branes transverse to the third complex plane $Y_3$,
denoted D7$_3$-branes in what follows. Open strings in the $37_3$ and
$7_33$ sectors contribute new fields in the D3-brane world-volume. In the
R sector, there are fermion zero modes in the NN and DD directions $Y_4$,
$Y_3$. Such states are labeled by  $\lambda|s_3;s_4\rangle$, with
$s_3=s_4=\pm\frac 12$, where $s_4$ defines the spacetime chirality. The
projection for left-handed fermions $s_4=-\frac 12$ are
\beqa
\lambda_{37_3} = e^{i\pi  b_3/N} \gamma_{\theta,3} \lambda
\gamma_{\theta,7_3}^{-1} \quad, \quad
\lambda_{7_33} = e^{i\pi  b_3/N} \gamma_{\theta,7_3} \lambda
\gamma_{\theta,3}^{-1}
\eeqa
Scalars arise from the NS sector, which contains fermion zero modes in
the DN directions ($Y_1$, $Y_2$). States are labeled as
$\lambda|s_1,s_2\rangle$, with $s_1= s_2=\pm 1/2$. The projections for
$\lambda|\frac 12,\frac 12\rangle$ are
\beqa
\lambda_{37_3} = e^{-i\pi (b_1+b_2)/N} \gamma_{\theta,3} \lambda
\gamma_{\theta,7_3}^{-1} \quad, \quad
\lambda_{7_33} = e^{-i\pi (b_1+b_2)/N} \gamma_{\theta,7_3} \lambda
\gamma_{\theta,3}^{-1}
\eeqa
We can give the resulting spectrum quite explicitly. Let us consider the
Chan-Paton embedding
\beqa
\gamma_{\theta,7_3} & = & \diag (\ \id_{u_0}, e^{2\pi i/N} \id_{u_1},\ldots,
e^{2\pi i(N-1)/N} \id_{u_{N-1}}) \quad \quad {\rm for}\;\;  b_3={\rm even}
\\ 
\gamma_{\theta,7_3} & = & \diag (e^{\pi i\frac 1N} \id_{u_0}, e^{2\pi i
\frac{3}{N}} \id_{u_1},\ldots, e^{2\pi i \frac{2N-1}{N}} \id_{u_{N-1}}) 
\quad \quad {\rm for}\;\;   b_3={\rm odd} \nonumber 
\eeqa
The resulting spectrum is
\beqa
\begin{array}{llll}
b_3={\rm even} & \to & {\rm Fermions} &
\sum_{i=0}^{N-1}\, [\, (n_i,{\ov u}_{i+\frac 12 b_3}) + 
(u_i,{\ov n}_{i+\frac 12 b_3}) \,] \\ 
 & & {\rm Complex}\;{\rm Scalars} & 
\sum_{i=0}^{N-1}\, [\, (n_i,{\ov u}_{i-\frac 12 (b_1+b_2)}) +
(u_i,{\ov n}_{i-\frac 12 (b_1+b_2)}) \,] \\
b_3={\rm odd} & \to & {\rm Fermions} &
\sum_{i=0}^{N-1}\, [\, (n_i,{\ov u}_{i+\frac 12(b_3-1)}) + (u_i,{\ov
n}_{i+\frac 12(b_3+1)}) \,] \\
 & & {\rm Complex}\;{\rm Scalars} & 
\sum_{i=0}^{N-1}\, [\, (n_i,{\ov u}_{i-\frac 12(b_1+b_2+1)}) +
(u_i,{\ov n}_{i-\frac 12(b_1+b_2-1)}) \,]
\end{array}
\label{spectwo}
\eeqa
The computation is identical for other D7$_r$-branes, transverse to the
$r^{th}$ complex plane, i.e. with world-volume defined by the
equation $Y_r=0$. Notice that for a general twist, D7-branes with
world-volume $\sum_{r=1}^3 \beta_r Y_r=0$, with arbitrary complex
coefficients $\beta_r$, are not consistent with the orbifold action (more
precisely, they are not invariant under the orbifold action, and suitable
$\IZ_N$ images should be included). For twists with several equal
eigenvalues, such D7-branes are possible (see footnote~5).

Notice that in the non-compact setting, fields in the 77 sector are
non-dynamical from the viewpoint of the D3-brane world-volume field
theory. For instance, 77 gauge groups correspond to global symmetries, and
77 scalars act as parameters of the D3-brane field theory. Only after
compactification of the transverse space, as in the models discussed in
Section 4, 77 fields become four-dimensional, and should be treated on an
equal footing with 33 and 37, 73 fields.

We conclude this section by restricting the above results to the case of
singularities $\IC^3/\IZ_N$ preserving $\NN=1$ supersymmetry on the
D3-brane world-volume. That is, for $a_4=0$, and hence
$a_1+a_2+a_3=0\,{\rm mod}\, N$. The spectrum is given by
\beqa
\begin{array}{cccc}
{\bf 33} & {\rm Vector\,\, mult.} & \prod_{i=0}^{N-1} U(n_i) & \cr
& {\rm Chiral\,\, mult.} & \sum_{i=0}^{N-1} \sum_{r=1}^3 (n_i,{\ov
n}_{i+a_r}) & \cr
{\bf 37_3}, {\bf 7_3 3} & {\rm Chiral\,\, mult.} & \sum_{i=0}^{N-1} \,
[\,(n_i,{\ov u}_{i-\frac 12 a_3}) + (u_i,{\ov n}_{i-\frac 12 a_3}) \, ] &
a_3\; {\rm even} \cr
& & \sum_{i=0}^{N-1} \, [\,(n_i,{\ov u}_{i-\frac 12 (a_3+1)}) + (u_i,{\ov
n}_{i-\frac 12 (a_3-1)}) \, ] & a_3\; {\rm odd} 
\label{specsusy}
\end{array}
\eeqa
We will denote $\Phi^r_{i,i+a_r}$ the $33$ chiral multiplet in the
representation $(n_i,{\ov n}_{i+a_r})$. We also denote (assuming
$a_3={\rm even}$ for concreteness) $\Phi^{(37_3)}_{i,i-\frac 12 a_3}$, 
$\Phi^{(7_3 3)}_{i,i-\frac 12 a_3}$ the $37_3$ and $7_3 3$ chiral
multiplets in the $(n_i,{\ov u}_{i-\frac 12 a_3})$, $(u_i,{\ov n}_{i-\frac
12 a_3})$. With this notation, the interactions are encoded in the
superpotential
\beqa
W & = & \sum_{r,s,t=1}^3\; \epsilon_{rst} \; \Tr (\,\Phi^r_{i,i+a_r}
\Phi^s_{i+a_r,i+a_r+a_s} \Phi^t_{i+a_r+a_s,i} \,) + 
\sum_{i=0}^{N-1}\; \Tr (\,\Phi^3_{i,i+a_3} 
\Phi^{(37_3)}_{i+a_3,i+\frac 12 a_3} \Phi^{(7_3 3)}_{i+\frac 12 a_3,i} \,)
\nonumber
\label{superp}
\eeqa

\subsection{Anomaly and Tadpole Cancellation}

With the fermionic spectrum at hand we can proceed to compute non-abelian
anomalies and establish the constraints for consistent anomaly-free
theories. Moreover, such constraints can be rephrased in terms of
twisted tadpole cancellation conditions \cite{leroz,abiu}. Let us address
the computation of the non-abelian anomaly for $SU(n_i)$, in a case with    
D7$_r$-branes, $r=1,2,3$. Let us assume $b_r={\rm even}$ for concreteness,
and denote by $u^r_j$ the number of entries with phase $e^{2\pi ij/N}$ in
$\gamma_{\theta,7_r}$. The $SU(n_j)$ non-abelian anomaly cancellation
conditions are
\beqa
\sum_{\alpha=1}^4 (n_{i+a_\alpha}-n_{i-a_\alpha}) + \sum_{r=1}^3  
(u^r_{i+\frac 12b_r}-u^r_{i-\frac 12 b_r}) = 0
\label{nonabanom}
\eeqa
These conditions are equivalent to the consistency conditions of the
string theory configuration, namely cancellation of RR twisted tadpoles.
To make this explicit, we use
\beqa
\begin{array}{lll}
n_j  =  \frac 1N \sum_{k=1}^N e^{-2\pi i \, kj/N} \Tr\gamma_{\theta^k,3}
& \quad ;\quad  & 
u^r_j  =  \frac 1N \sum_{k=1}^N e^{-2\pi i \,
kj/N}\Tr\gamma_{\theta^k,7_r}
\end{array}
\eeqa
and substitute in (\ref{nonabanom}). We obtain
\beqa
\frac {2i}N \sum_{k=1}^N e^{-2\pi i\, jk/N} [\, \sum_{\alpha=1}^4 \sin
(2\pi k a_\alpha/N) \Tr \gamma_{\theta^k,3} + \sum_{r=1}^3 
\sin (\pi k b_r/2) \Tr \gamma_{\theta^k,7_r} \,] \, = \, 0
\eeqa
Using the identity
\beqa
\label{identity}
\sum_{\alpha=1}^4 \sin(2\pi ka_\alpha/N) = 4\prod_{r=1}^3 \sin(\pi kb_r/N)
\eeqa
the Fourier-transformed anomaly cancellation condition is recast as
\beqa
[\, \prod_{r=1}^3 2\sin(\pi kb_r/N)\, ]\; \Tr \gamma_{\theta^k,3} +
\sum_{r=1}^3 2\sin(\pi k b_r/N)\; \Tr \gamma_{\theta^k,7_r} \, = \, 0
\label{tadpogen}
\eeqa
These are in fact the twisted tadpole cancellation conditions. Notice that
the contributions from the different disk diagrams are weighted by
suitable sine factors, which arise from integration over the string center
of mass in NN directions.

\subsection{Structure of $U(1)$ Anomalies and Non-anomalous $U(1)$'s}

An important property of systems of D3-branes at singularities leading
to chiral world-volume theories is the existence of mixed
$U(1)$-nonabelian gauge anomalies. These field theory anomalies are
canceled by a generalized Green-Schwarz mechanism mediated by closed
string twisted modes \cite{iru} (see \cite{sixdanom, dm, intri} for a
similar mechanism in six dimensions). 

Consider the generically non-supersymmetric field theory constructed from
D3- and D7$_r$-branes at a $\IZ_N$ singularity, with spectrum given in 
(\ref{specone}), (\ref{spectwo}). Assuming $b_r={\rm even}$ for 
concreteness, the mixed anomaly between the $j^{th}$ $U(1)$ (that within
$U(n_j)$) and $SU(n_l)$ is 
\beqa
A_{jl} =  \frac 12 n_j \sum_{\alpha=1}^4 (\delta_{l,j+a_\alpha}
-\delta_{l,j-a_\alpha}) + 
\frac 12 \delta_{j,l} [\, \sum_{\alpha=1}^4 (n_{j+a_\alpha}-n_{j-a_\alpha}) 
+ \sum_{r=1}^3 (u^r_{j-\frac 12 b_r}- u^r_{j+\frac 12 b_r})\, ]
\nonumber
\eeqa
The anomaly is not present if $n_j$ or $n_l$ vanish. After using the
cancellation of cubic non-abelian anomalies (\ref{nonabanom}) (i.e. the
tadpole cancellation conditions), the remaining piece is given by
\begin{equation}
A_{jl}=\frac 12 n_{j} \sum_{\alpha=1}^4
(\delta_{l,j+a_\alpha}-\delta_{l,j-a_\alpha})
\label{mixedone}
\end{equation}
This anomaly can be rewritten as
\beqa
A_{jl} & = & \frac {-i}{2N} \sum_{k=1}^{N-1} [n_j \exp(2i\pi \, kj/N)
\exp(-2i\pi kl/N) \prod_{r=1}^3 2\sin(\pi k b_r/N)\, ]
\label{mixedtwo}
\eeqa
which makes the factorized structure of the anomaly explicit. The anomaly
is canceled by exchange of closed string twisted modes \cite{iru}, which
have suitable couplings to the gauge fields on the D-brane world-volume
\cite{dm,dgm,abd,ss}.

An important fact is that anomalous $U(1)$'s get a tree-level mass of the
order of the string scale \cite{poppitz}, and therefore do not appear in
the low-energy field theory dynamics on the D3-brane world-volume. It is
therefore interesting to discuss the existence of non-anomalous $U(1)$'s
and their structure. Concretely, we consider linear combinations of the
$U(1)$ generators
\begin{equation}
\label{charge}
Q_{c}= \sum _{j=0}^{N-1} c_j { {Q_{n_j}}\over {n_j} }
\end{equation}
(we take $c_j=0$ if $n_j=0$). The condition for $U(1)$'s free of 
$Q_{c}-SU(n_l)^2$ anomalies reads
\begin{equation}
\frac 12 \sum_{\alpha=1}^4 \sum_{j=0}^{N-1} c_{j}
(\delta_{l,j+a_\alpha}-\delta_{l,j-a_\alpha})=
\sum_{\alpha=1}^4 (c_{l-a_\alpha}-c_{l+a_\alpha}) = 0
\label{afreecond}
\end{equation}
for all $l=0,\dots N-1$. It is clear that $c_{j}= const.$ lead to an
anomaly free combination
\beqa
Q_{diag.} & = & \sum_{i=0}^{N-1} {{Q_{n_i}} \over {n_i}}
\label{qdiag}
\eeqa
This generically non-anomalous $U(1)$ plays a prominent role in the
realistic models of Section~3. In general (\ref{afreecond}) gives $N-1$
independent conditions for the
 $N$ unknowns $c_j$, and (\ref{qdiag}) is the only non-trivial solution
(as long as no $n_j$ vanishes). In certain cases, however, the number of
independent equations may be smaller, and additional non-anomalous
$U(1)$'s appear. In order to compute the number of independent equations
we rewrite the unknowns $c_j$ in terms of the new variables  $r_k =
\sum_{j=0}^{N-1} e^{2i\pi kj/N} c_j$. The original variables are given by 
\begin{equation}
\label{ci}
c_j = \frac 1N \sum_{k=0}^{N-1} e^{-2\pi i kj/N} r_k.
\end{equation}
By replacing in (\ref{afreecond}) above we find $\sum_{\alpha=1}^4
\sin(2\pi k a_\alpha / N) r_k = 0$ and, by using the identity
(\ref{identity}) 
we obtain
\begin{equation}
[\; \prod_{r=1}^3 \sin(\pi k b_r / N)\; ]\; r_k = 0
\end{equation}
Thus, we have managed to diagonalize the matrix corresponding to eq.
(\ref{afreecond}). We see that, besides the diagonal solution (corresponding 
to $r_i=0$ for $i=0.\dots,N-1$) other non trivial solutions are possible
whenever there exist twists $\theta ^k$ that leave, at least, one unrotated 
direction. In other words, the rank of the matrix is given by the number
of twists rotating all complex planes. It is possible to describe
explicitly the non-anomalous $U(1)$'s, as follows. Let us consider a
$\IZ_M$ subgroup, generated by a certain twist $\theta^p$, leaving e.g.
the third complex plane fixed, hence $pa_1=-pa_2$ and $pa_3=-pa_4$. For
each value of $I=1,\ldots, M-1$, there is one non-anomalous $U(1)$,
defined by $c_j=\delta_{pj,I}$. This satisfies the condition
(\ref{afreecond}) by a cancellation of contributions from different values
of $\alpha$. Hence we obtain one additional non-anomalous $U(1)$ per twist
leaving some complex plane fixed \footnote{There are arguments
suggesting these additional non-anomalous $U(1)$'s are nevertheless
massive due to their mixing with closed string twisted modes \cite{mp}.
This observation however, would not change our analysis in the following
sections, since our model building involves only the diagonal combination
(\ref{qdiag}), for which such mixing vanishes.}.

Let us consider some explicit examples. For instance, for $\IZ_3$ with
twist $v=\frac 13(1,1,-2)$, equation (\ref{afreecond}) leads to
$b_0=b_1=b_2$ and the only anomaly free combination is (\ref{qdiag}).
Consequently, it is  not possible to have an anomaly free $U(1)$, unless
all $n_i\ne 0 $. This will always be the case for $\IZ_N$ orbifold actions
(supersymmetric or not) without twists with fixed planes. Consequently,
and as we discuss further in section 3, if we are interested in gauge
groups similar to the Standard Model one, with an anomaly free
(hypercharge) $U(1)$, then all $n_i$'s in (\ref{gamma3}) should be non
vanishing.

The situation is different when subgroups with fixed planes exist. For
instance, consider the $\IZ_6 \equiv \frac{1}6(1,1,-2)$ example. The ${\bf
33}$ spectrum reads 
\beqa
&U(n_0) \times U(n_1) \times U(n_2) \times U(n_{3}) \times U(n_4) \times
U(n_5) \times U(n_6) & \nonumber \\
&2[(n_0,\ov{n}_{1})+(n_1,\ov{n}_{2})+(n_2,\ov{n}_{3})+
(n_3,\ov{n}_{4})+(n_4,\ov{n}_{5})+(n_5,\ov{n}_{0}) +  
(n_3,\ov{n}_{0})] &\\
& +(n_0,\ov{n}_{4})+(n_2,\ov{n}_{0})+(n_4,\ov{n}_{2})+
(n_1,\ov{n}_{5})+(n_3,\ov{n}_{1})+(n_5,\ov{n}_{3}) & 
\eeqa
and the anomaly matrix (\ref{mixedone}) is given by
{\small
\begin{equation}
2T_{ij}^{\alpha\beta} =  \left ( \begin{array}{cccccc}
0 &  2n_0 & -n_0 & 0 &n_0   & -2n_0    \\
-2n_1 & 0 & 2n_1 & -n_1& 0 & n_1  \\
n_2 & -2n_2 & 0 & 2n_2& -n_2 & 0 \\ 
0 & n_3 & -2n_3 & 0 & 2n_3& -n_3  \\
-n_4 & 0 & n_4 & - 2n_4 & 0& 2n_4 \\
2n_5 & -n_5 &0 & n_5 & -2n_5 &0
\end{array}
\right )   
\end{equation}}
The search for anomaly-free combinations leads to eq.(\ref{afreecond})
above, which has two non-trivial independent solutions. In fact, (\ref{ci}) 
becomes $c_i=r_0+(-1)^i r_3$, namely  $c_0=c_2=c_4$ and $c_1=c_3=c_5$
indicating that two anomaly-free abelian factors can be present. In
particular, by choosing $n_1=n_3=n_5=0$ and $n_4=3, n_2=2, n_0=1$ we
obtain a field theory with Standard Model group $SU(3)\times SU(2)\times
U(1)$ and one generation $({\bf 3},{\bf \ov 2})_{1/6} +({\bf 1},{\bf
2})_{1/2}+({\bf \ov 3}, {\bf 1})_{-2/3}$, with subscripts giving $U(1)$
charges (fields in a suitable 37 sector should complete this to a full SM
generation).

\section{Particle Models from Branes at Singularities}

In this Section we discuss the embedding of the Standard Model (and
related Left-Right symmetric extensions) in systems of D3-branes at
$\IC^3/\IZ_N$ singularities, with $\IZ_N\subset SU(3)$. We also
discuss possible extensions to more general cases.

\subsection{Number of Generations and Hypercharge}

Let us start by recalling the structure of the field theory on D3-branes
at a $\IC^3/\IZ_N$ singularity, defined by the twist $v=(a_1,a_2,a_3)/N$.
It has $\NN=1$ supersymmetry, and the following field content: there are
vector multiplets with gauge group $\prod_{i=0}^{N-1} U(n_i)$, and $3N$
$\NN=1$ chiral multiplets $\Phi^r_{i,i+a_r}$, $i=0,\ldots,N-1$, $r=1,2,3$, 
transforming in the representation $(n_i,{\ov n}_{i+a_r})$. The
interactions are encoded in the superpotential 
\beqa
W=\epsilon_{rst} \,
\Tr (\Phi^r_{i,i+a_r} \Phi^s_{i+a_r,i+a_r+a_s} \Phi^t_{i+a_r+a_s,i}).
\label{superpagain}
\eeqa
In general, the configurations will also include D7-branes, but for the
moment we center of general features in the {\bf 33} sector.

We are interested in constructing theories similar to the standard model
or some simple extension thereof. In particular, we will be interested in
constructing models which explicitly contain an $SU(3)\times SU(2)$
factor, to account for color and weak interactions.  Besides simplicity,
this choice has the additional advantage (discussed in detail below) that
a non-anomalous $U(1)$ leading to correct hypercharge assignments arises
naturally. 

Hence we consider models in which two factors, which without loss of
generality we take $U(n_0)$ and $U(n_j)$, are actually $U(3)$, $U(2)$.
Since the matter content contains only bi-fundamental representations, the 
number of generations is given by the number of left-handed quarks, i.e. 
fields in the representation $({\bf 3},{\bf 2})$. This is given by
the number of twist eigenvalues $a_r$ equal to $j$ (minus the number of
$a_r$ equal to $-j$), which is obviously at most three, this maximum
value occurring only for the $\IC^3/\IZ_3$ singularity, with twist
$v=(1,1,-2)/3$. For this reason, this singularity will play a prominent
role in our forthcoming models.

Before centering on that concrete case, it will be useful to analyze the
issue of hypercharge. In order to obtain a theory with standard model 
gauge group, one needs the presence of at least one non-anomalous $U(1)$
to play the role of hypercharge. Happily, our analysis in section 2.3 has
shown that, as long as no $n_i$ vanishes, D3-branes at singularities always 
have at least one non-anomalous $U(1)$ generated by $Q_{diag}$ in
(\ref{qdiag}). Therefore, a possibility to obtain the standard model gauge
group, with no additional non-abelian factors would be to consider models
with group \footnote{Such choice for the Chan-Paton embedding for
D3-branes leads in general to non-vanishing tadpoles, which can be
canceled by the introduction of D7-branes. We leave their discussion for
the explicit examples below.} $U(3) \times U(2) \times U(1)^{N-2}$. In the
generic case, only the diagonal combination 
\beqa
Q_{diag}&=& - \left( \frac 13 Q_3 +\frac 12 Q_2 + \sum_{s=1}^{N-2}
Q_1^{(s)} \right)
\label{qdiagsm}
\eeqa
will be non-anomalous (the overall minus sign is included for later
convenience). In a generic orbifold all other $N-1$ additional $U(1)$
factors will be anomalous and therefore massive, with mass of the order of
the string scale. Of course, the fact that we have a non-anomalous $U(1)$
does not guarantee it has the right properties of hypercharge. Quite
surprisingly this is precisely the case for (\ref{qdiagsm}). For instance,
fields transforming in the $({\bf 3},{\bf 2})$ representation have
$Q_{diag}$ charge $-\frac 13+\frac 12=\frac 16$, as corresponds to
left-handed quarks. Fields transforming in the $({\bf 3},{\bf 1})$
(necessarily with charge $-1$ under one of the $Q_1^{(s)}$ generators) 
have a $Q_{diag.}$ charge $-\frac 13+1=-\frac 23$, as corresponds to
right-handed U quarks, etc. Analysis of the complete spectrum requires
information about the D7-brane sector, and is postponed until the
construction of explicit examples in section 3.3. Notice that correct
hypercharge assignments would not be obtained had our starting point been
e.g. $ SU(4)\times SU(2)$, hence our interest in the $SU(3)\times SU(2)$
structure. 

It is worth noticing that normalization of this hypercharge $U(1)$ depends
on $N$. In fact, by normalizing $U(n)$ generators such that $\Tr T_a
^2=\frac{1}2$ the normalization of $Y$ generator is fixed to be  
\begin{equation}
\label{Ynorm}
k_1= 5/3 +2(N-2)
\end{equation}
This amounts to a dependence on $N$ in the Weinberg angle, namely
\begin{equation}
\label{sw}
\sin^2 \theta _W= \frac1{k_1+1}= \frac3{6N-4}
\end{equation}

\medskip
Thus the weak angle decreases as $N$ increases. Notice that 
the  $SU(5)$ result $3/8$ is only obtained for a $Z_2$ singularity.
However in that case the (33) spectrum is necessarily vector-like 
and hence one cannot reproduce the SM spectrum. 

We will also be interested in constructing left-right symmetric extensions
of the standard model. In particular, we consider gauge groups with a
factor $SU(3)\times SU(2)_L\times SU(2)_R$, which are obtained by choosing
suitable values for three of the entries $n_i$ in the Chan-Paton embedding. 
As above, the corresponding tadpole must be canceled by additional
D7-branes, whose details we postpone for the moment. The number of
generations is again given by the number of representations $({\bf 3}, 
{\bf 2}, {\bf 1})$, and is equal to three only for the $\IZ_3$ orbifold.

To reproduce hypercharge after the breaking of the right-handed $SU(2)$
factor, an essential ingredient is the existence of a non-anomalous
$(\BL)$ $U(1)$ in the theory. In order to obtain at least one 
non-anomalous $U(1)$ in the D3-branes, we are led to consider models with
group $U(3)\times U(2)\times U(2) \times U(1)^{N-3}$. Generically, only
the diagonal combination 
\beqa
Q_{diag}&=& -2\; \left( \, \frac 13 Q^{(3)} +\frac 12 Q^{(2_L)} + \frac 12
Q^{(2_R)} + \sum_{s=1}^{N-3} Q^{(1)}_{s}\, \right)
\label{qdiaglr}
\eeqa
(the overall factor is included for convenience) is non-anomalous.
Interestingly, the charges under this non-anomalous $U(1)$ turn out to
have the correct $B{\rm -}L$ structure. For instance, fields transforming
in the $({\bf 3}, {\bf 2}, {\bf 1})$ or $({\bf 3},{\bf 1}, {\bf 2})$
representations have $Q_{diag}$ charge is $-2(\frac 13 -\frac 12)=\frac
13$, correct for quark fields.  Again, the discussion for the
complete spectrum is postponed to the explicit examples in section
3.4.

Notice, as above, that normalization of  $B{\rm -}L$
generator is $N$ dependent and leads to $k_ {B{\rm -}L}=8/3+8(N-2)$. Since
hypercharge is given by $Y=-T^3_R+Q_{ B{\rm -}L}$ (with $T^3_R$ the
diagonal
generator of $SU(2)_R$) the 
values (\ref{Ynorm}) and, thus, (\ref{sw}) are reobtained for hypercharge
normalization and Weinberg angle.

We find it is quite remarkable that the seemingly complicated hypercharge
structure in the standard model is easily accomplished by the structure of
the diagonal $U(1)$ in this class of orbifold models. 

We conclude by remarking that in cases with additional non-anomalous
$U(1)$'s $Q_c$ (\ref{charge}), they could be used as hypercharge or $\BL$
generators, as long as the $U(1)$ factors in $U(3)$ and $U(2)$ belong to
the corresponding linear combination in $Q_c$ (as in the $\IZ_6$ example
in section 2.3). However, since the presence of these $U(1)$'s is not
generic, we will not analyze this possibility in detail. Moreover, they
are not present in the case of $\IZ_3$ singularity, which is the only
candidate to produce three-generation models.

\subsection{Generalities for $\IC^3/\IZ_3$}

In the following we construct some explicit examples of standard model or
left-right symmetric theories based on the $\IC^3/\IZ_3$ singularity.
This is the most attractive case, since it leads naturally to three-family
models. It also illustrates the general technology involved in model
building using branes at singularities.

Consider a set of D3-branes and D7$_r$-branes at a $\IC^3/\IZ_3$ orbifold
singularity, generated by the twist $v=\frac 13(1,1,-2)$. Its action on
the Chan-Paton factors is in general given by the matrices
\beqa
\begin{array}{lcl}
\gamma_{\theta,3} = \diag (\id_{n_0}, \alpha \id_{n_2}, \alpha^2\id_{n_3}) 
& ; & \gamma_{\theta,7_1}= - \diag (\id_{u^ 1_0}, \alpha \id_{u^ 1_2},
\alpha^2 \id_{u^1_3}) \\
\gamma_{\theta,7_3}=\diag (\id_{u_0^3}, \alpha \id_{u_1^3}, \alpha^2
\id_{u_2^3}) & ; &
\gamma_{\theta,7_2}= - \diag (\id_{u^ 2_0}, \alpha \id_{u^ 2_2},
\alpha^2 \id_{u^2_3})
\label{cpz3}
\end{array}
\eeqa
with $\alpha=e^{2\pi i/3}$. The notation, slightly different from that in
section 2.1, is more convenient for $\IC^3/\IZ_3$.

The full spectrum is given by
\beqa
\begin{array}{cc}
{\bf 33} & U(n_0) \times U(n_1) \times U(n_2) \\  
   & 3\, [ (n_0, {\ov n}_1) + (n_1,{\ov n}_2) + (n_2,{\ov n}_0) \, ] \\
{\bf 37_r}, {\bf 7_r3} & (n_0,{\ov u^r}_1) + (n_1, {\ov u^r}_2) +
(n_2,{\ov u^r}_0)+\\
& +(u^r_0,{\ov n}_1) + (u^r_1, {\ov n}_2) + (u^r_2,{\ov n}_0) 
\end{array} 
\label{spec3}
\eeqa
The superpotential \footnote{It is possible to consider the generic case
of D7$^{\beta}$-branes, with world-volume defined by $\sum_r \beta_r Y_r=0$, 
which preserve the $\NN=1$ supersymmetry of the configuration for
arbitrary complex $\beta_r$ \cite{rotated}. The 37$^\beta$, 7$^\beta$3
spectra are as above, but the superpotential is $W= \sum_i \sum_r \beta_r
\Tr(\Phi^r \Phi^{37^{\beta}} \Phi^{7^{\beta}3})$, with fields from a
single mixed sector coupling to 33 fields from all complex planes.} terms
are
\beqa
W= \sum_{i=0}^2 \sum_{r,s,t=1}^3 \; \epsilon_{rst} \Tr (\Phi^r_{i,i+1}
\Phi^s_{i+1,i+2} \Phi^t_{i+2,i}) + \sum_{i=0}^2 \sum_{r=1}^3
\Tr (\Phi^r_{i,i+1} \Phi^{37_r}_{i+1,i+2} \Phi^{7_r3}_{i+2,i})
\label{superpz3}
\eeqa

The twisted tadpole cancellation conditions are
\beqa
\Tr \gamma_{\theta,7_3} - \Tr \gamma_{\theta, 7_1} - \Tr
\gamma_{\theta,7_2} + 3 \Tr \gamma_{\theta,3} = 0
\label{tadpoz3}
\eeqa
where the relative signs, coming from the sine prefactors, cancel those
in the definitions in (\ref{cpz3}), hence it is consistent to ignore
both. Eqs (\ref{tadpoz3}) are equivalent to the non-abelian
anomaly cancellation conditions.

\subsection{Standard Model and Branes at $\IC^3/\IZ_3$ Singularity}

\begin{figure}
\begin{center}
\centering
\epsfysize=8cm
\leavevmode
\epsfbox{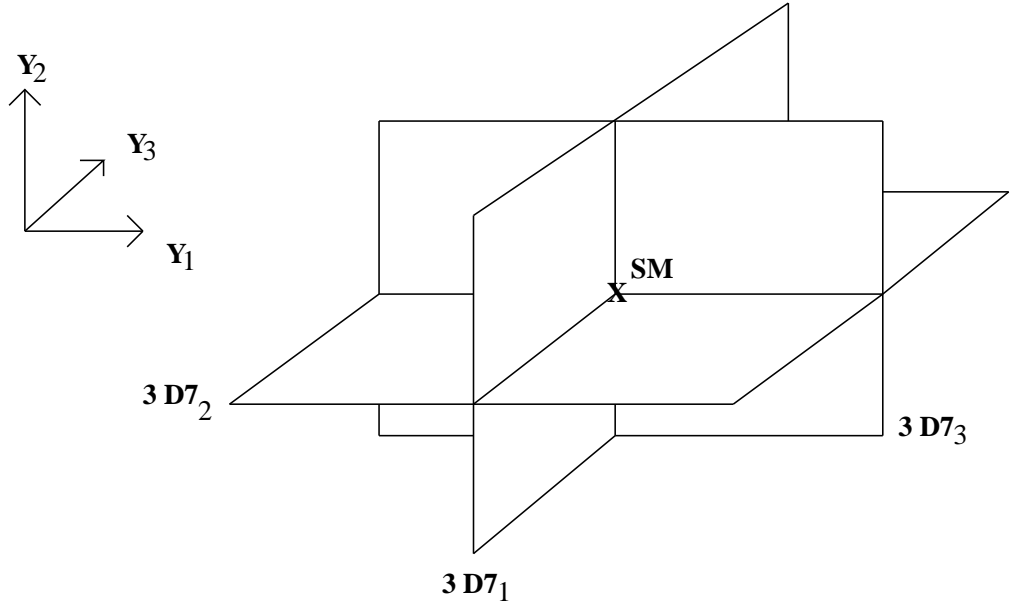}
\end{center}
\caption[]{\small A non-compact Type IIB $\IZ_3$ orbifold singularity
yielding SM spectrum. Six D3 branes sit on top of a $\IZ_3$ singularity at
the origin. Tadpoles are canceled by the presence of intersecting
D7-branes with their worldvolumes transverse to different complex planes.}
\label{fig2}
\end{figure}

Following the general arguments in section~3.1, the strategy to obtain a
field theory with standard model gauge group from the $\IZ_3$
singularity is to choose a D3-brane Chan-Paton embedding 
\beqa
\gamma_{\theta,3}&=&\diag(\id_3,\alpha \id_2,\alpha^2 \id_1)
\label{cpdthreesm}
\eeqa
The simplest way to satisfy the tadpole conditions (\ref{tadpoz3}) is to
introduce only one set of D7-branes, e.g. D7$_3$-branes, with
Chan-Paton embedding $u_0^3=0$, $u_0^1=3$, $u_0^2=6$. The gauge group on
the D3-branes is $U(3)\times U(2)\times U(1)$, whereas
in the D7$_3$-branes is $U(3)\times U(6)$ on each.
Note that, before compactification, the latter behave
as global symmetries in the worldvolume of the
D3-branes.  The D7$_3$-branes group  can be
further broken by global effects, since the corresponding branes are
extended along some internal dimensions. 

An alternative procedure to obtain a smaller group on the D7-branes
is to use all three kinds of D7-branes, as depicted in Figure~\ref{fig2}.
For instance, a very symmetrical choice consistent with (\ref{cpdthreesm})
is $u_0^r=0$, $u_1^r=1$, $u_1^r=2$, for $r=1,2,3$. Each kind of D7-brane
then carries a $U(1)\times U(2)$ group.

\begin{figure}
\begin{center}
\centering
\epsfysize=9cm
\leavevmode
\epsfbox{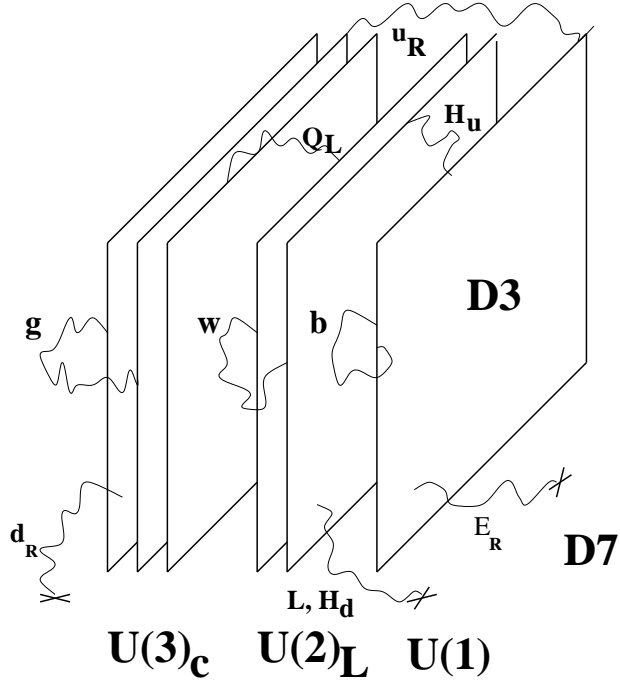}
\end{center}
\caption[]{\small D-brane configuration of a SM $\IZ_3$ orbifold model.
Six D3-branes (with worldvolume spanning Minkowski space) are located on a
$\IZ_3$ singularity and the symmetry is broken to $U(3)\times U(2)\times
U(1)$. For the sake of visualization the D3-branes are depicted at
different locations, even though they are in fact on top of each other.
Open strings starting and ending on the same sets of D3-branes give rise
to gauge bosons; those starting in one set and ending on different sets
originate the left-handed quarks, right-handed U-quarks and one set of Higgs
fields. Leptons, and right-handed D-quarks correspond to open strings
starting on some D3-branes and ending on the D7-branes (with
world-volume filling the whole figure).}
\label{pedagsm}
\end{figure}

The spectrum for this latter model is given in table~\ref{tabpssm} (for later 
convenience we have also included states in the $7_r7_r$ sectors; their
computation is analogous to the computation of the 33 sector).  
In the last column we give the charges under the anomaly-free combination
(\ref{qdiagsm}):
\beq
Y\ = -\left(\; \frac{1}{3}\ Q_{3}\ + \  \frac{1}{2}\ Q_{2}\ + \ Q_{1}\ \right)
\eeq
As promised, it gives the correct hypercharge assignments for standard
model fields. A pictorial representation of this type of models is given
in Figure~\ref{pedagsm}.


\begin{table}[htb] \footnotesize
\renewcommand{\arraystretch}{1.25}
\begin{center}
\begin{tabular}{|c|c|c|c|c|c|c|}
\hline Matter fields  &  $Q_3$  & $Q_2 $ & $Q_1 $ & $Q_{u_1^r}$ &
   $Q_{u_2^r}$
& $Y$   \\
\hline\hline {{\bf 33} sector} &  & & & & & \\
\hline $3(3,2)$ & 1  & -1 & 0 & 0 & 0 & 1/6  \\
\hline $3(\bar 3,1)$ & -1  & 0  & 1 & 0 & 0 & -2/3 \\
\hline $3(1,2)$ & 0  & 1  & -1 & 0 & 0 & 1/2  \\
\hline\hline {{\bf 37$_r$} sector} & & & & & & \\
\hline $(3,1)$ & 1 & 0 & 0 & -1  & 0 & -1/3 \\
\hline $(\bar 3,1;2')$ & -1 & 0 & 0 & 0 & 1  & 1/3 \\
\hline $(1,2;2')$ & 0 & 1 & 0 & 0 & -1 & -1/2 \\
\hline $(1,1;1')$ & 0 & 0 & -1 & 1 & 0 & 1 \\
\hline\hline {\bf 7$_r$7$_r$} sector & & & & & &  \\
\hline $3(1;2)'$ & 0 & 0 & 0 & 1 & -1 & 0 \\
\hline \end{tabular}
\end{center} 
\caption{\small Spectrum of $SU(3)\times SU(2)\times U(1)$ model. We 
present the quantum numbers  under the $U(1)^9$ groups. The first three
$U(1)$'s come from the D3-brane sector. The next two come from the
D7$_r$-brane sectors, written as a single column with the understanding 
that e.g. fields in the {\bf 37}$_r$ sector are charged under the 
$U(1)$ in the ${\bf 7_r7_r}$ sector.
\label{tabpssm} }
\end{table}

We find it remarkable that such a simple configuration produces a spectrum
so close to that of the standard model. In particular, we find encouraging
the elegant appearance of hypercharge within this framework, as the only
linear combination (\ref{qdiag}) of $U(1)$ generators which is naturally
free of anomalies in systems of D3-branes at orbifold singularities.

There is another interesting advantage in the fact that hypercharge arises
exclusively from the {\bf 33} sector. Notice that it allows for the fields
in the 77 sector to acquire nonvanishing vev's \footnote{Strictly
speaking, {\bf 77} fields are not dynamical from the point of view of the
D3-brane field theory in the non-compact context. This comment should be
regarded as applied to compact models with a local behaviour given by the
above system of D3- and D7$_r$-branes at a $\IZ_3$ singularity, like those
in Section~5.} without breaking hypercharge. These  vevs can be used to
further break the ${\bf 77}$ gauge groups, and produce masses for the
extra triplets in the {\bf 37} sectors. Then the only remaining light
triplets are those of the standard model. Notice also that the same
argument does not apply to the doublets in {\bf 37} sectors, which remain
massless even after the {\bf 77} fields acquire vevs. Hence the three
families of leptons remain light. This behavior is reminiscent of the
models in \cite{aiq1}, basically because (see Appendix~E) their
realistic sector has (in a T-dual version) a local structure very similar
to our non-compact $\IZ_3$ singularity. 

The model constructed above, once embedded in a global context, may
provide the simplest semirealistic string compactifications ever built.
Indeed, in Section~4 we will provide explicit compact examples of this
kind. Let us once again emphasize that, however, many properties of the
resulting theory will be independent of the particular global structure
used to achieve the compactification, and can be studied in the
non-compact version presented above, as we do in Section~5.

\subsection{Left-Right Symmetric Models and the $\IC^3/\IZ_3$ Singularity}

\begin{figure}
\begin{center}
\centering
\epsfysize=8cm
\leavevmode
\epsfbox{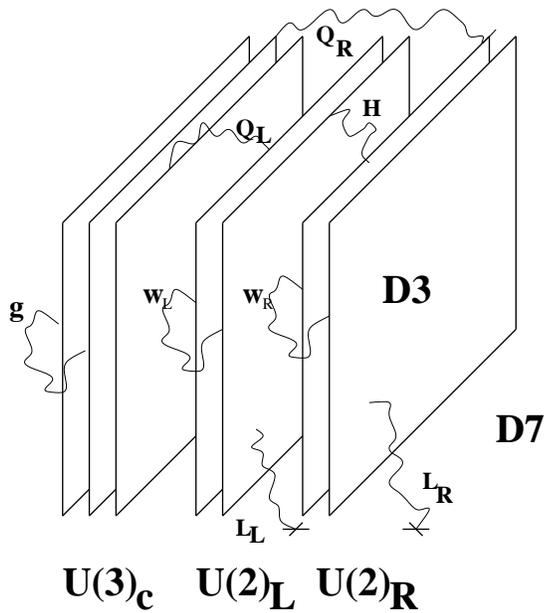}
\end{center}
\caption[]{\small D-brane configuration of a LR  $\IZ_3$ orbifold model.
Seven D3-branes (with worldvolume spanning Minkowski space) are located on
a $\IZ_3$ singularity and the symmetry is broken to $U(3)\times U(2)\times
U(2)$. For the sake of visualization the D3-branes are depicted at
different locations, even though they are actually coincident. Open
strings starting and ending on the same sets of D3-branes give rise
to gauge bosons; those starting and ending on different sets originate
the quarks and Higgs fields. Leptons correspond to open strings starting
on some D3-branes and ending on D7-branes.}
\label{pedaglr}
\end{figure}

One may use a similar approach to construct three-generation models with
left-right symmetric gauge group. Following the arguments in section 3.1,
we consider the D3-brane Chan-Paton embedding
\beqa
\gamma_{\theta,3}& = &\diag(\id_3,\alpha \id_2, \alpha^2 \id_2)
\eeqa
The corresponding tadpoles can be canceled for instance by D7$_r$-branes,
$r=1,2,3$ with the symmetric choice $u_0^r=0$, $u_1^r=u_2^r=1$. The gauge
group on D3-branes is $U(3)\times U(2)_L\times U(2)_R$, while each set of
D7$_r$-branes contains $U(1)^2$. As explained above, the combination
(\ref{qdiaglr})
\beq
Q_{B{\rm -}L}= - 2 \left(\frac 13\ Q_{3}\ +\frac 12  \ Q_{L}\ +
\frac12  \ Q_{R}\right)
\eeq
is non-anomalous, and in fact behaves as $\BL$. The spectrum for this
model, with the relevant $U(1)$ quantum numbers is given in
table~\ref{tabpslr}. A pictorial representation of this type of models
is given in Figure \ref{pedaglr}.

\begin{table}[htb] \footnotesize
\renewcommand{\arraystretch}{1.25}
\begin{center}
\begin{tabular}{|c|c|c|c|c|c|c|}
\hline Matter fields  &  $Q_3$  & $Q_L $ & $Q_R $ & $Q_{U_1^i}$ &
   $Q_{U_2^i}$
& $B-L$   \\  
\hline\hline {{\bf 33} sector} &  & & & & & \\
\hline $3(3,2,1)$ & 1  & -1 & 0 & 0 & 0 & 1/3  \\
\hline $3(\bar 3,1,2)$ & -1  & 0  & 1 & 0 & 0 & -1/3 \\
\hline $3(1,2,2)$ & 0  & 1  & -1 & 0 & 0 & 0  \\
\hline\hline {{\bf 37$_r$} sector} & & & & & & \\  
\hline $(3,1,1)$ & 1 & 0 & 0 & -1  & 0 & -2/3 \\
\hline $(\bar 3,1,1)$ & -1 & 0 & 0 & 0 & 1  & 2/3 \\
\hline $(1,2,1)$ & 0 & 1 & 0 & 0 & -1 & -1 \\
\hline $(1,1,2)$ & 0 & 0 & -1 & 1 & 0 & 1 \\ 
\hline\hline {\bf 7$_r$7$_r$} sector & & & & & &  \\
\hline $3(1)'$ & 0 & 0 & 0 & 1 & -1 & 0 \\
\hline \end{tabular}
\end{center} \caption{Spectrum of $SU(3)\times SU(2)_L\times SU(2)_R$
model. We present the quantum numbers  under the $U(1)^9$ groups. The
first three $U(1)$'s arise from the D3-brane sector. The next two come
from the D7$_r$-brane sectors, and are written as a single column with the
understanding that {\bf 37}$_r$ fields are charged under $U(1)$ factors in
the ${\bf 7_r7_r}$ sector.
\label{tabpslr} }
\end{table}

We can see that the triplets from the 37$_r$ sectors can become massive
after the singlets of the ${\bf 7_r7_r}$ sector acquire a nonvanishing
vev, leaving a light spectrum really close to left-right theories
considered in phenomenological model-building. 

Left-right symmetric models
have several interesting properties that were recently emphasized in
\cite{aiq2}, (besides the original discussions in \cite{mohapatra}).
As a practical advantage, it is the simplest to construct, and it allows
to distinguish easily the Higgs fields from the leptons, since they
transform differently under the gauge group. Phenomenologically, it allows
to have gauge unification at the intermediate scale. Most of the
properties studied in \cite{aiq2}, regarding gauge unification, proton
stability, fermion masses, etc, are inherited by theories based on the
above construction. We will leave the phenomenological discussion to
Section~5.

\medskip

We conclude this section with a comment that applies to models both of SM
and LR type. A familiar property of D3-branes at singularities is that
sets of D3-branes with traceless Chan-Paton embedding can combine and move
off the singularity into the bulk. From the viewpoint of the world-volume
field theory this appears as a flat direction along which the gauge group
is partially broken, and which is parametrized by a modulus associated to
the bulk position of the set of branes. This process is possible in the
models discussed above, due to the existence of sets of D3-branes with
Chan-Paton embedding $\diag(1,\alpha,\alpha^2)$ (one such set for SM type
configurations, and two for LR type models). In fact, it is easy to find
the corresponding flat directions in the field theory we computed. This
fact will be important in the discussion of some compactified models in
Section~5. 

\medskip

\subsection{Non-Supersymmetric Singularities}

In the supersymmetric case the $\IC^3/\IZ_3$ singularity is singled out.
It is the only case leading to three-generation models, due to the fact
that three complex directions are equally twisted. The situation is quite
different for  non-supersymmetric singularities. In fact, by looking at
the {\bf 33} fermionic spectrum  (\ref{specone}), namely $\sum_{\alpha=1}^4
\sum_{i=0}^{N-1}  (n_i,{\ov n}_{i+a_\alpha})$, we notice that for
$a_1=a_2=a_3=a$ (then $a_4=-3a$) we have  
\begin{equation}
\label{3g}
3(n_i,{\ov n}_{i+a}) + (n_i,{\ov n}_{i-3a})
\end{equation}
and therefore a potential triplication. This singularities are therefore
well-suited for model building of non-supersymmetric realistic spectra. We
would like to point out that, despite the lack of supersymmetry, these
models do not contain tachyons, neither in open nor in closed (untwisted
or twisted) string sectors.

Without loss of generality we can
choose $a=1$ \footnote{For a $\IZ_N$ singularity, we must have
${\gcd(a,N)}=1$, hence there exists $p$ such that $pa=1\;{\rm mod}\; N$,
and we may choose $\theta^p$ to generate $\IZ_N$. This only implies a
harmless redefinition on the $n_i$'s in $\gamma_{\theta,3}$.}
corresponding to $\IZ_N$ twist given by $(1,1,1,-3)/N$. Thus, we observe
that $N=2$ leads to non-chiral theories, $N=4$ to four-generation models,
while each model with $N \ge 5$ leads to three-generations models. For
instance, by choosing $n_0=3, \, n_1=2$ and $n_i=1$ for $i=3,\dots, N-1$,
a Standard Model gauge group $SU(3)\times SU(2) \times U(1)_Y \times
U(1)^{N-1} $ is found with ${\bf 33}$ fermions transforming as
\begin{equation}
\label{ns3gen}
3[({\bf 3},{\bf 2})_{1/6} +({\bf \ov 3},{\bf 1})_{-2/3}+ ({\bf
1},{\bf 2})_{1/2}] + (N-3) \, {\rm singlets}.
\end{equation}
Thus, we obtain three generations of left-handed quarks and right-handed U
type quarks. Since such a matter content is anomalous, extra contributions
coming from D7-branes are expected to complete the spectrum to the three
generations of SM quarks and leptons. Notice that presence of D7-branes
only produce fundamentals of ${\bf 33}$ groups in ${\bf 37+73}$ sectors
and therefore the number of generations is not altered.

Interestingly enough, the correct SM hypercharge assignments above correspond 
to the anomaly-free diagonal $U(1)$ combination (\ref{qdiag})
\begin{equation}
\label{Yns}
Y= -Q_{diag}= -\left(\;\frac 13 Q_{n_0} + \frac 12 Q_{n_1} + 
\sum _{j=3}^{N-1}Q_{n_j}\; \right)
\end{equation}
Hence hypercharge can arise by the same mechanism discussed in section 3.1
for supersymmetric models. 

It is interesting to study the Weinberg angle prediction for these models.
The computation follows that in section~3.1, leading to the result
(\ref{sw}), which gives $\sin^2 \theta _W= 0.214, 0.115 ...$ for $N=3,
N=5, \dots $ respectively. Hence we see that, even though many
non-supersymmetric singularities lead to three-generation models, in
general they yield too low values of the Weinberg angle. It is interesting
that the value gets worse as the order of the singularity increases,
suggesting that simple configurations are better suited to reproduce
realistic particle models.

\medskip

\noindent{\bf A $\IZ_5$ example}

As a concrete example of the above discussion let us consider the $\IZ_5$ 
singularity acting on the ${\bf 4}$ with twist $(a_1,a_2,a_3,a_4)=(1,1,1,-3)$. 
Hence the action on the ${\bf 6}$ is given by $(b_1,b_2,b_3)=(2,2,2)$. 
The general D3-brane Chan-Paton matrix has the form
\beq
\gamma_{\theta,3} 
=\diag(\id_{n_0},\alpha \id_{n_1},\alpha ^2 \id_{n_2},\alpha ^3 \id_{n_3},
\alpha ^4 \id_{n_4})
\eeq
with $\alpha = e^{2\pi i/5}$. Anomaly/tadpole cancellation conditions
require 
\beqa
\label{tadanz5}
4(\alpha-\alpha^4)\Tr \gamma_{\theta,3} + \sum_{r=1}^3 \gamma_{\theta,7_r}
= 0
\eeqa
and, therefore, D7-branes must be added. Let us consider the case with 
only D7$_3$-branes, with Chan-Paton action
\beq
\gamma_{\theta,7_3}=\diag(\id_{u_0},\alpha \id_{u_1},\alpha ^2 \id_{u_2},
\alpha ^3 \id_{u_3}, \alpha ^4 \id_{u_4})
\eeq
The  massless spectrum reads 
\beqa
{\bf 33} & {\rm Vectors} & U(n_0)\times U(n_1)\times U(n_2)\times
U(n_3)\times U(n_4) \nonumber \\
   & {\rm Fermions} & 3 [ (n_0,{\ov n}_1) + (n_1,{\ov n}_2) + (n_2,{\ov n}_3) 
      + (n_3,{\ov n}_4)+ (n_4,{\ov n}_0) ] +\nonumber\\
& &   (n_0,{\ov n}_2) + (n_1,{\ov n}_3) + (n_2,{\ov n}_4) + 
   (n_3,{\ov n}_0)+ (n_4,{\ov n}_1) ]\nonumber\\
   & {\rm Cmplx. Sc.} & 3 [ (n_0,{\ov n}_3) + (n_1,{\ov n}_4) + 
     (n_2,{\ov n}_0) + (n_3,{\ov n }_1) + (n_4,{\ov n }_2) ,] \nonumber\\
{\bf 37_3}  & {\rm Fermions} & 
   (n_0,{\ov u }_1) + (n_1,{\ov u }_2)+(n_2,{\ov u }_3)+ (n_3,{\ov u }_4)+
   (n_4,{\ov u}_0)\nonumber \\ 
 & {\rm Cmplx. Sc.} & (n_0,{\ov u }_3) + (n_1,{\ov u }_4)+(n_2,{\ov u
}_0)+ (n_3,{\ov u }_1) + (n_4,{\ov u}_2)\nonumber\\
{\bf 7_33} & {\rm Fermions} & (u_0,{\ov n }_1)   + (u_1, {\ov n }_2) + 
(u_2,{\ov n }_3) + (u_3,{\ov n }_4) + (u_4,{\ov n }_0)\nonumber \\
   & {\rm Cmplx. Sc.} &  (u_0,{\ov n }_3) +  (u_1, {\ov n }_4) 
+ (u_2,{\ov n }_0) + (u_3,{\ov n }_1)  + (u_4,{\ov n }_2)\nonumber
\eeqa
where $n_i$, $u_i$'s are constrained by (\ref{tadanz5}). 

To be more specific, let us build up a Left-Right model, by choosing
$n_0=3, n_1=n_4=2$ and $n_2=n_3=1$, which leads to a gauge group
$SU(3)\times SU(2) \times SU(2) \times U(1)_{B-L} \times [U(1)^5]$. 
The choice $u_0=0, u_1=3, u_2= u_3= 7, u_4=3$ ensures tadpole cancellation. 
The $\BL$ charge is provided by the anomaly-free diagonal combination
(\ref{qdiag}). From the generic spectrum above we find {\bf 33} fermions
transform, under LR group as 
\begin{eqnarray}
& & 3 [ (3,2,1)_{\frac 13} + (1,2,1)_{1} + (1,1,1)_{0} +
(1,1,2)_{-1} + ({\ov 3},1,2)_{-\frac 13} ]\nonumber\\
& &(3,1,1)_{\frac 43} + ({\ov 3},1,1)_{-\frac 43} + (1,2,1)_{+1}
+ (1,1,2)_{-1} + (1,2,2)_0
\end{eqnarray}
while scalars live in 
\begin{eqnarray}
& & 3 [ (3,1,1)_{\frac 43} + ({\ov 3},1,1)_{-\frac 43} + (1,2,1)_{-1}
+(1,1,2)_{-1} + (1,2,2)_0]\nonumber
\end{eqnarray}
Fermions in {\bf 37 +73} sectors transform as 
\begin{eqnarray}
&(3,1,1;{\ov 3},1,1,1)_{-\frac 23} + ({\ov 3},1,1;1,1,1,3)_{\frac 23}+&
\nonumber \\
&+ (1,2,1;1,{\ov 7},1,1)_{-1} + (1,1,2;1,1,7,1)_{1}+{\rm LR \, singlets} &
\end{eqnarray}
while scalars do as 
\begin{eqnarray}
&& (3,1,1;1,1,{\ov 7},1)_{-\frac 23} + (1,2,1;1,1,1,{\ov 3})_{-1} + 
(1,1,2;1,{\ov 7},1,1)_{-1} +  \nonumber \\
&&+(1,1,2;3,1,1,1)_{1} + ({\ov 3},1,1;1,7,1,1)_{\frac 23} + 
(1,2,1;1,1,7,1)_{1} 
+ {\rm LR \,  singlets} \quad
\end{eqnarray}
We obtain three quark-lepton families from the 33 sector. The remaining
fields, vector-like with respect to the LR group could acquire masses 
by breaking of the 77 groups.

\subsection{Other Possibilities}

It is interesting to compare the kind of models we have constructed, using
abelian orbifold singularities, with the field theories arising from
D3-branes at more complicated singularities
\footnote{We have not discussed $\IC^3/(\IZ_N\times \IZ_M)$ singularities.
However, it is a simple exercise to check they lead to models similar to
those of $\IC^3/\IZ_N$ singularities. In particular, realistic SM and 
LR gauge groups are easily achieved by choosing suitable Chan-Paton
actions, and hypercharge (or the $\BL$ $U(1)$ in LR models) arises as a
diagonal combination, straightforward generalization of (\ref{qdiag}).
However, no three-generation models exist within this class. Also, the
Weinberg angle, given by (\ref{sw}) by replacing $N$ by $NM$, is too small
for any singularity of this type.}. In this section we discuss
some relevant cases. The details for the construction of the corresponding
field theories can be found in appendices A, B, C and D.

\subsubsection{Orbifold Singularities with Discrete Torsion}

Orbifold singularities $\IC^3/(\IZ_{M_1}\times\IZ_{M_2})$ lead to
different models depending on their discrete torsion. In appendix A we
review the field theory on singularities $\IC^3/(\IZ_N\times\IZ_M\times
\IZ_M)$, with $\IZ_N$ twist $(a_1,a_2,a_3)/N$, and discrete torsion
$e^{2\pi i/M}$ between the $\IZ_M$ twists. The final spectrum on the
D3-brane world-volume is identical to that of the $\IC^3/\IZ_N$
singularity (\ref{specsusy}), but the superpotential is modified to
(\ref{supdt}). 

Hence it follows that the phenomenologically most interesting models in
this class are those obtained from the $\IZ_3$ orbifold by a further
$\IZ_M\times \IZ_M$ projection with discrete torsion. The resulting
spectrum coincides with (\ref{spec3}), and can lead to three-generation SM
or LR models. The superpotential in the 33 sector is modified to
\beqa
W & = & \ \Tr [\; \Phi^1_{i,i+1} \Phi^2_{i+1,i+2} \Phi^3_{i+2,i} ] -
e^{2\pi i\frac 1M} \Tr [ \Phi^1_{i,i+1} \Phi^3_{i+1,i+2} \Phi^2_{i+2,i}\;]
\eeqa
Since the spectra we obtain from such singularities are identical to those
in the simpler case of $\IC^3/\IZ_3$, we may question the interest of
these models. They have two possible applications we would like to mention. 
The first, explored in section~5, is that discrete torsion enters as a
new parameter that modifies the superpotential of the theory, and
therefore the pattern of Yukawa couplings in phenomenological models. The
second application is related to the process of moving branes off the
singularity into the bulk, a phenomenon that, as discussed in Section 3
can take place in the realistic models there constructed. In
$\IZ_N\times\IZ_M\times \IZ_M$ singularities with discrete torsion, the
process can also occur, but the minimum number of branes allowed to move
into the bulk is $NM^2$. In particular, in SM theories constructed from
the $\IC^3/(\IZ_3\times \IZ_M\times \IZ_M)$ singularity, there are only
six D3-branes with traceless Chan-Paton embedding, so motion into the bulk
is forbidden for $M\geq 2$. For LR models, motion into the bulk of the
twelve traceless D3-branes is forbidden for $M\geq 3$. In fact, the
trapping is only partial, since branes are still allowed to move along
planes fixed under some $\IZ_M$ twist \cite{douglasdt}. This partial
trapping will be exploited in section 4.1 in the construction of certain
compact models.

\subsubsection{Non-Abelian Orbifold Singularities}

We may also consider the field theories on D3-branes at non-abelian
orbifold singularities. The rules to compute the spectrum, along with the
relevant notation, are reviewed in Appendix B. They allow to search for
phenomenologically interesting spectra.
The classification of non-abelian discrete subgroups of $SU(3)$ and
several aspects of the resulting field theories have been explored in
\cite{hh,gremuto}. An important feature in trying to embed the
standard model in such field theories is that of triplication of families.
Seemingly there is no non-abelian singularity where the spectrum appears
in three identical copies. A milder requirement with a chance of leading
to phenomenological models would be the appearance of three copies of at
least one representation, i.e. $a^{\bf 3}_{ij}=3$ for suitable
$i\neq j$. Going through the explicit tables in \cite{hh} there is one
group with this property, $\Delta_{3n^2}$ for $n=3$, on which we center in
what follows. The 33 spectrum \footnote{The gauge group in page 16 of
\cite{hh} corresponds to the particular case of Chan-Paton embedding given
by the regular representation. Here we consider a general choice.} for this 
case is
\beqa
& \prod_{i=1}^9 U(n_i)\times U(n_{10})\times U(n_{11}) & \nonumber \\
&\sum_{i=1}^9 (n_i,{\ov n}_{10}) + 3\, (n_{10},{\ov n}_{11}) +
\sum_{i=1}^9 (n_{11},{\ov n}_i) &
\eeqa

If the triplicated representation is chosen to give left-handed quarks, a
potentially interesting choice is given by $n_i=1$, $n_{10}=3$, $n_{11}=2$. 
However, and due to the additional factors $r_i$ (in our case $r_i=1$ for
$i=1,\dots, 9$, $r_{10}=r_{11}=3$) in (\ref{qdiagnonab}), the diagonal
combination does not lead to correct hypercharge assignments. Besides
(\ref{qdiagnonab}) there are eight non-anomalous $U(1)$'s given by
combinations $Q_{c}= \sum_{i=1}^9 c_i Q_{n_i} + Q_{n_{10}} + \frac 32
Q_{n_{11}}$, with $\sum_{i=1}^9 c_i=9$. The charge structure under the
combination $c_1=c_2=c_3=3$, $c_i=0$ for $i=4,\ldots,9$ is particularly
interesting. We obtain
\beqa
& SU(3) \times SU(2) \times U(1) \; (\, \times U(1)^8 \,) & \nonumber \\
&3(3,2)_{1/6} + 3(1,2)_{1/2} + 6 (1,2)_{-1/2} + 
3({\ov 3},1)_{-2/3} + 6({\ov 3},1)_{1/3} &
\eeqa
Hence, the spectrum under this $U(1)$ contains fields present in the
standard model, and with the possibility of leading to three net copies
(if suitable 37, 73 sectors are considered), even though they would still
be distinguished by their charges under the additional $U(1)$'s. It is
conceivable that this model leads to phenomenologically interesting field
theories by breaking the additional $U(1)$ symmetries at a large enough
scale. However, it is easy to see that the Weinberg angle, which, taking
into account (\ref{relcoupl}), is still given by (\ref{sw}) where now 
$N=11$, is
exceedingly too small $\sin^2 \theta_W=3/62$. Nevertheless we hope the
model is illustrative on the type of field theory spectra one can achieve
using non-abelian singularities.

\subsubsection{Non-Orbifold Singularities}

In appendix C we discuss the construction of field theories on D3-branes
at some non-orbifold singularities, which is in general rather involved.
We also discuss that a promising spectrum is obtained from a partial
blow-up of a $\IZ_3$ quotient of the conifold. The general spectrum is
given in (\ref{nonorbione}), (\ref{nonorbitwo}). There are several
possibilities to construct phenomenologically interesting spectra from
this field theory. For the purpose of illustration, let us consider one
example with $n_2=3$, $n_0=2$, $n_1'=2$, $n_1'=1$ and $v_2=1$, $x_2=3$,
$x_3=5$ (all others vanishing). This choice leads to the spectrum
\beqa
& SU(3)\times SU(2)\times SU(2)\times U(1)_{diag}\times U(1)' & \nonumber
\\
{\bf 33}\qquad & 3(3,2,1)_{\frac 13} + ({\ov 3},1,1)_{-\frac 43} + 2({\ov
3}, 1,2)_{-\frac 13} + (1,2,2)_0 + 2(1,2,1)_{1} + (1,1,2)_{-1} &\nonumber \\
{\bf 37}, {\bf 73}\qquad & (3,1,1;1,1)_{-\frac 23} + (1,1,1;1,1)_{2} +
(3,1,1;{\ov 3},1)_{-\frac 23} + & \nonumber \\
&(1,1,2;3,1)_1 + (1,2,1;1,{\ov 5})_{-1} + ({\ov 3},1,1;1,5)_{\frac 23}&
\eeqa
with subindices giving charges under $U(1)_{diag}$. We obtain three net
generations, but the right-handed quarks have a different embedding 
into the left-right symmetry: whereas two generations of right-handed
quarks are standard $SU(2)_R$ doublets the other generation 
are $SU(2)_R$ singlets. 
Notice that an interesting property this model
illustrates is that one can achieve three quark-lepton generations without
triplication of the $(1,2,2)$  Higgs multiplets. This could be useful
in order to suppress 
 flavour changing neutral currents for these models.

\subsubsection{Orientifold Singularities}

There is another kind of singularities that arises naturally in string
theory, which we refer to as orientifold singularities. They arise from
usual geometric singularities which are also fixed under an action $\Omega
g$, where $\Omega$ reverses the world-sheet orientation, and $g$ is an
order two geometric action. Orientifolds were initially considered in
\cite{oldorient} and have recently received further attention
\cite{neworient,moreorient,afiv}. 

Starting with the field theory on a set of D3-branes at a usual singularity, 
the main effect of the orientifold projection is to impose a $\IZ_2$
identification on the fields. Most examples in the literature deal with
orientifolds of orbifolds singularities \cite{kakush,iru}, on which we
center in the following (see \cite{pru} for orientifold of some simple
non-orbifols singularities). To be concrete, we consider orientifolds of
$\IC^3/\IZ_N$ singularities with odd $N$ \cite{iru}. The spectrum before
the orientifold projection is given in (\ref{specsusy}). This configuration 
can be modded out by $\Omega R_1 R_2 R_3 (-1)^{F_L}$ (where $R_r$ acts as
$Y_r\to-Y_r$, and $F_L$ is left-handed world-sheet fermion number),
preserving $\NN=1$ supersymmetry. The action of the orientifold projection
on the 33 spectrum amounts to identifying the gauge groups $U(n_i)$ and
$U(n_{-i})$, in such a way (due to world-sheet orientation reversal) that
the fundamental representation $n_i$ is identified with the
anti-fundamental ${\ov n}_{-i}$. Consequently, there is an identification
of the chiral multiplet $\Phi^r_{i,i+a_r}$ and $\Phi^r_{-i-a_r,-i}$.
Finally, since $\Omega$ exchanges the open string endpoints, 37$_r$ fields 
$\Phi^{(37_r)}_{i,i-\frac 12b_r}$ map to 7$_r$3 fields
$\Phi^{7_r3}_{N-i+\frac 12 b_r, N-i}$.

Notice that the gauge group $U(n_0)$ is mapped to itself and, similarly,
chiral multiplets $\Phi^r_{i,i+a_r}$ are mapped to themselves if
$i+a_r=N-i$. There exist two possible orientifold projections, denoted
`SO' and `Sp', differing in the prescription of these cases. The SO
projection projects the $U(n_0)$ factor down to $SO(n_0)$, and the
bi-fundamental $\Phi^r_{i,i+a_r}(= \Phi^r_{i,-i})$ to the two-index
antisymmetric representation of the final $U(n_i)$. The Sp projection 
chooses instead $USp(n_0)$, and two-index symmetric representations. 

It is easy to realize that D3-branes at orientifold singularities will
suffer from a generic difficulty in yielding realistic spectra. The
problem lies in the fact that the orientifold projection removes
from the spectrum the diagonal $U(1)$ (\ref{qdiag}) (as is obvious, since
the $U(1)$ in $U(n_0)$ is automatically lost), which was crucial in
obtaining correct hypercharge in our models in section 3.

For illustration, we can check explicitly that, in the only candidate to
yield three-family models, the orientifold of the $\IC^3/\IZ_3$ 
singularity, no realistic spectra arise. The general 33 spectrum for this
orientifold (choosing, say the $SO$ projection) is
\beqa
& SO(n_0)\times U(n_1) & \nonumber \\
& 3\; [\; (n_0,{\ov n}_1) + (1,\frac 12n_1(n_1-1)) \;] &
\eeqa
The $U(1)$ factor is anomalous, with anomaly canceled by a GS mechanism
\cite{iru}.

A seemingly interesting possibility would be $SO(3)\times U(3) \simeq 
SU(3)\times SU(2)\times U(1)$. However, the $U(1)$ factor does not
provide correct hypercharge assignments. Moreover, it is anomalous and
therefore not even present in the low-energy theory. A second possibility,
yielding a Pati-Salam model $SO(4)\times U(4) \simeq  SU(2)\times SU(2)
\times SU(4)\times U(1)$ is unfortunately vector-like. In fact, the most
realistic spectrum one can construct is obtained for $n_0=1$, $n_1=5$,
yielding a $U(5)$ (actually $SU(5)$) gauge theory with chiral multiplets
in three copies of ${\bf \ov 5}+{\bf 10}$. This GUT-like theory, which
constitutes a subsector in a compact model considered in \cite{lpt}, does
not however contain Higgs fields to trigger breaking to the standard model
group.

We hope this brief discussion suffices to support our general
impression that orientifold singularities yield, in general, field
theories relatively less promising than orbifold singularities. 

\subsubsection{Non-Supersymmetric Models from Antibranes}

We would like to conclude this section by considering a further set of
field theories, obtained by considering branes and antibranes at
singularities. The rules to compute the spectrum are reviewed in Appendix 
C. For simplicity we center on systems of branes and antibranes at
$\IC^3/\IZ_N$ singularities, with $\IZ_N\subset SU(3)$, even though the
class of models is clearly more general. Notice that, due to the presence
of branes and antibranes, the resulting field theories will be
non-supersymmetric. Let us turn to the discussion of the generic features
to be expected in embedding the standard model in this type of D3/$\Dtb$
systems.

The first possibility we would like to consider is the case with the
standard model embedded on D3- and $\Dtb$-branes. From
(\ref{projtachyon}), tachyons arise whenever the Chan-Paton embedding
matrices $\gamma_{\theta,3}$, $\gamma_{\theta,{\bar 3}}$ have some common
eigenvalue. Denoting $n_j$, $m_j$ the number of eigenvalues $e^{2\pi i
j/N}$ in $\gamma_{\theta,3}$, $\gamma_{\theta,{\bar 3}}$, tachyons are
avoided only if one considers models where $n_{i}$ vanishes when $m_{i}$
is non-zero, and vice-versa. This corresponds, for instance, to embedding
$SU(3)$ in the D3-branes and $SU(2)$ in the ${\ov D3}$-branes.

One immediate difficulty is manifest already at this level, regarding the
appearance of hypercharge. As we remark in Appendix D, it is a simple
exercise to extend the analysis of $U(1)$ anomalies of section 2.3 to the
case with antibranes, with the result that (generically) there are two
non-anomalous diagonal combinations (\ref{qdiag}) if no $n_{i}$,
$m_{i}$ vanish. But this case is excluded by the requirement of
absence of tachyons. Hence tachyon-free models lack the diagonal
linear combination of $U(1)$'s and in general fail to produce correct
hypercharge. It is still possible that in certain models, suitable
tachyon-free choices of $n_{i}$, $m_i$ may produce hypercharge out of
non-diagonal additional $U(1)$'s. Instead of exploring this direction
(which in any case would not be available in the only three-family case of
the $\IC^3/\IZ_3$ singularity), we turn to a different possibility. 

An alternative consists in embedding the standard model in, say D3-branes,
but to satisfy the tadpole cancellation conditions using $\Dsb$-branes
(and possibly D7-branes as well). The resulting models are closely related
to those in Section~3, differing from them  only in the existence of $3{\ov
7}$, ${\ov 7}3$ sectors. Let us consider a particular example, with
a set of D3-, D7$_3$- and $\Dsb_3$-branes at a $\IC^3/\IZ_3$ singularity
with twist $v=(1,1,-2)/3$, with Chan-Paton embeddings
\beqa
\gamma_{\theta,3}=\diag(\id_3,\alpha \id_2,\alpha^2 \id_1) \quad , \quad
\gamma_{\theta,7_3}=\diag(\alpha^2 \id_3) \quad ,\quad
\gamma_{\theta,{\ov 7}_3}=\diag(\id_3)
\eeqa
which satisfy the tadpole condition
\beqa
\Tr \gamma_{\theta,7_3} -\Tr \gamma_{\theta,{\ov 7}_3} + 3
\Tr\gamma_{\theta,3} = 0
\eeqa
The resulting spectrum on the D3-branes is
\beqa
         & SU(3)\times SU(2)\times U(1)_Y &\nonumber\\
{\bf 33}\qquad\qquad
 & 3(3,2)_{\frac 16} +3 (1,2)_{\frac 12} + 3({\ov 3},1)_{-\frac
23} &\nonumber\\
{\bf 37_3}, {\bf 7_33}\qquad 
& (1,2;{\ov 3},1)_{-\frac 12} + ({\ov 3},1;3,1)_{\frac 13} \nonumber \\
{\bf 3{\ov 7}_3}, {\bf {\ov 7}_33}\qquad 
& (1,1;1,3)_{1} + (1,2;1,{\ov 3})_{\frac 12} &  
{\rm (left\;handed\;\, fermions)} \nonumber \\
& (3,1;1,{\ov 3})_{-\frac 13} + ({\ov 3},1;1,3)_{\frac 13} & 
{\rm (complex\;\, scalars)}
\eeqa
In the supersymmetric sectors, 33, 37 and 73, the above representations
correspond to chiral multiplets, while in the non-supersymmetric 3${\ov
7}$, ${\ov 7}$3 the spectrum for fermions and scalars is different. The
complete spectrum corresponds to a three-generation model. Notice that
models of this type can be obtained from our constructions in section 3
simply by adding an arbitrary number of $\Dsb$-branes with traceless
Chan-Paton factors, and annihilating fractional D7- and $\Dsb$-branes with
identical Chan-Paton phase (this corresponds to the condensation of the
corresponding tachyons). Since many of the relevant properties of the
models in section 3 are inherited by the non-supersymmetric theories with
branes and antibranes, we do not pursue their detailed discussion here.

\section{Embedding into a Compact Space}

As discussed in the introduction, even though SM gauge interactions
propagate only within the D3-branes even in the non-compact setup, gravity 
remains ten-dimensional. In order to reproduce correct four-dimensional
gravity the transverse space must be compact. In this section we present
several simple string compactifications which include the local structures
studied in section~3, as particular subsectors. 

Since these local structures contain D3- and D7-branes, the corresponding
RR charges must be cancelled in the compact space. A simple possibility is
to cancel these charges by including antibranes (denoted $\Dtb$-,
$\Dsb$-branes) in the configuration. General rules to avoid tachyons and
instabilities against brane-antibrane annihilation have been provided in
\cite{au} (see \cite{sugi,ads} for related models), which we exploit in
section 4.1 to construct compact models based on toroidal orbifolds. A
second possibility is to use orientifold planes. In section 4.2 we provide
toroidal orientifolds without $\Dtb$-branes, the D3-brane charge being
cancelled by orientifold 3-planes (O3-planes).
All of the above models contain some kind of antibranes, breaking
supersymmetry in a hidden sector lying within the compactification space.
It may be possible to find completely supersymmetric embeddings of our
local structure by using toroidal orientifolds including O3- and
O7-planes, even though we have not found such examples within the class of
toroidal abelian orientifolds. This is hardly surprising, since this class
is rather restricted. We argue that suitable $\NN=1$ supersymmetric
compactifications of our local structures exist in more general
frameworks, like compactification on curved spaces. In fact, in section
4.3 we present an explicit F-theory compactification of this type.

\subsection{Inside a Compact Type IIB Orbifold}

A simple family of Calabi-Yau compactifications with singularities is
given by toroidal orbifolds $T^6/\IZ_N$. The twist $\IZ_N$ is constrained
to act cristalographycally, and a list of the possibilities preserving
supersymmetry is given in \cite{dhvw}. Several such orbifolds include
$\IC^3/\IZ_3$ singularities, the simplest being $T^6/\IZ_3$, with 27
singularities arising from fixed points of $\IZ_3$. Parameterizing the
three two-tori in $T^6$ by $z_r$, $r=1,2,3$, with $z_r\simeq z_r+1\simeq
z_r+e^{2\pi i/3}$, the 27 fixed points are located at $(z_1,z_2,z_3)$ with
$z_r=0$, $\frac{1}{\sqrt 3} e^{\pi i/6}$, $\frac{1}{\sqrt{3}}e^{-\pi
i/6}$. For simplicity, we denote these values by $0$, $+1$, $-1$ in what
follows. We now present several compact versions of the models in section
3, based on this orbifold.

\subsubsection{Simple Examples}

The simplest possibility, in order to embed the local structures of
section 3 in a compact orbifold, is to consider models with only one type
of D7-brane. We will also consider the addition of Wilson lines, a
modification which, being a global rather than a local phenomenon, was
not available in the non-compact case. We first construct an example of a
left-right symmetric model, which illustrates the general technique, and
then briefly mention the construction of a model with standard model group.
\bigskip

\noindent{\bf A Left-Right Symmetric Model}
\bigskip

Consider a compact $T^6/\IZ_3$ orbifold, and locate seven D3-branes with
\beqa
\gamma_{\theta,3}=\diag(\id_3,\alpha \id_2, \alpha^2 \id_2)
\label{cpzero}
\eeqa
at the origin. To cancel twisted tadpoles at the origin we will add  six
D7$_1$-branes at the origin in the first complex plane, with Chan-Paton
embedding
\beq
\gamma_{\theta,7} =\diag(\alpha \id_3, \alpha^2 \id_3)
\label{cpone}
\eeq
The local structure around the origin is exactly as in the LR models
studied in section 3.4, hence the final model will automatically contain a
$SU(3)\times SU(2)_L\times SU(2)_R\times U(1)_{\BL}$ gauge group, with
three quark-lepton generations. That one can make this statement at such
an early stage in the construction of the model is the main virtue of the
bottom-up approach we are presenting. In the following, we simply discuss
how to complete the model satisfying the string consistency conditions,
and will not bother computing the detailed spectrum from the additional
sectors.

The choice (\ref{cpone}) cancels the twisted tadpoles at the origin. But
the worldvolume of these D7$_1$-branes also overlap with other eight fixed
points $(0,n,p)$, where $n,p=0,\pm 1$, so we will also have to ensure
twisted tadpole cancellation at them. A simple option would be to add
one single D3-brane with $\gamma_{\theta,3}=1$ at each. However we
consider the following more interesting possibility, which in addition
breaks the (too large) D7$_1$-brane $U(3)\times U(3)$ gauge group.
Consider adding a Wilson line e.g. along the second complex plane, and
acting on seven-branes as
\beq
\gamma_{W,7}=\diag( \id_1 , \alpha \id_1 , \alpha ^2 \id_1 , \id_1 ,
\alpha \id_1 , \alpha ^2 \id_1 )
\eeq
The D7$_1$-brane gauge group is broken down to $U(1)^6$ (some of which
will actually be anomalous and therefore not really present). The
fixed points $(0,\pm 1 , p)$, $p=0,\pm 1$ feel the presence of the Wilson 
line \cite{inq,afiv}, hence the D7$_1$ contribution to twisted tadpoles is
given by $\tr
\gamma_{\theta,7}\gamma_{W,7}$ or $\tr\gamma_{\theta,7}\gamma_{W,7}^2$, 
which vanish. Hence twisted tadpole cancellation does not require any
additional branes at those fixed points. On the other hand the points
$(0,0, \pm 1)$ do not feel the presence of the Wilson line and require
e.g. the presence of one D3-brane with $\gamma_{\theta,3}=1$ on each of
them to achieve tadpole cancellation \footnote{An amusing possibility is
to add, instead of these single D3-branes, `new brane-worlds' of seven
3-branes with (\ref{cpzero}), so that the model contains three different
but analogous universes.}.

\begin{figure}
\begin{center}
\centering
\epsfysize=8cm
\leavevmode
\epsfbox{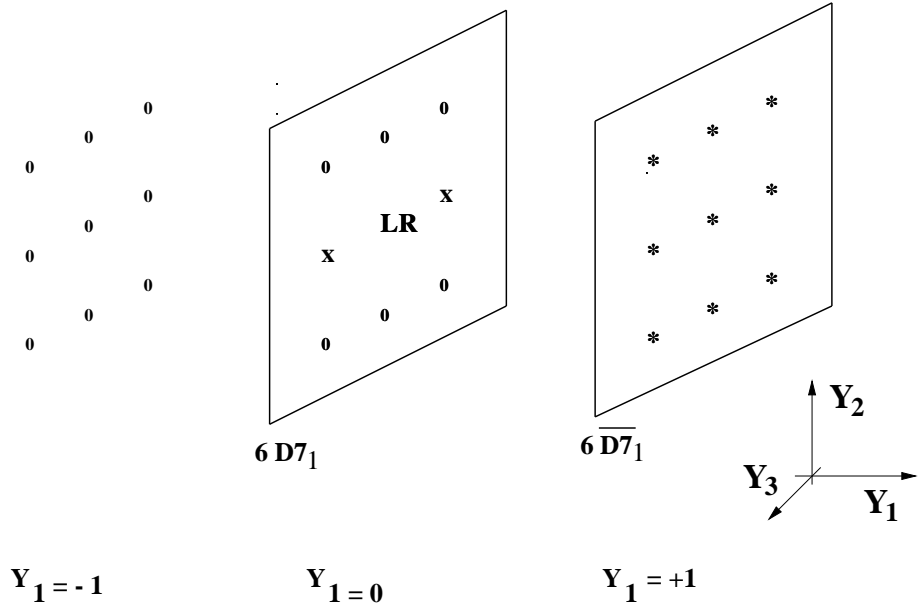}
\end{center}
\caption[]{\small A compact Type IIB $T^6/\IZ_3$ orbifold model with a
LR subsector. The points marked $(0)$ correspond to fixed points 
without D3-branes; those marked $(*)$ contain anti-D3-branes and those
marked $(x)$ have D3-branes located on them. Seven D3-branes, leading to 
a LR model of the type studied in section 3.4, reside at the origin. The
overall RR charge cancels.}
\label{fig4}
\end{figure}

The above configuration has a total of nine D3-branes and six
D7$_1$-branes. The simplest possibility to cancel untwisted tadpoles is
to locate six $\Dsb_1$-branes at a different point, say $Y_1=1$, in the
first complex dimension, with the same Chan-Paton twist matrix
(\ref{cpone}). Twisted and untwisted tadpoles are cancelled by
adding one $\Dtb$-brane at each of the nine fixed points $(1,m,p)$,
$m,p=0,\pm 1$, with Chan-Paton embedding $\gamma_{\theta,{\bar3}} 
=\diag(\id_1)$. The number of branes minus that of antibranes vanishes.
The complete configuration is schematically depicted in Figure~\ref{fig4}

A comment on the stability of this brane configuration is in order. In
this compact configuration the D7-branes and $\Dsb$-branes are trapped at
the singular points, which they cannot leave. The flat directions
associated to motions into the bulk are not present in their world-volume
field theory. Moreover such motions would violate the twisted tadpole 
cancellation conditions. Hence, these objects cannot leave the fixed
points and annihilate each other into the vacuum. On the other hand,
groups of three D3-branes in the LR subsector may in principle abandon the
origin and travel into the bulk in $\IZ_3$-invariant configurations,
attracted by the $\Dtb$-branes living at the fixed points $(1,m,p)$.
Eventually the D3-branes could reach the $\Dtb$-branes, and partial
brane-antibrane annihilation would follow. This annihilation cannot be
complete, since again the tadpole cancellation conditions would be
violated. Since the complete configuration is non-supersymmetric, a more
detailed analysis of the forces involved would be needed to find out
whether the emission of D3-branes from the origin to the bulk is actually
dynamically preferred or not, but we expect this specific model to suffer
from such instability. On the other hand, as we will argue below, there
are other ways to embed the set of SM or LR D3-branes into a compact space
in which brane-antibrane annihilation is not possible.  

\bigskip

\noindent{\bf One Standard-like Model}
\bigskip

From the above construction, it is clear how to construct a compact
orbifold model with an SM subsector, as we sketch in the following. We place 
the six D3-branes at the origin with
\beqa
\gamma_{\theta,3}=\diag(\id_3,\alpha \id_2, \alpha^2 \id_1)
\eeqa
and nine D7$_1$-branes with
\beq
\gamma_{\theta,7} =\diag(\alpha \id_3, \alpha^2 \id_6)
\eeq
The local structure at the origin is of the type considered in section
3.3, and will lead to a three-generation $SU(3)\times SU(2)_L\times
U(1)_Y$ subsector. In order to break the $U(3)\times U(6)$ D7$_1$-brane
gauge group, we add a Wilson line along the second complex plane, with
\beq
\gamma_{W,7}=\diag( \id_1 , \alpha \id_1 , \alpha ^2 \id_1 , \id_2 ,
\alpha \id_2 , \alpha ^2 \id_2 )
\eeq
which break the D7$_1$-brane gauge group to $U(2)^3\times U(1)^3$. Again,
the fixed points $(0,\pm 1 , p)$, $p=0,\pm 1$, feel the Wilson lines and
receive no contribution to twisted tadpoles, hence do not require the
presence of D3-branes. On the other hand the fixed points $(0,0, \pm 1)$
require additional D3-branes for tadpole cancellation. The simplest option
is to add three 3-branes at each of these with 
$\gamma_{\theta,3}=\diag(\id_2,\alpha \id_1)$.

Let us add nine $\Dsb_1$-branes at e.g. $Y_1=+1$, with 
\beqa
\gamma_{\theta,{\ov 7}} = \diag(\id_3, \alpha \id_3, \alpha ^2 \id_3).
\eeqa
We also add a Wilson line $W'$ along the second complex plane, with
embedding
\beqa
\gamma_{W',{\ov 7}} =\diag(\id_3,1,\alpha,\alpha^2,1,\alpha,\alpha^2).
\eeqa
Twisted tadpoles are generated only at points of the form $(+1,\pm 1,p)$,
$p=0,\pm 1$. They can be cancelled by adding two $\Dtb$-branes with
$\gamma_{\theta, {\ov 3}} = \diag(\alpha,\alpha^2)$ at each. This
completes the model, which satisfies the requirements of twisted and
untwisted tadpole cancellation.

\subsubsection{Trapping of Branes through Discrete Torsion}

As noted in section 3.6,  orbifold models with discrete torsion may lead
to partial trapping of D3-branes along certain directions. In this section
we present a model with D3-branes and $\Dtb$-branes where this partial
trapping is used to avoid brane-antibrane annihilation.

Consider the compact orbifold $T^6/(\IZ_3\times \IZ_2\times \IZ_2)$, with 
discrete torsion in the $\IZ_2\times \IZ_2$ factor. We place a set of
twelve D3-branes at the origin $(0,0,0)$, with Chan-Paton embedding
\footnote{Actually, one should specify the embeddings of the $\IZ_2$
twists. They are given, up to phases, by the choice (\ref{cpdistor}), and
we do not give them explicitly to simplify the discussion.}  
\beqa
\gamma_{\theta,3} = \diag(\id_6, \alpha \id_4, \alpha \id_2)
\eeqa
As discussed in appendix A, the additional $\IZ_2\times \IZ_2$ twist
breaks the gauge group to $U(3)\times U(2)\times U(1)$. In order to
ensure cancellation of twisted tadpoles at the origin, we introduce
D7$_1$-branes at $Y_1=0$, with Chan-Paton embedding
\beqa
\gamma_{\theta,7_1} = \diag(\alpha \id_6, \alpha^2 \id_{12})
\eeqa
The local structure around the origin is precisely that of the SM theories
in section 3.6.1, hence the final model will contain a three-generation
$SU(3)\times SU(2)_L\times U(1)_Y$ sector. 

Due to the $\IZ_2\times \IZ_2$ twist, the D7$_1$-brane gauge group would
be $U(3)\times U(6)$. To reduce it further, we introduce a Wilson line
$W_2$ along the second complex plane, embedded as
\beqa
\gamma_{W_2,7} = \diag(\id_2,\alpha \id_2, \alpha^2 \id_2 ; 
\id_4,\alpha \id_4,
\alpha^2 \id_4 )
\eeqa
The surviving D7$_1$-brane gauge group is $U(1)^3\times U(2)^3$. The
D7$_1$-brane contribution to twisted tadpoles at fixed points
$(0,\pm 1, p)$ for $p=0,\pm 1$, vanishes, hence no D3-branes are required
at them. At fixed points of the form $(0,0,\pm 1)$, tadpoles generated by
the D7$_1$-branes can be cancelled by placing D3-branes with 
$\gamma_{\theta,3} = \diag(\id_4 , \alpha \id_2)$. Notice that these
D3-branes are stuck at the fixed points.

\begin{figure}
\begin{center}
\centering
\epsfysize=8cm
\leavevmode
\epsfbox{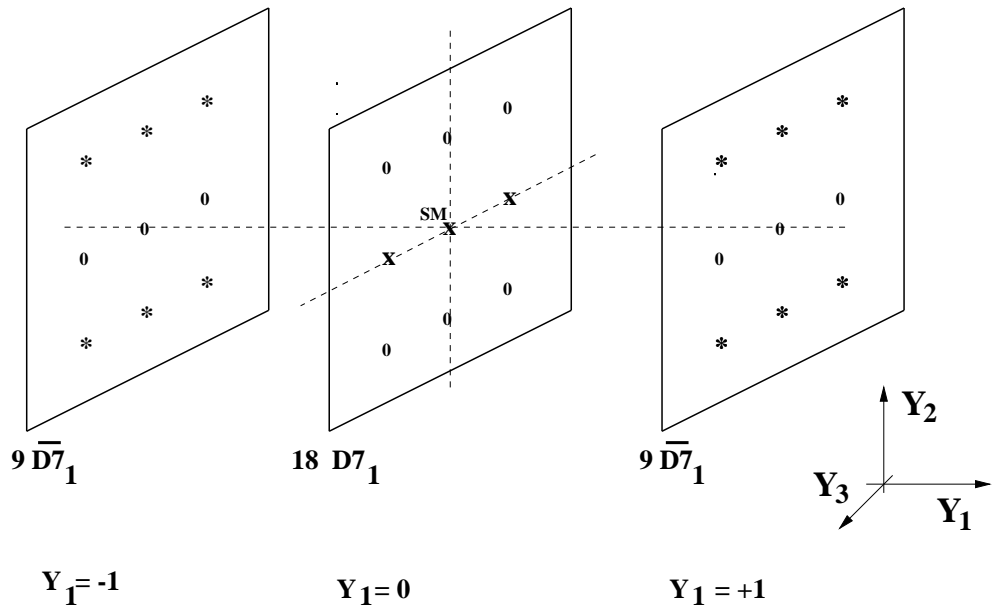}
\end{center}
\caption[]{\small A compact $\IZ_3\times \IZ_2\times \IZ_2$ orbifold
model with discrete torsion. Twelve D3-branes reside at the origin giving
rise to the SM. In addition there are six D3-branes at the points marked
$(x)$ and two anti-D3-branes at those marked $(*)$. The D3-branes in the
SM cannot travel to the bulk and annihilate the anti-D3-branes because they 
can only move along the straight lines shown.}
\label{fig5}
\end{figure}

So far we have placed twelve D3-branes at $(0,0,0)$ and six D3-branes at 
each of the points $(0,0,\pm 1)$. Also, there are eighteen D7$_1$-branes
at $Y_1=0$. In order to cancel the D7$_1$-brane untwisted tadpole, we
introduce nine anti-D7$_1$-branes at $Y_1=\pm 1$, with 
\beqa
\gamma_{\theta,{\ov 7}} = \diag(\id_3, \alpha \id_3, \alpha ^2 \id_3).
\eeqa
We also add a Wilson line $W'$ along the second complex plane, with
embedding
\beqa
\gamma_{W_2',{\ov 7}} = \diag(\id_3,1,\alpha,\alpha^2,1,\alpha,\alpha^2).
\eeqa
They induce no tadpoles at the points of the form $(+1,0,p)$, $p=0,\pm1$.
The non-vanishing tadpoles at the points of the form $(+1,\pm 1,p)$,
$p=0,\pm 1$, can be cancelled by adding two anti-D3-branes at each, with
$\gamma_{\theta, {\ov 3}} = \diag(\alpha,\alpha^2)$. The $\Dsb_1$-branes
at $(-1,n,p)$, with $n,p=0,\pm 1$, and the corresponding $\Dtb$-branes
are mapped to the above by the $\IZ_2\times \IZ_2$ operations, hence their
embedding is determined by the above. The final configuration contains 18
D7$_1$-branes and 18 $\Dsb_1$-branes, as well as 24 D3-branes and 24
$\Dtb$-branes, and is shown in figure~\ref{fig5}. All RR charges are
suitably cancelled.

One important virtue of this model (and others of its kind) is that all
$\Dtb$-branes are stuck at fixed points, and so are most D3-branes. In
fact, the only D3-branes not completely trapped are those in the SM
sector at the origin $(0,0,0)$. As explained in section~3.6.1, they are
however partially trapped, being able to move away at most in only one
complex direction. Hence, they can never reach the points $(\pm 1, \pm 1,
p)$ where anti-D3-branes sit, and brane-antibrane annihilation is
therefore not possible. This example illustrates how to employ partial
trapping by discrete torsion to improve the stability properties of the
models.

\subsection{Inside a Compact Type IIB Orientifold}

In this section we will present an explicit type IIB orientifold model
with the following properties:
\begin{enumerate}
\item{}The standard model is embedded on a D3-brane at a $\IC^3/\IZ_3$
singularity.
\item{}There are no anti-D3-branes that could destabilize the
system.
\item{}There are equal number of D7-branes and anti-D7-branes,
breaking supersymmetry, trapped at different $\IZ_3$ fixed points,
providing a clean example of gravity mediated supersymmetry breaking.
\end{enumerate}
Models of this type are similar to the ones constructed in \cite{aiq1,aiq2} 
(based on the general framework in \cite{au}, see also \cite{ads}),
but in those cases the standard model was inside a higher dimensional
brane (D7 or D9-brane) and supersymmetry was broken by the presence of
lower dimensional anti-branes ($\Dtb$- or $\Dfb$-branes). Furthermore
those examples did not include models where all the anti-branes were
completely trapped, and it was left as a dynamical question if the
configuration was stable. The class of models we are going to present 
succeeds in trapping all antibranes. 
 Nevertheless, it is well known that $T$-dualizing these
models with respect to the different extra dimensions we should get to 
models where the standard model is embedded into D7 or D9-branes and
therefore models like those considered in \cite{aiq1,aiq2}. We will
then find the $T$ duals of these models which happen to be complicated
versions of the models in \cite{aiq1,aiq2} in terms of several Wilson
lines. We will submit the reader to the appendices where we discuss in
detail the $T$ duality between the models of this section and those of
\cite{aiq1}.

\subsubsection{Explicit Models}

As an illustration to this class of models let us consider in detail
the following orientifold of the $T^6/\IZ_3$ orbifold. The orientifold
action is $\Omega (-1)^{F_L}\ R_1\  R_2\ R_3$ with $R_i$ reflection on the
$i^{th}$ plane. There are 64 orientifold three-planes (O3-planes), which
are localized at points in the internal space. To cancel their RR charge
we need a total of 32 D3 branes. We will distribute them among the 27 
orbifold fixed points $(m,n,p)$, $m,n,p=0,\pm 1$. We will also introduce an
equal number of D7 and $\Dsb$-branes. Among these 27 points, only the
origin $(0,0,0)$ is fixed under the orientifold action, hence it is an
orientifold singularity, of the type mentioned in section 3.6.4. The
cancellation of tadpoles at this point requires 
\beq
3\ \Tr\gamma_{\theta,3} + (\Tr\gamma_{\theta,7}
- \Tr\gamma_{\theta,{\bar{7}}}) = -12
\label{tadpoin}
\eeq
 As shown in appendix C, this orientifold singularity is not suitable
 to incorporate the standard model.  Fortunately, $\IZ_3$ fixed points other than the origin
are not fixed under the orientifold action, and therefore are
$\IC^3/\IZ_3$ {\em orbifold} singularities, which we may employ to embed
local configurations as those in section 3.3. 

The strategy is then to concentrate at an orbifold point different
from the origin, say $(0,1,0)$ where we want to have the standard model
(its $\IZ_2$ mirror image $(0, -1, 0)$ will have identical spectrum).
We choose a suitable twist matrix for the D3 branes such as that in
section 3.3. At these points the tadpole condition is simply:
\beq
\Tr\gamma_{\theta,7}-\Tr\gamma_{\theta,{\bar 7}}+3\Tr\gamma_{\theta,3}=0
\label{tadpoout}
\eeq
\bigskip
The challenge to construct a model is to be able to accommodate the 32
D3-branes among the different orbifold/orientifold points such that the
corresponding twist matrices satisfy the tadpole conditions (\ref{tadpoin}), 
(\ref{tadpoout}) and of course having the same number of D7 and 
$\Dsb$-branes. The other degree of freedom we have is the possibility of
adding Wilson lines. They play an important role because at a given plane,
a Wilson line differentiates among the different orbifold fixed points.
\bigskip

\noindent{\bf A Left-Right Symmetric Model}
\bigskip

We can satisfy the tadpole condition (\ref{tadpoin}) 
in many ways, but for future convenience we
choose having eight D3-branes at the origin with 
\beq
\gamma_{\theta,3}\ = \ \diag \left(\alpha \id_4, 
\alpha^2 \id_4\right)
\label{cporientone}
\eeq
Let us now discuss the rest of the points $(0,n,p)$.  Since these are
only orbifold 
points the condition for tadpole cancellation is (\ref{tadpoout}).
If we want to have a standard-like model at a D3-brane on one of these
points we must also have to introduce the D7 or $\Dsb$-branes. Let us
introduce D7$_1$-branes at $Y_1=0$.
In order not to alter the tadpole cancellation at the point $(0,0,0)$ the 
twist matrix for the 7-branes has to be traceless, so we take six 
D7$_1$-branes with:
\beq
\gamma_{\theta,7}\ =\ \diag\left(\id_2, \alpha\ \id_2, \alpha^2\ \id_2\right)
\label{cporienttwo}
\eeq
In order to have nontrivial impact in the other points $(0, n, p)$ we add
a Wilson line $\gamma_W$ on the second complex plane. We choose
\beq
\gamma_{W,7}\ = \ \diag\left(\alpha, \alpha^2, \id_2, \id_2\right)
\label{cporientthree}
\eeq
The addition of this Wilson line implies that $\Tr\gamma_{\theta,7}
\gamma_W=\Tr\gamma_{\theta,7}\gamma_W^2=-3$ so we need to have D3-branes
only at the points $(0,n,p)$ with $n\neq 0$ with twist matrix satisfying
$\Tr\gamma_{\theta,3}=1$. We achieve this at each of the points $(0,\pm 1,
0)$ by adding seven D3 branes with twist matrix
\beq
\gamma_{\theta,3}\ = \diag \left( \id_3, \alpha\  \id_2, \alpha^2\
\id_2\right)
\label{cporientfour}
\eeq
The local structure at these points is like that of left-right models
 in section
3.3. Hence they will lead to a $SU(3)\times SU(2)_L\times SU(2)_R$ gauge
group with three quark-lepton generations.

Next consider the points $(0, \pm 1,\pm 1)$. We can easily cancel
tadpoles by introducing one single D3 brane at each of the four points,
with $\gamma_{\theta,3}=1$ 

\begin{figure}
\begin{center}
\centering 
\epsfysize=8cm
\leavevmode
\epsfbox{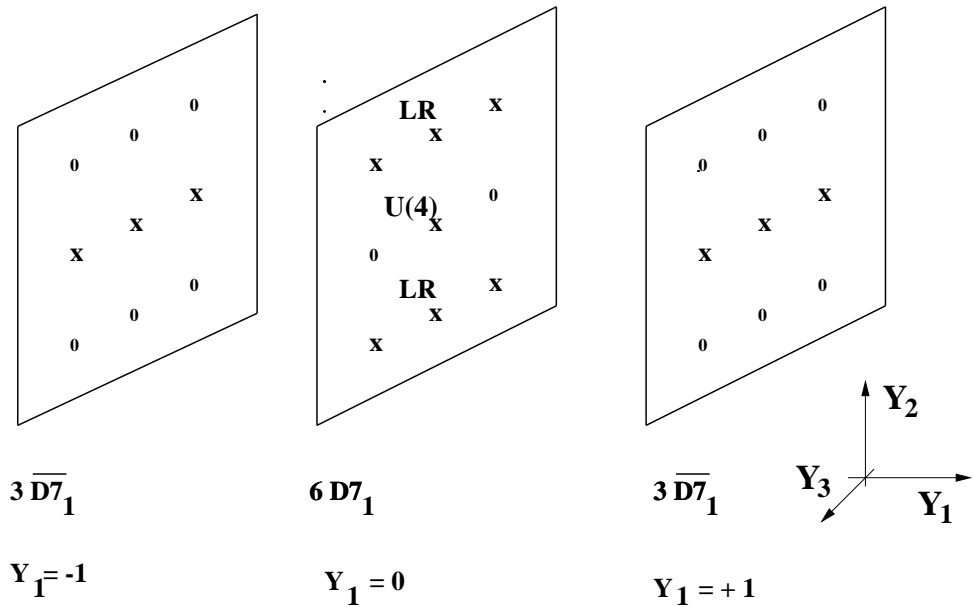}
\end{center}
\caption[]{\small A compact $\IZ_3$ orientifold model with no anti-D3-branes.
A left-right symmetric sector is located at the fixed point $(0,1,0)$
and its orientifold mirror. There is a single D3-brane at each of the fixed
points marked $(x)$. At the origin, which is an orientifold point, there
are eight D3-branes giving rise to a $U(4)$ gauge group. The
anti-D7-branes are trapped and cannot annihilate the D7-branes.}
\end{figure}

Finally we have to consider the points $(m,n,p)$ with $m\neq 0$. So far we
have six D7$_1$-branes and $8+7+7+4=26$ D3-branes. Therefore we need to
introduce six more D3 branes to complete 32, and six $\Dsb$-branes. We can 
easily achieve this by locating three $\Dsb_1$-branes at $Y_1=\pm 1$
with twist matrix and Wilson lines given by
\beq
\gamma_{\theta,{\bar 7}}\ =\ \id_3\qquad\qquad \gamma_{{\bar W}{\bar 7}}\
=\ \left(1,\alpha, \alpha^2 \right)
\eeq 
We then have $\Tr\gamma_{\theta,{\bar 7}}\gamma_{{\bar W},{\bar 7}} =
\Tr\gamma_{\theta,{\bar 7}}\gamma_{{\bar W},{\bar 7}}^{2}=0$ and do not
need to add D3 branes at the points $(1, \pm 1, p)$. At the points $(1,0,p)$ 
we can put a single D3-brane with $\gamma_{\theta,3}=1$, canceling the
tadpoles from $\gamma_{\theta,\bar{7}}$. At the points $(-1,n,p)$ we have
a similar structure, consistently with  the orientifold projection. In
total, we have added precisely the six $\Dsb$-branes and six D3-branes, as
needed for consistency.

Since this model provides an interesting example of a realistic
compactification with gravity mediated supersymmetry breaking, we give
some details on the computation of the spectrum in the visible sector. 

The D7-branes, with Chan-Paton embedding (\ref{cporienttwo}) would
(before the orientifold projection) lead to a group $U(2)^3$ with matter
in three copies of $(2,2,1)+(1,2,2 ) + (2 ,1, 2)$. The Wilson line
(\ref{cporientthree}) breaks the gauge symmetry to $U(1)^2\times U(2)\times
U(2)$ with matter in three copies of $(2, \bar 2)$ with zero charge under
the $U(1)^2$. A simple way to read the spectrum is by using the shift
vectors defined in the Cartan-Weyl basis (in analogy with \cite{afiv})
for the twist matrix and Wilson line which in this case are:
\beqa
V_{7_0}\ &  =\ & \frac{1}{3}\left(0,0,1,1,-1,-1\right)\nonumber \\
W_{7_0}\ &  =\ &  \frac{1}{3}\left(1,1, 0, 0, 0, 0\right)
\eeqa
We keep states of the form $P=(1,-1,0,...)$ (and permutations) satisfying
$P V=1/3$ and $P W=0$. The orientifold projection identifies two pairs of
groups and further breaks the symmetry to $U(1)\times U(2)$ with the
simple matter consisting of three singlets (antisymmetrics of $SU(2)$)
with charge +2 under the $U(1)$ inside $U(2)$.

The D3-branes at the origin, with Chan-Paton embedding (\ref{cporientone}) 
lead, before the orientifold projection, to an $U(4)\times U(4)$ gauge
group with matter on three copies of $(4,\bar {4})$. After the orientifold 
projection this becomes a single $U(4)$ group with matter on three copies
of a {\bf 6}. For states in the {\bf 37} sector for D3-branes at the
origin, we use the extended shift  vector
\beq
\tilde{V}\ =\ \frac{1}{3}\ \left(1, 1, 1, 1, -1, -1,  -1,
-1\right)\otimes \ V_{7_0}
\eeq
with the first eight  entries corresponding to the original $U(4)^2$ on the 
D3-branes. We then keep the vectors \footnote{The {\bf 73} sector would
have the opposite relative signs between the D7 and D3 branes; but it is
identified to the {\bf 37} sector by the orientifold projection and need
not be considered independently.} $P=(-1,0,\cdots;1,0,\cdots)$ satisfying
$P \tilde{V}=1/3$. We obtain the matter fields transforming as
$(4,2)_0+(\bar 4,1)_{1}+(\bar 4,1)_{-1}$ under the final $U(4)\times
U(2)\times U(1)$.
 
For D3-branes at $(0,\pm 1,0)$ the Chan-Paton twist is (\ref{cporientfour}),
we obtain the spectrum of the LR model. The gauge group is $U(3)\times
U(2)\times U(2)$, with two anomalous $U(1)$'s being actually massive, and
the diagonal combination giving $\BL$. In the 33 sector, we obtain
matter fields 

{\bf 33} sector:
\beq
3\ \left[ (3,\bar 2 , 1)+ (1, 2, \bar 2 )
+ (\bar 3 , 1, 2)\right]
\eeq
These can be explicitly seen by using the effective shift vector
$V_3= \frac{1}{3} \left( 0, 0, 0, 1, 1, -1, -1\right)$.

For matter in {\bf 37}, {\bf 73} sectors, the gauge quantum numbers
are easily computed by considering the shift vectors $V_3\otimes (V_7\pm
W_7)$, and keeping the state vectors $P=(-1,0,\cdots 0)\otimes (1,0\cdots
,0)$ and permutations for the {\bf 37} sector, and the opposite sign for
the {\bf 73} sector. We get:

{\bf 37} sector:
\beq
(3, 1, 1;1)_{-1}+(3,1,1; 2)_0+(1,2,1;1)_{1}+ (1,2,1;2)_0
\eeq

{\bf 73} sector:
\beq
(\bar 3, 1,1; 1)_1 + (\bar 3,1,1; 2)_0 + (1,1,2;1)_1+(1,1,2;2)_0
\eeq
The orientifold projection map the sets of branes at $(0, 1, 0 )$ and $
(0, -1, 0)$ to each other, so only one copy of the LR model is obtained.
Notice that these sector contain some extra vector-like
chiral fields beyond those obtained in section 3.4. Six  pairs of
colour triplets will in general become massive once $(7_i7_i)$ 
states get vevs. The remaining extra vector-like fields beyond
the leptons transform like $(3,1,1)+(1,2,1)+(1,1,2)+h.c.$.
As discussed in chapter 6, the presence of extra $SU(2)_R$ doublets
is in fact welcome in order to give rise of the required
$SU(2)_R$ gauge symmetry breaking.

For D3-branes at fixed points $(0, \pm 1,\pm 1)$, with
$\gamma_{\theta,3}=1$, we have one  $U(1)$ gauge group, with no
33 matter, at each brane. The orientifold projection relates two
pairs of these points reducing the group to $U(1)^2$. The matter on each
of these two 37 sectors can be easily computed as above (shift
vector $0\otimes V_7$) and transforms as $1_{-1,1}+2_{0,1}$ under the
$\left(U(2)\times U(1)\right)\times U(1)$ (where the first $U(1)$ arises
from D7-branes and the second one from D3-branes). The 73 sector
gives multiplets $1_{-1,-1}+2_{0,-1}$.

\bigskip\vfill\eject
\noindent{\bf A Standard-like Model}
\bigskip

Let us now briefly present a model with a SM subsector. We put twelve
D7-branes at $Y_1=0$ and four D3-branes at the origin with
\beqa
\gamma_{\theta,3}=\diag(\alpha \id_2, \alpha^2 \id_2) \quad, \quad
\gamma_{\theta,7}=\diag(\alpha \id_6, \alpha^2 \id_6)
\eeqa
satisfying the tadpole condition (\ref{tadpoin}) at the origin. We also
introduce a Wilson line along the second plane, with action
\beqa
\gamma_{W,7}=\diag(1,\alpha,\alpha^2, \id_9)
\eeqa
The contribution to twisted tadpoles for the orbifold fixed points $(0,\pm
1, 0)$ is given by $\tr\gamma_7\gamma_W=\tr\gamma_7\gamma_W^2=6\alpha^2+
3\alpha$. It allows us to put six D3-branes at each of the two points
with 
\beqa
\gamma_{\theta,3}=\diag(\id_3,\alpha \id_2, \alpha^2 \id_1)
\eeqa
The structure at these points is of the SM type considered in section 3.3
(it differs from it by an irrelevant overall phase in the Chan-Paton
embeddings). Hence we obtain one three-generation SM sector. 

At each of the points $(0,\pm1,\pm 1)$, we cancel the tadpoles by placing
three D3-branes with $\gamma_3=(\id_2,\alpha I_1)$
Finally, we put two D3-branes with $\gamma_3=\id_2$ at each of the points
$(0,0,\pm 1)$. The total number of D3-branes we have introduced is
$4+6+6+12+4=32$, and there are also twelve D7-branes. 

The twelve $\Dsb$-branes required to cancel the untwisted tadpole
introduced by the D7-branes, could be evenly distributed at the positions
$Y_1=\pm 1$, with $\gamma_{\theta,\bar 7}=(\id_2,\alpha \id_2, \alpha^2
\id_2)$. Being traceless, this embedding does not require to introduce
more D3-branes nor Wilson lines. Unfortunately, these $\Dsb$-branes can
move off the fixed points into the bulk, and annihilate partially against
the D7-branes at the origin. The result of such process is not necessarily
disastrous for the SM sector we had constructed. In particular, it leads
to a local structure of the type studied in section 3.6.5, leading to a
non-supersymmetric SM with three generations. Hence the model would 
provide a realistic non-supersymmetric model, but necessarily requiring a
TeV string scale.

Alternatively, in order to avoid the partial annihilation of  branes and
anti-branes and obtain a gravity mediated supersymmetry breaking scenario
we can modify the model in  a simple way. Let us add, for instance, a
second Wilson line, now in the third plane, acting on the D7-branes with 
\beq
\gamma_{W',7}\ =\ \left(\id_3; 1,\alpha, \alpha^2, 1, \alpha, \alpha^2,
1, \alpha, \alpha^2 \right)
\eeq
This Wilson line has no effect on the points $(0, \pm 1, 0)$. Since
$\Tr\gamma_{\theta,7}\gamma_{W',7}^n= 3\alpha$ for $n=1,2$, we need to add
two D3 branes at the points $(0, 0, \pm 1)$ with $\gamma_{\theta,3}=
\diag(1, \alpha^2)$ to cancel tadpoles. However, since
$\Tr\gamma_{\theta,7}\gamma_{W,7}^m\gamma_{W',7}^n =0$ for $m,n=1,2$, we
do not need to add any D3-brane at the points $(0,\pm 1,\pm 1)$. Therefore 
the remaining 12 D3-branes must sit at some of the fixed points $(\pm 1,
m, n)$. We can achieve that having at each of the two planes $Y_1=\pm 1$
six $\Dsb_1$-branes with $\gamma_{\theta,{\ov 7}}= (\alpha\id_3, \alpha^2
\id_3)$ with one Wilson line in the second plane with $\gamma_{W,\ov
7}=(1,\alpha, \alpha^2, 1, \alpha, \alpha^2)$ which precisely requires two
D3 branes at each of the three fixed points points $(1, 0, m)$ with
$\gamma_3=(\alpha, \alpha^2)$ to cancel tadpoles. Putting a similar
distribution at the plane $Y_1=-1$, we end up with a model in which the
anti-branes are also trapped, they cannot leave the $Y_1=\pm 1$ planes.
The modulus parameterizing this motion has been removed by the Wilson line
We have then succeeded in constructing a standard-like model with
antibranes trapped at a hidden sector located at $Y_1=\pm 1$ without
any danger to annihilate with the D7 branes of the visible sector, so
it is a genuine gravity-mediated scenario.

\subsection{Inside a Calabi-Yau}

Since our local singularities reproducing the semirealistic spectra 
in Section 3 contain D7-branes, the general framework to discuss compact
models is F-theory \cite{vafafth}. F-theory describes compactification of
type IIB string theory on curved manifolds in the presence of seven-branes.
Cancellation of magnetic charge under the RR axion implies the compact
models should in general include a set of $(p,q)$ seven-branes, not
mutually local, and around which the type IIB complex coupling constant
$\tau$ suffers non-trivial $SL(2,\IZ)$ monodromies. Compactification to
four dimensions on a three-complex dimensional manifold $B_3$, with a
set of $(p,q)$ seven-branes wrapped on two-complex dimensional
hypersurfaces in $B_3$, can be encoded in the geometry of a four-fold
$X_4$, elliptically fibered over $B_3$, with $(p,q)$ seven-branes
represented as the two-complex dimensional loci in $B_3$ over which the
elliptic fiber degenerates by shrinking a $(p,q)$ one-cycle. In order to
preserve $\NN=1$ supersymmetry in four dimensions, $X_4$ must be
Calabi-Yau, while $B_3$ in general is not.

An important feature of four-dimensional compactifications of F-theory is
that they generate a non-zero tadpole for the type IIB 4-form $A_4^+$
\cite{svw}, its value (in the absence of 3-form fluxes \cite{gflux}) being
given, in units of D3-brane charge, by $-\frac {1}{24}\chi(X_4)$, where
$\chi(X_4)$ is the Euler characteristic of $X_4$. This tadpole may be
cancelled by the addition of D3-branes (or instantons on the 7-brane
gauge bundles, see below).

Hence, F-theory compactifications contain the basic ingredients employed
in our local structures (namely D7-branes, D3-branes, and a non-trivial
geometry $B_3$) in a completely supersymmetric fashion. Clearly, local
structures of the type studied in Section~3 will appear embedded in
F-theory compactification where the base $B_3$ of the fourfold contains
$\IC^3/\IZ_3$ singularities. In the following we construct a particular
simple model containing a sector of D7-branes and D3-branes located at a
$\IC^3/\IZ_3$ singularity, and reproducing the spectrum of a LR model of
the type studied in Section 3.4.

\begin{figure}
\begin{center}
\centering
\epsfysize=9cm
\leavevmode
\epsfxsize= 5.5in
\epsfbox{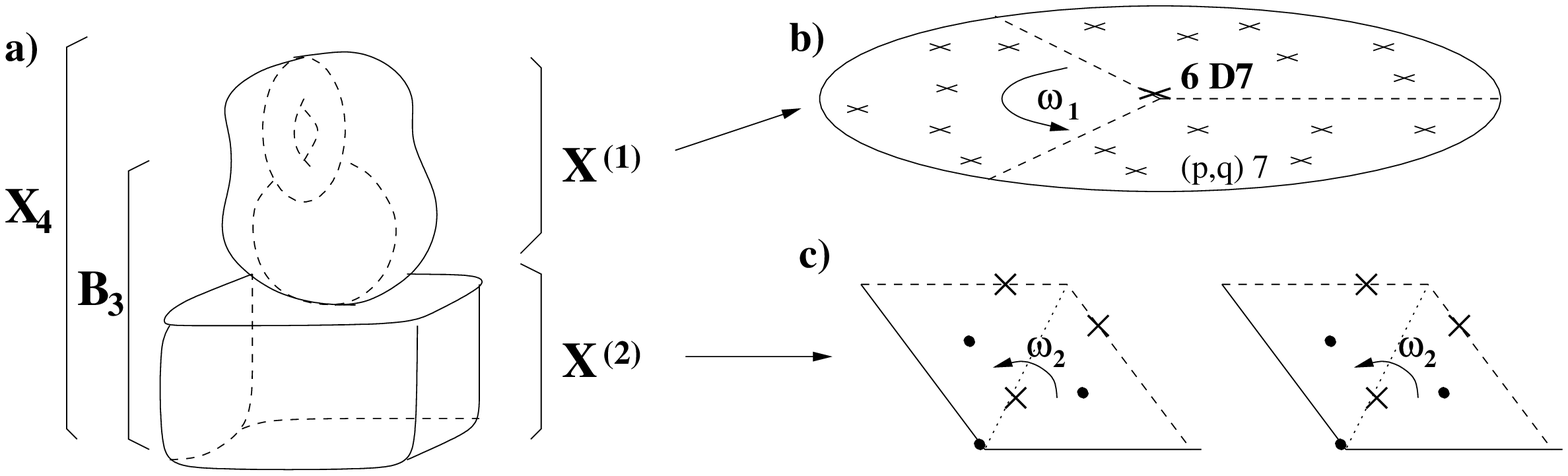}
\end{center}
\caption[]{\small The fourfold $X_4$ is elliptically fibered over $B_3$,
which is the product of $\IP_1$ times the K3 $X^{(2)}$. Figure b) depicts
the structure of the elliptic fibration over $\IP_1$, with crosses
denoting points at which the torus fiber degenerates. In type IIB
language, they correspond to seven-branes at points in $\IP_1$ and wrapped
over $X^{(2)}$. Six {\bf D}7-branes sit at a $\IZ_3$ invariant point,
while the remaining 18 $(p,q)$ seven-branes form a $\IZ_3$ symmetric
arrangement. Figure c) depicts the structure of $X^{(2)}$, which is a
$T^4/\IZ_3$ orbifold, with fixed points denoted by dots. The final
fourfold is obtained by quotienting $X_4$ by the combined $\IZ_3$ action
$\omega_1$ and $\omega_2$. Fixed points of $w_2$ in $X^{(2)}$ are denoted
as crosses in figure c). The model yields a three-generation LR model
arising from D3-branes sitting at the origin in $\IP_1$ and at one of the
$\omega_2$ fixed points in $X^{(2)}$.}
\label{fourfold}
\end{figure}

In order to keep the model tractable, we will take a simple fourfold as our
starting point, namely $X_4=X^{(1)}\times X^{(2)}$, where $X^{(1)}$ is an
elliptically fibered K3, and $X^{(2)}$ is also a K3, which we will take to
be a $T^4/\IZ_3$ orbifold. Unfortunately K3$\times$K3 cannot contain
$\IC^3/\IZ_3$ singularities, hence our final model will be a $\IZ_3$
quotient of $X_4$. The following construction of the model is pictured in
figure~\ref{fourfold}

Let us start by describing $X_4$. The K3 manifold $X^{(1)}$ is a fibration
of $T^2$ over a $\IP_1$, which can be defined in the Weierstrass form
\beqa
y^2 & = & x^3 + f_8(z,w) x + g_{12}(z,w)
\label{weierstrass}
\eeqa
where $x$, $y$ parameterize $T^2$, and $[z,w]$ are projective coordinates
in $\IP_1$. The functions $f_8$, $g_{12}$ are of degree $8$ and $12$
respectively in their arguments. The elliptic fibration degenerates at
the 24 points $[z,w]$ given by zeroes of the discriminant
\beqa
\delta_{24}& = & 4f_8(z,w)^3 + 27 g_{12}(z,w)^2
\eeqa
These degenerations signal the presence of 24 $(p,q)$ seven-branes sitting
at a point in $\IP_1$ and wrapping $X^{(2)}$ completely. We will be
interested in a particular family of K3 manifolds, containing six
coincident {\bf D}7-branes sitting at $z=0$. Geometrically, we require the
fibration (\ref{weierstrass}) to have an $I_6$ Kodaira type degeneration 
\cite{mv} at $z=0$, meaning $f_8$ and $g_{12}$ are non-zero at $z=0$, but
$\delta_{24}$ vanishes as $z^6$. We will be more explicit below.

For $X_4=K3\times K3$, $\chi(X_4)=24\times 24$, and the compactification
leads to an $A_4^+$-tadpole of $-24$ units. It will be useful to rederive
this result from a different point of view. On the
world-volume $M_7$ of each seven-brane (of any $(p,q)$ type) there exist
couplings 
\beqa
-\frac{1}{24} \int_{M_7} R\wedge R\wedge A_4^+ + \int_{M_7} F\wedge
F\wedge A_4^+
\label{coupl}
\eeqa
where $R$ is the Ricci tensor of the induced curvature, and $F$ is the
field strength tensor for the world-volume gauge fields. In the absence of
world-volume gauge instantons the second term drops, and the contribution
of the first for each seven-brane wrapped on the K3 $X^{(2)}$ (recall
$\chi(K3)=\int_{K3} R\wedge R=24$) is $-\int_{M_4} A_4^+$, where $M_4$
is four-dimensional spacetime. Namely, we get an $A_4^+$-tadpole of $-1$ 
for each of the 24 seven-branes in the model, hence a total tadpole of
$-24$. This tadpole can be cancelled by adding 24 D3-branes sitting at a
point in $B_3=\IP_1\times X^{(2)}$, or by turning on instantons on the
seven-brane gauge bundles.

We will be interested in considering $X^{(2)}$ in the $T^4/\IZ_3$ orbifold
limit. We start with $T^4$ parametrized by two coordinates $z_1$, $z_2$
with identifications $z_i\simeq z_i+1$, $z_i\simeq z_i+e^{2\pi i/3}$ (we
set the radii to unity for the sake of simplicity), and mod out by the
$\IZ_3$ action
\beqa
\theta: \quad z_1 \to e^{2\pi i/3} z_1 \quad, \quad z_2 \to e^{-2\pi i/3}
z_2
\label{actheta}
\eeqa
The quotient is a K3 manifold, flat everywhere except at the nine fixed
points of $\theta$ (located at points $(z_1,z_2)$ with $z_1$, $z_2$ = $0$,
$\frac 13(1+e^{\pi i/3})$ or $\frac 13(e^{\pi i/3} + e^{2\pi i/3})$),
which descend to singularities $\IC^2/\IZ_3$ in the quotient. The
curvature is concentrated at those points, and gives a contribution of
$\int_{C^2/Z_3} R\wedge R=8/3$ at each. Hence the $-1$ $A_4^+$-tadpole
at each seven-brane is split in nine $-1/9$ $A_4^+$-tadpole, arising from
the nine points $\IC^2/\IZ_3$ wrapped by the seven-brane world-volume.
There are no contributions from world-volume gauge instantons if the gauge
bundle over these $\IC^2/\IZ_3$ is trivial; in more familiar terms, if the
$\IZ_3$ action is embedded trivially on the seven-brane indices. Given
this $\theta$-embedding, one may worry about twisted tadpoles. However,
since the $\IZ_3$ twist leaves fixed the complete $\IP_1$, such tadpoles
receive contributions from all seven-branes, and they cancel by the same
reason the overall axion RR-charge cancels, namely by cancellation among
contributions of different $(p,q)$ seven-branes.

As mentioned above, in order to obtain $\IC^3/\IZ_3$ singularities we are
forced to consider a quotient of the four-fold $X_4$ considered
above, by $\IZ_3$, with generator $\omega$ acting simultaneously on
$\IP_1$ and $X^{(2)}$. We choose the action of $\omega$ on $\IP_1$ to be
given by 
\beqa
\omega_1: \quad  z \to  e^{2\pi i/3} z \quad ,\quad  w \to  w
\label{actomone}
\eeqa
For the configuration to be invariant under this action, the K3 $X^{(1)}$
must be given by (\ref{weierstrass}) with the functions $f_8$, $g_{12}$
depending only on $z^3$ 
\beqa
f_8(z,w) & = & A w^8 + B z^3 w^5 + C z^6 w^2 \nonumber \\
g_{12}(z,w) & = & D w^{12} + E z^3 w^9 + F z^6 w^6 + G z^9 w^3 + H z^{12}
\eeqa
The requirement that there are six {\bf D}7-branes at $z=0$ amounts to the
vanishing of the coefficients of order $1$ and $z^3$ in $\delta_{24}$
\beqa
4A^3+27D^2  =  0 \quad, \quad 12 A^2 B + 54 D E  =  0
\eeqa
It will be convenient to require no seven-branes to be present at
$[z,w]=[1,0]$, hence we require $H\neq 0$. The configuration we have
constructed therefore contains six {\bf D}7-branes at $z=0$, and the
remaining 18 seven-branes distributed on $\IP_1$ in a $\IZ_3$ symmetric
fashion. Notice that not only the locations, but the $(p,q)$ types of
these seven-branes are consistent with the symmetry, since the F-theory K3
$X^{(1)}$ is $\IZ_3$ invariant. Also, notice that no $(p,q)$ seven-branes
(other than the six at $z=0$) are fixed under $\omega$.

We have to specify also the action of $\omega$ on $X^{2}$, which we take
to be
\beqa
\omega_2:\;  z_1 \to  e^{2\pi i/3} z_1 + \frac 13 (1+e^{\pi i/3})
\quad, \quad
z_2 & \to  e^{2\pi i/3} z_2 + \frac 13 (1+e^{\pi i/3})
\label{actomtwo}
\eeqa
(The shifts have been included to avoid $\omega$-fixed points to coincide
with $\theta$-fixed points).
That this is a symmetry of $X^{(2)}=T^4/\IZ_3$ follows from the fact that
$\omega_2$ is a symmetry of $T^4$, and that it commutes with the
action $\theta$ (\ref{actheta}) required to form $T^4/\IZ_3$. The action
$\omega_2$ has nine fixed points in $T^4$, given by points $(z_1,z_2)$
with $z_i=\frac 13 e^{\pi i/3}$, $\frac 13(1+2e^{\pi i/3})$, $\frac
13(e^{2\pi i/3}+2e^{\pi i/3})$. The action of $\theta$ maps them to each
other so that in the quotient they yield just three singular points.
Similarly, $\omega_2$ maps the nine fixed points of $\theta$ to each
other, so that in the quotient they give rise to just three singularities.
Notice that in the final model the $T^4$ is modded out by $\theta$ and
$\omega_2$, and consequently by all other twists they generate. In this
respect it is important to notice that $\theta\omega_2$ and $\theta^2
\omega_2$ (and their inverses) are freely acting due to the shifts in
(\ref{actomtwo}), and do not generate new singularities.

The final model is obtained by quotienting $\IP_1\times X^{(2)}$ by
$\omega$, the combined action of (\ref{actomone}), (\ref{actomtwo}). One
can see that it preserves $\NN=1$ supersymmetry in the quotient by 
noticing that the holomorphic 4-form in $X_4$, given by $(dx/y)
(dz/w) dz_1 dz_2$ is invariant, hence the quotient is still a
four-fold. Notice that the fixed points of $\omega$ correspond to
$[z,w]=[0,1], [1,0]$ in $\IP_1$ times the three fixed points in $X^{(2)}$.

We must also specify the action of $\omega$ on the seven-branes. For the 
$(p,q)$ seven-branes at $z\neq 0$, the action is fully specified by the
geometrical action in $\omega_1$. The six {\bf D}7-branes at $z=0$,
sit at a fixed point of $\omega$, and hence may suffer a non-trivial
action on their Chan-Paton indices, which we choose to be given by
\beqa
\gamma_{\omega,7}=\diag(e^{2\pi i\frac 13} \id_3, e^{2\pi i\frac 23}
\id_3)
\label{cpfth}
\eeqa
This amounts to choosing a non-trivial $U(6)$ bundle on {\bf D}7-brane
world-volume. This will lead to a non-zero
contribution to world-volume instanton number, and hence influence the
computation of the $A_4^+$-tadpole in the quotient. Moreover, the
embedding (\ref{cpfth}) also leads to non-zero $\omega$-twisted tadpoles
which must be properly cancelled.

Let us turn to the computation of the $A_4^+$ tadpole. Since the $\omega$
action has fixed points, the Euler characteristic of the quotient
four-fold is {\em not} simply $\chi(X_4)/3$. Direct computation seems
rather involved, hence we prefer to compute the $A_4^+$ tadpole by using
the seven-brane world-volume couplings (\ref{coupl}). The 18 $(p,q)$
seven-branes at $z\neq 0$ are not invariant under $\omega$, hence their
world-volume $R\wedge R$ contribution comes only from the nine
$\theta$-fixed points, yielding a total $A_4^+$ tadpole of $-18$. This is
reduced to $-6$ by the $\omega$ action (equivalently, the 18 seven-branes
in $X_4$ descend to just six in the quotient by $\omega$). The six {\bf
D}7-branes at $z=0$ sit at a $\omega$ fixed point, hence their $R\wedge R$
coupling receives contributions from three $\IC^2/\IZ_3$ (from nine
$\theta$-fixed $\omega$-identified points), and three $\IC^2/\IZ_3$ (from
nine $\omega$-fixed $\theta$-identified points). The total contribution is
$-\frac{1}{24} \times \frac 83 \times 3 \times 2 \times 6 =-4$. There is a
further contribution, from the instanton number of the
nontrivial bundle implied by (\ref{cpfth}). This can be computed (see
\cite{intri,afiuv} for a discussion of instanton numbers for orbifold
spaces) to be $+1$ per $\IC^3/\IZ_3$ generated by $\omega$, yielding an
overall contribution $+3$ to the $A_4^+$ tadpole. Hence, the total
$A_4^+$ tadpole in the model is $-7$, and one must introduce $7$
D3-branes (as counted in the quotient) to achieve a consistent model.

Recall that the choice of the twist (\ref{cpfth}) generates non-zero
$\omega$-twisted tadpoles at the nine fixed points of $\omega_2$ (three in
the quotient), at $z=0$. Happily, these are $\IC^3/\IZ_3$ fixed points of
the type studied in section 3.2, for which the twisted tadpole condition
reads
\beqa
\Tr \gamma_{\omega,7} + 3 \Tr \gamma_{\omega,3} & = & 0
\eeqa
The tadpoles are easily cancelled by choosing $\gamma_{\omega,3}=(I_3,
e^{2\pi i\frac 13} I_2, e^{2\pi i\frac 23} I_2)$ at three of them (one
after the $\theta$-identification) and $\gamma_{\theta,3}=I_1$ at six (two
in the quotient by $\theta$). This employs 3 D3-branes (as counted in the
quotient) out of the 7 we had available. The remaining 4 can be placed at
a generic, smooth, point in the quotient (so that they would be described
by $4\times 3\times 3$ in the covering space of the $\theta$ and $\omega$
actions).

This concludes the construction of the model. We have succeeded in
constructing a $\NN=1$ supersymmetric compact model with a local
singularity of the form $\IC^3/\IZ_3$, on which D3-branes and D7-branes
sit. This local structure is of the kind considered in section 3.4, and
leads to a left-right symmetric model $SU(3)\times SU(2)\times
SU(2)\times U(1)_{B-L}$ with three generations of standard model
particles. Notice that in this concrete example the {\bf D}7-brane gauge
group is relatively large, but it should not be difficult to modify the 
construction (for instance, by adding Wilson lines on $T^4$) to reduce it.
We prefer not to complicate the discussion for the moment.

We would like to conclude with a remark. F-theory model building (see e.g.
\cite{ack}) has centered on embedding the standard model gauge group on
the seven-branes, while D3-branes are basically a useless sector. In our 
framework, the situation is inverted, with the D3-branes playing the main
role, even if seven-branes also contribute to the spectrum. One particular
advantage of our alternative embedding, which we have stressed throughout
the paper, is that the relevant sector can be determined by using just the
local behaviour of the compactification manifold. In this respect it is
clear that many other consistent supersymmetric embeddings of our local
structures may be achieved by considering more general four-folds with
$\IC^3/\IZ_3$ singularities on the base, on which D3-branes and D7-branes
(with gauge bundles locally reproducing the required monodromy) are
located. 

\section{ Some Phenomenological Aspects}

Obtaining  $SU(3)\times SU(2)_L\times U(1)_Y$ (or its left-right symmetric
extension $SU(3)\times SU(2)_L\times SU(2)_R\times U(1)_{B-L}$) and three
quark-lepton generations in a simple manner is already remarkable.
However we would like also to know if the class of models discussed in the
previous sections is sufficiently rich to accommodate other important
phenomenological features like gauge coupling unification, a hierarchical 
structure of quark/lepton masses and proton stability. In this chapter we
would like to address these issues. Let us make clear from the beginning
that we do not intend here to give a full account of these questions but
rather to evaluate the phenomenological potential of these models without 
compromising ourselves with a particular one.

We will limit ourselves to the case of models obtained from D-branes
at $\NN=1$ supersymmetric singularities. As already remarked above, some
properties like number of generations, gauge group and gauge coupling
normalization will  mostly be controlled by the local D-brane
configurations introduced in section 3.  On the other hand other
aspects like some Yukawa couplings or the $\NN=1$ Kahler metrics will be
in general dependent on the particular compact space chosen in order to
embed the realistic D-brane set. Thus, for example, the Kahler metric of
the matter fields will in general be different for an embedding inside a
compact $\IZ_3$ orbifold compared to an embedding inside an F-theory
model. Thus, the complete effective action will certainly depend on the
specific model.

We will try to discuss certain features which seem generic among the class
of models considered in the previous chapters. In particular, that is the
case of the question of gauge coupling unification. We will also study
some aspects of the Yukawa couplings both in the SM and LR type of models.
As a general conclusion we believe that the structure of the models seems
to be sufficiently rich to be able to accommodate the main phenomenological
patterns of the fermion mass spectrum. Finally, we comment on the 
structure of mass scales and in particular the value of the string scale
$M_s$ that should be considered in these constructions. Since this
depends on the compactification scale, the choice of $M_s$ is also model
dependent. 

\bigskip

\noindent{\bf i) Gauge Coupling Unification }

\bigskip 

This question obviously depends on whether we are considering a SM like
the one considered in section 3.3 or a left-right symmetric model as in
section 3.4. As we will see, from the gauge coupling unification point of
view the left-right models look more interesting. Let us consider first
the case of the SM. As mentioned in section 3, the tree-level gauge
coupling constants at the string scale $M_s$ are in the ratios $g_3^2 :
g_2^2 : g_1^2 =  1 : 1: 11/3 $. The one-loop running between the weak
scale $M_Z$ and the string scale $M_s$ is governed by the equations
\beqa
\sin^2\theta _W(M_Z)\ =&\ {3\over {14}}\left(1\ +\ {{11}\over {6\pi }}
\alpha(M_Z)\ (b_2-{3\over {11}}b_1)\ \log({{M_s}\over
{M_Z}})\ \right) \nonumber  \\
{1\over {\alpha _3(M_Z)}}\ =&\ {3\over
{14}}\left({1\over {\alpha (M_Z)}}\ -\ {1\over {2\pi }}\
\left(b_1+b_2-{{14}\over 3}b_3\right)\ \log\left({{M_s}\over {M_Z}}
\right)\ \right)
\label{senos}
\eeqa
where $b_i$ are the $\beta$-function one-loop coefficients. Notice that
the tree level result for $\sin^2\theta _W$$=3/14=0.215$ is only slightly
below the measured value at $M_Z$, $\sin^2\theta _W(M_Z)=0.231$. Thus in
order to get the correct sign for the one-loop correction we need to have
$b_2>3b_1/11$. With the massless spectrum in Table 1 we have $b_3=0$ ,
$b_2=3$, $b_1=15$, although as mentioned in section 3.3, generically some
three pairs of colour triplets will be heavy, leading to $b_3=-3$, $b_2=3$, 
$b_1=13$. In both cases the one-loop correction has the wrong sign and one
gets $\sin^2\theta_W(M_Z)=0.18$, $M_s=2.2\times 10^{15}$ GeV 
($\sin^2\theta_W(M_Z)=0.204$, $M_s= 10^{10}$ GeV) respectively. Thus 
standard logarithmic gauge coupling unification within the particular SM
configurations of section 3.3 does not seem to work. We cannot exclude,
however that a more sophisticated configuration of D-branes (yielding, 
in particular, some extra massless doublets) could work
\cite{aiq2,ross}.

Let us now move to the left-right model case. In this case the gauge
coupling unification question works remarkably well, as already pointed
out in ref.\cite{aiq2}. Indeed, the massless spectrum of the $SU(3)\times
SU(2)_L\times SU(2)_R\times U(1)_{B-L}$ model of section 3.4 is
essentially identical to the model considered in that reference. There are
now two regions for the running, $M_R<Q<M_s$, where the gauge group is the
left-right symmetric one and $M_Z<Q<M_R$ where the gauge group is the SM
one. Thus $M_R$ is the scale of breaking of the left-right symmetry. If
the $\beta$-function coefficients of the left-right gauge group are
denoted by $B_3,B_L,B_R$ and $B_{B-L}$, the one-loop running yields
\beqa
& \sin^2\theta _W(M_Z)\ =&\ {3\over {14}}\left(1\ +\ {{11\alpha _e
(M_Z)}\over {6\pi }}\ \left[\left(B_L-{3\over {11}}B_1'\right)\
 \log\left({{M_s}\over
{M_R}}\right) \  \right. \right.
\nonumber \\ & &   \left. \left. 
+\ \left(b_2-{3\over {11}}b_1\right)\ \log\left(
{{M_R}\over
{M_Z}}\right) \ \right]\right)
   \\
& { 1\over {\alpha _e(M_Z)} }\ -\  
{{14}\over {3\alpha _3(M_Z)}}\ =&\
 {1\over {2\pi }}\ \left[
\left(b_1+b_2-{{14}\over 3}b_3\right)\ \log\left({{M_R}\over
{M_Z}}\right) \
\right.
\nonumber \\
& & \left. +\
\left(B_1'+B_L-{{14}\over 3}B_3\right)\ \log\left({{M_s}\over
{M_R}}\right)\ 
\right]
\label{senostwo}
\eeqa
where one defines
\beq
B_1'\ =\ B_R\ + \ {1\over 4} B_{B-L}
\eeq
and $b_i$ are the $\beta$-function coefficients with respect to the SM 
group. For the generic (i.e., no extra  triplets) massless spectrum found
in section 3.4  one has $B_3=-3$, $B_L=B_R=+3$ and $B_{B-L}=16$. To get a
numerical idea let us assume that the left-right symmetry is broken not
far away from the weak scale
\footnote{This is in fact the most natural assumption if both
electroweak and $SU(2)_R$ symmetry breaking are  triggered 
by soft SUSY-breaking soft terms.}
, e.g. at  $M_R\propto $ 1 TeV, and that the
spectrum below that scale  is given by that of the MSSM (so that $b_3=-3$,
$b_2 =1$ and $b_1=11$). \footnote{A more general quantitative analysis for
different values of $M_R$ and SUSY-breaking soft terms  may be found in
ref.\cite{aiq2}.  The general agreement is in general found as long as
$M_R<1$ TeV or so.} Then one  finds the results:
\beq
\sin^2\theta _W(M_Z)\ = \ 0.231  \ \ ;\ \   M_s \ =\  9\times 10^{11} \
{\rm GeV}
\eeq
in remarkable agreement with low-energy data. This agreement requires 
$SU(2)_R$ to be broken not much above 1 TeV. In this connection, notice
that the massless spectrum of the left-right model in section 3.4 does not
have the required fields in order to do this breaking. However, 
slight variations like the explicit model in section 4.2.1 
do have extra $SU(2)_R$ doublets which can produce the breaking.
 On the other hand it is easy to check
that the presence of the additional vector-like 
 chiral fields in that model does not affect the
one-loop coupling unification and hence the agreement remains.

In summary, although logarithmic gauge coupling unification in the SM
would require some modification of the models, couplings unify nicely
(with equal precision than in the MSSM) in the case of the $SU(3)\times
SU(2)_L\times SU(2)_R\times U(1)_{B-L}$ models. In this case the string
scale coincides with the unification scale and should be  of order
$10^{12}$ GeV.

\bigskip

\noindent{\bf ii) Yukawa Couplings }

\bigskip

 Many phenomenological
issues depend on the Yukawa coupling structure of the models. At the
renormalizable level there are essentially three type of superpotential
couplings involving physical chiral fields : a) $(33)^3$ , b) 
$(33)(73)(37)$ and c) $(73)(37)(77)$. Whereas the couplings of types b)
and c) depend on the structure of D7-branes and hence are more sensitive
to global effects, that is not the case of couplings of type a) which
involve couplings among different $(33)$ chiral matter fields. Those are
expected to depend mostly in the structure of the orbifold singularity on
top of which the D3-branes reside. Indeed, for the more general case of
a $\IZ_3\times \IZ_M\times \IZ_M$ singularity in the presence of discrete
torsion (as in appendix A) the structure of those superpotentials has the
form:
\beqa
W & = & \sum_{i=0}^2\  \epsilon ^{abc}  [\; \Phi^a_{i,i+1}
\Phi^b_{i+1,i+2} \Phi^c_{i+2,i} ]
+ (1 - e^{2\pi i\frac 1M} )s^{abc}[ \Phi^a_{i,i+1} \Phi^b_{i+1,i+2}
\Phi^c_{i+2,i}\; ]
\eeqa
where (see equation (\ref{supdt}))
 $s^{abc}$ is such that $s^{321}=s^{132}=s^{213}=1$ and all other
components vanish.
 The gauge group is $U(n_0)\times U(n_1)\times
U(n_2)$ and the $\Phi^a_{i,j}$ are bifundamental representations
$(n_i,{\overline {n_j}})$. In the absence of discrete torsion the second
piece vanishes and the superpotentials are purely antisymmetric. In the
particular case of the SM, these include superpotential couplings
corresponding to Yukawa couplings giving masses to U-type quarks, i.e.,
couplings of type $h_{abc}Q_L^aU_R^bH^c$. The corresponding Yukawa
couplings have thus the form:
\beq
h_{abc}\ =\ \epsilon_{abc}\ +\ (1 - e^{2\pi i\frac 1M} )s_{abc}
\eeq
with $a,b,c=1,2,3$ are family indices. However, to make contact
with the low energy Yukawa couplings we have to recall that in a general
$\NN=1$ supergravity theory the kinetic terms of these chiral fields will
not be canonical but will be given by a a Kahler metric $K_{ab}$ which
will in general be a function of the compactification moduli. Consider, to
be specific, the case in which the compact case is a standard $\IZ_3$
toroidal 
orbifold or orientifold, like the examples shown in chapter 4. In this
case one has $K_{ab}= (T_{a{\bar b}}+T_{a{\bar b}}^*)^{-1}$, where
$T_{a{\bar b}}$ are the nine untwisted Kahler moduli of the $\IZ_3$
orbifold. Now, one can easily check that the low-energy physical Yukawa
couplings corresponding to canonically normalized standard model fields
$h_{abc}^0$ will be related with the supergravity base ones $h_{abc}$ as
\cite{bim} :
\beq
h_{abc}^0 \ =\ h_{lmn} \ \exp(K/2) \ (K_{al}K_{bm}K_{cn})^{-1/2}   
\eeq
where we have neglected a  complex phase irrelevant for our purposes and
$K$ denotes the full Kahler potential. The latter is given in the weak
coupling limit by the expression $K=-\log(S+S^*)-\log \  \det(T_{a{\bar
b}}+T_{a{\bar b}}^*)$, where $S$ is the complex dilaton field. Thus
altogether, the low energy physical Yukawa couplings of U-type quarks in
these models will have the general form:
\beq 
h_{abc}^0 \ =\ (S+S^*)^{-1/2}[ \  \epsilon_{abc} \ + \ (1 - e^{2\pi i\frac
1M})s^{lmn} \det(t_{jk})^{-1/2}  (t_{al}t_{bm}t_{cn})^{1/2} \ ]
\eeq     
where we define $t_{ab}=T_{a{\bar b}}+T_{a{\bar b}}^*$. Notice that the
Yukawa couplings are proportional to the gauge coupling constant $g$,
since $ReS= 2/\lambda$, $\lambda $ being the Type IIB dilaton. Also, 
in the absence of discrete torsion ($s^{abc}=0$) these
 Yukawa couplings
do not depend on the moduli. 

To see
whether this kind of structure has a chance to describe the observed
pattern of U-quark masses, let us consider the case in which only one of
the three Higgs fields, say $H_1$, eventually gets a vev of order the
weak scale, $\langle H_1 \rangle=v$. The U-quark mass matrix will be thus
given by $M^U_{bc}=vh^0_{1bc}$. Let us assume also that the moduli with
largest vevs are $t_{11}, t_{33}$ and the off-diagonal ones  $t_{23},
t_{32}$. Then the U-quark mass matrix would have the general form in
leading order:
 \begin{equation}
M^U_{bc}\  = \ (S+S^*)^{-1/2}v\  \left ( \begin{array}{ccc}
0 &  0 & 0  \\
0  & 0 &   h^0_{123} \\
0 & h^0_{132}  & h^0_{133}
\end{array}
\right )
\end{equation}
Plugging the above expressions for the Yukawa couplings one finds for
$|t_{33}|>>|t_{32}|$  three U-quark eigenstates with masses of order 
$gv(0, 1/\delta, \delta)$, with $\delta =|t_{33}/t_{32}|^{1/2}$, yielding
 a hierarchical structure for the U-quarks. Thus the structure of the
U-quarks Yukawa couplings for this class of models is in principle
sufficiently rich to accommodate the observed hierarchy, at least as long
as off-diagonal moduli have modest hierarchical vevs. In the particular
setting of toroidal $Z_3$ orbifolds discrete torsion seems also
to be required.

The same discussion applies directly to the case of the  left-right
symmetric model of section 3.4, the only difference being that in this
case the Yukawa couplings $h_{abc}Q_L^aQ_R^bH^c$ give masses both to
U-quarks and D-quarks. For both a hierarchical structure should be
possible to be accommodated.

The rest of the phenomenologically relevant renormalizable Yukawa
couplings are of type b) and c), i.e., they involve  directly the D7-brane
sectors and are thus more compactification dependent. Let us describe some
qualitative features for the SM and the LR models in turn.

In the case of the SM of section 3.3 (and its compact versions), there are
renormalizable couplings of type $(33)_i(7_i3)(37_i)$, where $(33)_i$
denotes the chiral fields in the $(33)$ sector associated to the i-th
complex plane. Looking at the quantum numbers in Table 1 one sees that in
particular there are couplings of type $(3,2)({\bar 3},1;2')(1,2;2')$
which provide Yukawa couplings for the D-quarks. One of the doublets 
inside $(1,2;2')$ should be identified with the normal left-handed leptons
and the other with Higgs fields. By definition, the doublet appearing in
the D-quark Yukawa coupling will be identified with the Higgs field. As
pointed out in the  footnote at the end of section 3.2,
 in generic D7-brane configurations fields
from a given $37$ sector can couple to $33$
fields from all three complex planes. Thus this flexibility will also in
general allow for a hierarchy of D-quark masses. Concerning lepton masses,
they do not appear at the renormalizable level since all left- and
right-handed leptons as well as the Higgs field belong to $(37_i)$ sectors
and there are no $(37)^3$ type of couplings on the disk. They may however
appear from non-renormalizable couplings of type $(37)^n$ in compact
models if some standard model singlets in some other $(37)$ sectors get a
vev. This will be very model dependent.

In the case of the left-right symmetric model of section 3.4 (and its
compact versions), the $(33)^3$ type of couplings already give masses to
both U- and D-quarks. Concerning leptons, all anomaly free gauge 
symmetries allow for couplings of type $(33)(37)(73)$ including terms
transforming like $(1,2,2)(1,2,1)(1,1,2)$ which would give rise to
standard Dirac masses for leptons. Looking at Table 2 one observes that
those couplings are in fact forbidden by some anomalous $U(1)$ charges.
Nevertheless one expects those couplings to be present once appropriate
twisted closed string insertions (which are charged under anomalous
$U(1)$'s) are made \cite{aiq2}. Again, for different geometrical
configurations of D7-branes there will be enough flexibility to obtain a
hierarchy of Yukawa couplings also for leptons.

In the case of the left-right model we also have to worry about neutrino
masses. Generically they will get Dirac masses of the order of the charged
lepton masses. However, once $SU(2)_R$ symmetry-breaking takes place
through the vevs of the extra $SU(2)_R$ doublets discussed above, the 
right-handed neutrinos may combine with some singlets in $(7_i7_j)$
sectors and get masses of order $M_R$. We will not give details here which
would be very model dependent. Let us note however that the effective
left-handed neutrino masses would thus be of order $m_l^2/M_R$, where
$m_l$ are the neutrino Dirac masses. For values of $m_l$ of order the
electron mass and $M_R$ of order 1 TeV, they would have masses of order 1
eV, compatible with experimental bounds.

Let us finally comment about proton stability in these schemes. This is an
important question if, as seems to be indicated  by gauge  coupling
unification, the string scale is chosen to be well below the Planck mass
(e.g. at scales of order $10^{12}$ GeV or below). Since for this question
higher dimensional operators are in principle important, it is not
possible to make model independent statements about sufficient proton
stability. We would like to underline however, that from this point of
view the left-right symmetric models seem more promising since the
symmetry $\BL$ is gauged and hence dimension four baryon or lepton
violating operators are forbidden to start with. Furthermore,
B/L-violating  dimension 5 operators like F-terms of the form $Q_LQ_LQ_LL$
(which respect $\BL$) are of type $(33)^3(73)$  and hence are forbidden on
the disk. In fact, it was shown in ref. \cite{aiq2}, that in a left-right model
analogous to the compact orientifold of chapter 5, the proton is
absolutely stable due to the presence of induced baryon and lepton parity
$\IZ_2$ symmetries. Thus we believe that, at least in the left-right class
of models, a sufficiently stable proton may naturally be achieved. 

\bigskip
 
\noindent{\bf iii) String Scale versus Planck Mass}
\bigskip

In our approach, four-dimensional gravity arises once we embed our SM or
LR D-brane systems into a finite compact space. The precise relationship
between the string scale $M_s$ and four-dimensional Planck scale $M_p$
will depend on the compactification. In toroidal orbifold 
compactifications with factorized two-tori they are related by 
\beq
M_p\ =\ { {2\sqrt{2} M_s^4} \over {\lambda M_1M_2M_3} }
\eeq
where $M_i$ are the compactification mass scales of the three tori. Thus,
as is well known \cite{witten,lykken,adv,aadv}, values for $M_s$ much below
the 
Planck scale may be possible if some of the compact dimensions are very
large. Until we get a handle on  the dynamics fixing the radii, the value
of the string scale is only constrained  by accelerator limits to be
$M_s> 1$ TeV or so. In particular, one can accommodate intermediate scale
values \cite{b,biq} of order $M_s\propto 10^{10}-10^{12}$ GeV, as
suggested by the logarithmic gauge coupling unification arguments above.

Similar qualitative statements can be made concerning the Planck versus
string scales in F-theory models with the SM sector living on the
world-volume of D3-branes, like those discussed in chapter 5. Thus, e.g.
one can match the observed size of the Planck scale by making very large
the volume of the base $P_1$ in the first K3 $X^{(1)}$, while maintaining
the volume of the second  K3, $X^{(2)}$ of order the string scale.

It is quite interesting to remark that such intermediate values for the
string scale turn out to be also suggested in a completely independent
manner in models \cite{aiq1,aiq2} 
containing anti-branes like the orientifolds in chapter
5. Indeed, in those models supersymmetry is broken at the string scale
$M_s$ due to the presence of anti-D7-branes. However the physical SM or LR
sectors are away from these antibranes in the $Y_1$ transverse dimension.
Let us suppose that e.g. one has $M_2,M_3=M_s$ and $M_1=M_s^2/(\lambda M_p
)$ so that one obtains the correct Planck scale. This means taking very
large $Y_1$ radii which will also mean that the susy-breaking effects from
the anti-D7-branes will be felt in a Planck mass suppressed manner in the
visible SM world. Thus one expects effective SUSY-breaking contributions
of order $M_s^2/M_p$ in the SM sector. If $M_s$ is of order 
$10^{10}-10^{12}$ GeV, then these SUSY-breaking contributions would be of
order the weak scale, as required. The structure of these models would be
very similar to hidden-sector supersymmetry-breaking models, the main
difference being that in the present case the string and SUSY-breaking
scales do coincide. 

Let us also comment that the approach here presented is also consistent 
with low (i.e. 1-10 TeV) string scale scenarios \cite{adv,aadv} , just by
choosing larger
compact radii. In these cases an alternative understanding of the gauge
coupling unification problem should be found. 

Finally the $F$-theory constructions of the previous section allows for
the possibility of a more standard scenario. Being supersymmetric, we
do not have to identify the string scale with the supersymmetry
breaking scale. We may in principle conceive models of this type for 
which the string scale is closer to the Planck or GUT
scales. The left-right symmetric model presented in that section
still requires unification at the intermediate scale but in general, 
achieving or not gauge coupling unification 
will clearly depend on the spectrum of each model.

\section{Final Comments and Outlook}

We have presented what we consider to be the first steps on a new way
to deal with string model building. The fact that D3-brane worlds can
appear naturally on type IIB string theory suggests that looking directly
to the physics on these D3-branes may tell us many interesting
phenomenological aspects that depend very weakly on the details of the
compactified space. This is so since only gravitational strength
interactions feel the extra dimensions. It would be interesting to extend
this bottom-up construction of realistic models to other string theories,
for instance using type 0B D3-branes on singularities \cite{typezero}, or 
to M-theory. The latter framework was explored in \cite{west}, where an
explicit M5-brane configuration led to a $N=2$ supersymmetric (and
therefore non-chiral) toy model with SM group and correct gauge
non-abelian and $U(1)$ quantum numbers for some of the standard model
fields.

It is remarkable how easy it is to  obtain realistic models from this 
approach and how powerful the structure of singularities turn out to
be in order to achieve interesting phenomenological properties, such
as the number of families, the ubiquity of hypercharge and the value of
Weinberg's angle. For instance a concrete prediction of the class of
models  considered is that the number of families does not exceed 3 for
supersymmetric models and 4 for non-supersymmetric ones. The reason being
essentially the  $\NN = 4$ symmetry underlying the {\bf 33} spectrum,
which limits the number of replicated fermions after the orbifold 
projection. This is a very small (and realistic) number compared with more
standard compactifications in which the number of families depends on the
topology of the compact manifold and is naturally very large.

It is worth remarking that the consideration of the non-compact models
constructed here may have an importance {\it per se}. At present there is
increasing interest on non-compact extra dimensions raised after the work
of Randall and Sundrum \cite{rs} . It would be tempting to speculate that
a similar mechanism to localize gravity on the D3-branes may be present in
models similar to ours. However it is not clear if this could be
achievable. Furthermore, during the past few years, non-compact brane
models were used in order to obtain information on the dynamics of gauge
theories from the structure of branes. Much work has been done in this
direction, \cite{giveon}, although most of the activity was based on
extended supersymmetric models and non-chiral $\NN=1$ models. Chiral
theories were constructed in \cite{branebox}, and seen to be related to
theories on D3-branes at singularities \cite{hu}. Our models extend this
program by addressing (and actually succeeding in) the construction of
realistic models using brane configurations.

Independent of these issues, the fact that in compact models we have been
able to obtain realistic $\NN=0$ and $\NN=1$ models from type IIB strings
at singularities may be the most important concrete result of this work.
We have explored several mechanisms and generalizations to current model
building, each of them deserving further study. The new approach, while
sharing some of the good properties of the constructions of
\cite{aiq1,aiq2} (gauge unification, supersymmetry breaking, proton
stability), has several clear advantages over previous discussions,
besides the fact that it is more general: from the stability of the
models, by trapping anti-branes avoiding brane/anti-brane annihilation, to
the inclusion of new degrees of freedom, such as discrete torsion, which
allow more flexibility on the structure of fermion masses and helps to
stabilize the models.  

A scenario with intermediate fundamental scale is clearly favoured by
the non supersymmetric models which can also have  gauge coupling
unification at the same scale, especially in the case of the left-right
symmetric models. On the other hand, the fact that we have also obtained
quasi-realistic $\NN=1$ supersymmetric models allows for the possibility
of realistic scenarios with  fundamental scales as large as the Planck or
GUT scales. It is worth pointing out however that, as usual in string
theory, grand unified gauge models do not seem to appear naturally in this
scheme, whereas the standard model and its left-right symmetric extension
are very easy to obtain.

There are clearly many avenues that deserve to be explored in the future,
starting from our approach. Probably the most pressing one being the
understanding of moduli and dilaton stabilization after supersymmetry
breaking and the corresponding value of the vacuum energy.
\bigskip
\bigskip

\centerline{\bf Acknowledgements}
\bigskip

We thank  M.~Cvetic, S.~Franco,  M.~Klein, P.~Langacker, J. Maldacena,
B.~Ovrut, M.~Pl\"umacher and R.~Rabad\'an for useful discussions. A.~M.~U.
thanks the Department of Applied Mathematics and Theoretical Physics of
Cambridge University, and the Department of Physics and Astronomy of
University of Pennsylvania for hospitality, and M.~Gonz\'alez for
encouragement and support. G.~A. thanks the High Energy  Theory Group of
Harvard University for hospitality. This work has been partially supported
by CICYT (Spain), the European Commission (grant ERBFMRX-CT96-0045)
and PPARC. G.A work is partially supported by ANPCyT grant 03-03403. 

\vfill\eject

\begin{center}
{\Large {\bf Appendices}}
\end{center}

\appendix

\section{Orbifold Singularities with Discrete Torsion}

Consider a $\IC^3/(\IZ_{M_1}\times \IZ_{M_2})$ orbifold singularity, with
the two factors of the orbifold group generated by twists $g_1$, $g_2$.
The effect of discrete torsion in the closed string sector was analyzed in
\cite{vafadt} (see also \cite{fiq,vw}). It amounts to the introduction of
an additional phase $\epsilon(g_1,g_2)$ in the action of $g_1$ on
$g_2$-twisted sector states. This phase must be an $s^{th}$ root of unity,
where $s={\rm gcd} (M_1,M_2)$. Hence discrete torsion modifies the twisted
sector spectrum, and introduces relative phases among the different
contributions to the torus partition function.

The effect of discrete torsion on open string sector, and consequently on
systems of D-branes at singularities, has been analyzed only recently
\cite{douglasdt,bereleigh} (see \cite{raul} for orientifold examples). In
the presence of discrete torsion the embedding of the orbifold twists in
the Chan-Paton indices forms a projective representation of the orbifold
group \cite{douglasdt}, i.e. the Chan-Paton embedding matrices
$\gamma_g$ obey the group law up to phases, concretely
\beqa
\gamma_{g_1}\gamma_{g_2}=\epsilon(g_1,g_2)\; \gamma_{g_2}\gamma_{g_1}
\label{dtopen}
\eeqa
The field theory on a set of D3-branes at such an orbifold singularity
with discrete torsion has been discussed in detail in
\cite{douglasdt,bereleigh}. In the following we center on a particular
case which illustrates the main features.

Consider a $\IC^3/(\IZ_N\times \IZ_M\times \IZ_M)$ singularity, where the
generator $\theta$ of $\IZ_N$ acts on $\IC^3$ through the twist
$(a_1,a_2,a_3)/N$, and the generators $\omega_1$, $\omega_2$ of the
$\IZ_M$'s act through the twists $v_1=(1,0,-1)/M$, $v_2=(0,1,-1)/M$
\footnote{In certain cases, for instance when $N$ and $M$ are coprime, the
group is actually $\IZ_{NM}\times \IZ_M$. However, it will be convenient
to make the $\IZ_M\times \IZ_M$ action more manifest.}. Let us
consider the case of discrete torsion $\epsilon(\omega_1,\omega_2)=
e^{-2\pi i/M}$ in the $\IZ_M\times \IZ_M$ part. Consequently we choose the
following embedding in the Chan-Paton indices
\beqa
\gamma_{\theta,3} & = & \diag(I_{Mn_0},e^{2\pi i\frac 1N} I_{Mn_1},\ldots,
e^{2\pi i\frac{N-1}{N}} I_{Mn_{N-1}}) \nonumber \\
\gamma_{\omega_1,3} & = & \diag(1,e^{2\pi i\frac 1M},\ldots,e^{2\pi
i\frac{M-1}{M}}) \otimes \diag(I_{n_0},I_{n_1},\ldots,I_{n_{N-1}})
\nonumber \\
\gamma_{\omega_2,3} & = & {\pmatrix{
 & 1 & & \cr
 &   & \ldots & \cr
 &   & & 1 \cr
1 &  & & }} \otimes \diag(I_{n_0},I_{n_1},\ldots,I_{n_{N-1}})
\label{cpdistor}
\eeqa
Notice that $\gamma_{\omega_1,3}\gamma_{\omega_2,3}=e^{-2\pi i\frac 1M}
\gamma_{\omega_2,3} \gamma_{\omega_1,3}$, in agreement with
(\ref{dtopen}).

The spectrum is obtained after imposing the projections
\beqa
\begin{array}{llll}
{\rm Vector} & \lambda=\gamma_{\theta,3} \lambda \gamma_{\theta,3}^{-1}
             & \lambda=\gamma_{\omega_1,3}\lambda\gamma_{\omega_1,3}^{-1}
             & \lambda=\gamma_{\omega_2,3}\lambda\gamma_{\omega_2,3}^{-1}
\cr
{\rm Chiral}_1 & \lambda= e^{2\pi i \frac {a_1}{N}}\gamma_{\theta,3}
\lambda \gamma_{\theta,3}^{-1}
             & \lambda= e^{2\pi i \frac 1M}\gamma_{\omega_1,3}\lambda
\gamma_{\omega_1,3}^{-1}
             & \lambda=\gamma_{\omega_2,3}\lambda\gamma_{\omega_2,3}^{-1}
\cr
{\rm Chiral}_2 & \lambda=e^{2\pi i\frac{a_2}{N}}\gamma_{\theta,3} \lambda 
\gamma_{\theta,3}^{-1}
             & \lambda=\gamma_{\omega_1,3}\lambda\gamma_{\omega_1,3}^{-1}
             & \lambda= e^{2\pi i\frac 1M}
\gamma_{\omega_2,3}\lambda\gamma_{\omega_2,3}^{-1}
\cr
{\rm Chiral}_3 & \lambda=e^{2\pi i\frac{a_3}{N}}\gamma_{\theta,3} \lambda 
\gamma_{\theta,3}^{-1}
             & \lambda= e^{-2\pi i\frac 1M}\gamma_{\omega_1,3}\lambda
\gamma_{\omega_1,3}^{-1}
             & \lambda= e^{-2\pi i\frac 1M}\gamma_{\omega_2,3}\lambda
\gamma_{\omega_2,3}^{-1}
\cr
\end{array}
\eeqa
The projection process is relatively simple. For vector multiplets, the
$\theta$ projection yields a gauge group $\prod_{i=0}^{N-1} U(Mn_i)$. The
$\omega_1$ projection splits it into $\prod_{i=0}^{N-1} U(n_1)^N$.
Finally, the $\omega_2$ action identifies the different gauge factors with 
equal rank, leaving a remaining gauge group $\prod_{i=0}^{N-1} U(n_i)$.
Proceeding analogously, the complete spectrum is 
\beqa
{\rm Vector} & \prod_{i=0}^{N-1} U(n_i) \nonumber\\
{\rm Chiral} & \sum_{r=1}^3 \sum_{i=0}^{N-1} (n_i,{\ov n}_{i+a_r})
\eeqa
Notice this is identical to the spectrum of the $\IC^3/\IZ_N$ singularity
with twist $(a_1,a_2,a_3)/N$, in section 2.1 or 3.1. However, the effect
of the additional $\IZ_M\times \IZ_M$ twist and of the discrete torsion
are manifest in the superpotential, which reads
\beqa
W & = &  Tr\ [ \Phi^1_{i,i+a_1} \Phi^2_{i+a_1,i+a_1+a_2}
\Phi^3_{i+a_1+a_2,i} - e^{-2\pi i\frac 1M} \Phi^1_{i,i+a_1}
\Phi^3_{i+a_1,i+a_1+a_3} \Phi^2_{i+a_1+a_3,i} ]
\label{supdt}
\eeqa
which differs from the {\bf 33} piece in (\ref{superpagain}).

We spare the reader the detailed discussion of the introduction of
D7-branes in the configuration. The resulting spectrum is identical to
that obtained for $\IZ_N$ singularities in Section~3.1. 

Notice that, since the twists $\omega_1$, $\omega_2$ have traceless
Chan-Paton embeddings, the corresponding disk tadpoles vanish. The only
constraints on the integers $n_i$ arise from cancellation of tadpoles
in $\IZ_N$ twisted sectors. The corresponding conditions are those for
$\IZ_N$ singularities (as expected, since they are basically the anomaly
cancellation conditions).

\section{Non-Abelian Orbifolds}

The rules to compute the spectrum of D3-branes at non-abelian orbifold
singularities have been discussed in \cite{jm} for $\Gamma\subset SU(2)$,
and generalized in \cite{lnv}. In the following we center on $\NN=1$
supersymmetric field theories in the case $\Gamma \subset SU(3)$, studied
in detail in \cite{hh,gremuto} (see \cite{nonabnonsusy} for the general case 
$\Gamma\subset SU(4)$).

The computation of the spectrum is just a natural group-theoretical
generalization of the abelian case. Let $\{ {\cal R}_i \}_{i=1,\ldots, N}$
denote the irreducible representations of $\Gamma$, with $N$ the number of
conjugacy classes of $\Gamma$. We are interested in the field theory on a
set of D3-branes at a $\IC^3/\Gamma$ singularity, where the action of
$\Gamma$ on $\IC^3$ is specified by a three-dimensional representation
${\cal R}^{({\bf 3})}$, and its action on the D3-brane Chan-Paton indices
is given by a representation ${\cal R}^{CP}$, which decomposes as
\beqa
{\cal R}^{CP} & = & \sum_{i=1}^N n_i {\cal R}_i
\eeqa
The field theory on D3-branes at $\IC^3/\Gamma$ is obtained by projecting
the field theory on D3-branes on $\IC^3$ onto states invariant under the
combined (geometric plus Chan-Paton) action of $\Gamma$. The resulting
gauge group \footnote{In the literature on D-branes on non-abelian
orbifold, the representation ${\cal R}^{CP}$ is usually taken to be $k$
copies of 
the regular representation, $n_i=k {\rm dim} \; {\cal R}_i$, which
satisfies tadpole cancellation conditions without the need of additional
branes. We will consider more general choices of $n_i$, with the
understanding that additional D7-branes will cancel the corresponding
tadpoles. Unfortunately, the inclusion of D7-branes has not been discussed
in the literature, hence we will not be explicit about the 37 sector in
our models, and simply assume it provides additional chiral multiplets to
cancel gauge theory anomalies from the 33 sector.} is $\prod_{i=1}^N
U(n_i)$. There are also $a^{\bf 3}_{ij}$ chiral multiplets in the
$(n_i,{\ov n}_j)$ representation, where the adjacency matrix $a^{\bf 3}_{ij}$
is defined by the decomposition
\beqa
{\cal R}^{({\bf 3})}\otimes {\cal R}_i & = & \sum_{j=1}^N a^{\bf 3}_{ij}
{\cal R}_j
\eeqa
The superpotential is obtained by substituting the surviving chiral
multiplets $\Phi_{ij}$ in the $\NN=4$ superpotential (\ref{supnfour}), 
but we will not need it explicitly.

An important observation is that the gauge couplings for the different
factors are not equal in field theories from non-abelian orbifolds. The
gauge coupling for the $i^{th}$ group is given by \cite{lnv}
\beqa
\tau_i={{r_i \tau}\over {|\Gamma|}}
\label{relcoupl}
\eeqa
with $r_i={\rm dim}\; {\cal R}_i$, and $\tau$ an overall value independent
of $i$.

The analysis of $U(1)$ anomalies has not been performed in the literature,
but can be easily generalized from the abelian case \footnote{We thank
A.~Hanany for conversations on this point.}. Assuming the 37, 73
sectors contribute a set of bifundamental multiplets which cancel the
non-abelian anomaly in the 33 sector, the mixed $Q_{n_i}$-$SU(n_j)^2$
anomaly reduces to $A_{ij}= \frac 12 n_i (a^{({\bf 3})}_{ij}- a^{({\bf
3})}_{ji})$. It is possible to check that the diagonal combination
\beqa
Q_{diag} & = & \sum_{i=1}^N {{r_i} \over n_i} Q_{n_i}
\label{qdiagnonab}
\eeqa
is automatically non-anomalous. Also, as long as no $n_i$ vanishes, there
is one additional non-anomalous $U(1)$'s for each non-trivial conjugacy
classes leaving some complex plane invariant. An important difference with
respect to the abelian case is that, due to the presence of the factors
$r_i$ above, the structure $SU(3)\times SU(2)$ does {\em not} in general
guarantee the correct hypercharge assignments for the non-anomalous
$U(1)$'s.

\section{Non-Orbifold Singularities}

In this section we briefly discuss systems of D3-branes at non-orbifold
singularities. The only known systematic approach to construct the
world-volume field theory is to realize the non-orbifold singularity as a
partial resolution of a suitable orbifold singularity \cite{mp}
\footnote{Alternative approaches include the direct construction of field
theories with the right symmetries and the correct moduli space 
\cite{kw,gencon}, and the use of T-duality \cite{uraconi,keshav} to 
configurations of NS-branes and D4-branes \cite{bcooking,giveon}.}.
However, the identification of the field-theory interpretation of the
partial blowing-up is very involved beyond the simplest examples 
\cite{mp,greene,fhh}. Consequently, it is difficult to make general
statements about the phenomenological potential of these systems.

We would like to center on a particular family of singularities, which can
be constructed as orbifolds of the conifold singularity \cite{uraconi}
(see also \cite{ungeradu}), and their partial blow-ups. They have
potential interest since generically they lead to chiral $\NN=1$
supersymmetric field theories on the world-volume of D3-branes, and
include abelian orbifold singularities as particular cases. Moreover,
they have a T-dual representation in terms of configurations of
NS-fivebranes and D5-branes, known as `brane diamonds' \cite{aklm},
generalizing the brane box constructions in \cite{branebox,hu}, which
allow a systematic search of interesting field theories within this class.

Instead of extending on such analysis, we point out that the
only singularity in this family, leading to full triplication of chiral
multiplet content in the 33 sector is in fact the $\IC^3/\IZ_3$ 
singularity. A milder requirement would be to have triplication not for 
the full 33 sector, but at least for some representation. Besides the
$\IZ_3$ orbifold, there is only one singularity in this family fulfilling
this milder requirement. Let us briefly discuss its construction.

Consider the conifold, given by the hypersurface $xy=zw$ in $\IC^4$. 
The field theory on D3-branes at a conifold was determined in \cite{kw}.
The gauge group is $U(N_1)\times U(N_2)$ and there are chiral multiplets
$A_1$, $A_2$ in the $(\fund,\antifund)$, and $B_1$, $B_2$ in the
$(\antifund,\fund)$. There is also a superpotential
\beqa
W=\Tr(A_1 B_1 A_2 B_2) - \Tr (A_1 B_2 A_2 B_1)
\eeqa

We are going to consider a quotient of this variety by a $\IZ_3$ twist
with generator $\theta$ acting as
\beqa
x\to e^{2\pi i/3} x \quad ; \quad z \to e^{2\pi i/3} z \quad ; \quad
y\to e^{-2\pi i/3} y \quad ; \quad  w\to e^{-2\pi i/3} w
\eeqa
which preserves $\NN=1$ supersymmetry on the D3-brane world-volume. The
geometric action is reflected in the field theory as
\beqa
A_1 \to e^{\pi i} A_1 \;\; ; \;\; B_1 \to e^{-\pi i/3} B_1 \;\; ; \;\; 
A_2 \to e^{-\pi i/3} A_2 \;\; ; \;\; B_2 \to e^{-\pi i/3} B_2 
\eeqa
Finally, the action of $\theta$ may be embedded on the $U(N_1)$, $U(N_2)$
gauge degrees of freedom, through the matrices
\beqa
\gamma_{\theta,3}^{(1)} =  \diag(I_{n_0}, e^{2\pi i\frac 13} I_{n_1},
e^{2\pi i\frac 23} I_{n_2}) & ; &
\gamma_{\theta,3}^{(2)} = \diag(e^{\pi i} I_{n_0'}, e^{\pi i\frac 53}
I_{n_1'}, e^{\pi i\frac 13} I_{n_2'})
\eeqa
The field theory on D3-branes at the $\IZ_3$ orbifold of the conifold is
obtained by projecting the conifold field theory onto states invariant
under the combined (geometric plus Chan-Paton) action of $\theta$. The
resulting spectrum is
\beqa
&U(n_0) \times U(n_0') \times U(n_1) \times U(n_1') \times U(n_2) \times 
U(n_2') &\nonumber \\ 
&\begin{array}{cccc}
a_0 : (n_0,{\ov n}_0') \quad, & b_2 : (n_2,{\ov n}_0') \quad, & 
c_0 : ({\ov n}_1,n_0') \quad, & d_0 : ({\ov n}_1,n_0') \\
a_1 : (n_1,{\ov n}_1') \quad, & b_0 : (n_0,{\ov n}_1') \quad, &
c_1 : ({\ov n}_2,n_1') \quad, & d_1 : ({\ov n}_2,n_1') \\
a_2 : (n_1,{\ov n}_1') \quad, & b_1 : (n_1,{\ov n}_2') \quad, &
c_2 : ({\ov n}_0,n_2') \quad, & d_2 : ({\ov n}_0,n_2')
\end{array}&
\eeqa
with superpotential
\beqa
W & = & \Tr [\; a_0 c_0 b_1 d_2 - a_0 d_0 b_1 c_2 + a_1 c_1 b_2 d_0 - 
a_1 d_1 b_2 c_0 + a_2 c_2 b_0 d_1 - a_2 d_2 b_0 c_1 \;] 
\eeqa
In general the gauge anomalies in this theory are non-vanishing, but a
suitable set of D7-branes can be introduced to yield the configuration
consistent.

Assuming a 37 sector with an arbitrary set of bifundamental 
representations, it is possible to perform the analysis of $U(1)$
anomalies. In general, mixed anomalies do not vanish, but have a nice
factorized structure which suggests they are cancelled through a
Green-Schwarz mechanism mediated by closed string twisted modes,
generalizing the observation in \cite{iru} to these non-orbifold
singularities. We also conclude that the only non-anomalous $U(1)$ in
this example is the diagonal combination $Q_{diag}=\sum_{i=0}^2
(Q_{n_i}/n_i + Q_{n_i'}/n_i')$.

The singularity leading to field theories with triplication of at least
some chiral multiplet is obtained as a partial blow-up of this $\IZ_3$
quotient of the conifold. The partial resolution is manifested in the
field theory as an F-flat direction, which is not D-flat by itself, but
requires turning on a Fayet-Illiopoulos term, i.e. turning on a vev
for blow-up modes. For instance, when $n_0=n_0'$, $n_2=n_2'$ there is one
such direction corresponding to diagonal vevs $\langle a_0 \rangle=v_0$,
$\langle a_2 \rangle = v_2$, along which the field theory is reduced to
\beqa
&U(n_0) \times U(n_1) \times U(n_1') \times U(n_2) &\nonumber \\ 
& \begin{array}{ccc}
a_1 : (n_1,{\ov n}_1') \quad ;& b_0 : (n_0,{\ov n}_1') \quad ;&
c_0, d_0 : ({\ov n}_1,n_0) \\
b_1 : (n_1,{\ov n}_2) \quad ;& c_1,d_1 : ({\ov n}_2,n_1') \quad ;&
b_2, c_2, d_2 : (n_2,{\ov n}_0) 
\end{array} &
\label{nonorbione}
\eeqa
with superpotential
\beqa
W = v_0\; \Tr (c_0 b_1 d_2 - d_0 b_1 c_2) + \Tr(a_1 c_1 b_2 d_0 - 
a_1 d_1 b_2 c_0) + v_2 \;\Tr(c_2 b_0 d_1 - d_2 b_0 c_1)
\eeqa
This corresponds to a partial blow-up to a certain singular variety $X$,
whose form could be determined from the above choice of expectation
values. It is possible to follow the above construction in the presence of
D7-branes, but since the computation is rather involved, we just quote
the result. One may introduce four kinds of D7-branes, containing a total
of ten unitary group factors, with ranks denoted $u_1$, and $v_i$, $w_i$,
$x_i$. The 37, 73 sectors are
{\small
\beqa
\begin{array}{ccccccccccc}
(n_1',{\ov u}_1) &,& (u_1,{\ov n}_1) & \to & a_1 & ; &
(n_1',{\ov v}_1) &,& (v_1,{\ov n}_0) & \to & b_0 \\
(n_2,{\ov v}_2) &,& (v_2,{\ov n}_1) & \to & b_1 & ; &
(n_0,{\ov v}_3) &,& (v_3,{\ov n}_2) & \to & b_2 \\
(n_1,{\ov w}_1) &,& (w_1,{\ov n}_0) & \to & c_0 & ; &
(n_2,{\ov w}_2) &,& (w_2,{\ov n}_1') & \to & c_1 \\
(n_0,{\ov w}_3) &,& (w_3,{\ov n}_2) & \to & c_2 & ; &
(n_1,{\ov x}_1) &,& (x_1,{\ov n}_0) & \to & d_0 \\
(n_2,{\ov x}_2) &,& (x_2,{\ov n}_1') & \to & d_1 & ; &
(n_0,{\ov x}_3) &,& (x_3,{\ov n}_2) & \to & d_2
\label{nonorbitwo}
\end{array}
\eeqa}
we have also indicated by an arrow the 33 field to which the corresponding
37, 73 fields couple in the superpotential.
Tadpole conditions cannot be computed directly, but we assume they amount
to cancellation of non-abelian anomalies. This should also guarantee that
$U(1)$ anomalies cancel by the Green-Scharz mechanism. In this case there
are two non-anomalous $U(1)$'s generated by
\beqa
Q_{diag}={{Q_{n_0}}\over {n_0}} + {{Q_{n_1}}\over {n_1}} +
{{Q_{n_1'}}\over{n_1'}} + {{Q_{n2}}\over{n_2}}\quad ;\quad
Q'= {2\over{n_1}} Q_{n_1} -{{Q_{n_1'}}\over {n_1'}} -{{Q_{n_2}}\over{n_2}}
\eeqa

When $n_1=n_1'$, a further diagonal vev $\langle a_1\rangle =v_1$ 
corresponds to a blow-up of $X$ to the $\IC^3/\IZ_3$ singularity. Indeed,
the resulting field theory agrees with (\ref{spec3}), after the
replacements $b_i,c_i,c_i \to \Phi^r_{i,i+1}$, $v_i,w_i,x_i\to u_i^r$. The
33 piece of the superpotential 
\beqa
W = v_0\; \Tr (b_1 d_2 c_0 - b_1 c_2 d_0) + v_1\; \Tr (b_2 d_0 c_1 - 
b_2 c_0 d_1) + v_2\; \Tr(b_0 d_1 c_2 - b_0 c_1 d_2) 
\eeqa 
is more flexible than the 33 piece in (\ref{superpz3}), a fact that may
have phenomenological applications. 

\section{Non-Supersymmetric Models from Antibranes}

The computation of the spectrum in a set of D3/${\Dtb}$-branes at orbifold
singularities in the presence of D7/$\Dsb$-branes can be extracted from
\cite{au} (for a recent review on non-supersymmetric brane configurations
with references to previous material see \cite{matthias}\ ). The only
differences with respect to the case without antibranes is the opposite
GSO projections in the brane-antibrane sectors. For instance,
in ${\bf \bar{3}\bar{3}}$ sectors, GSO projection is as usual and the
spectrum is analogous to that in 33 sectors.

We consider a generically non-susy $\IZ_N$ singularity.
Let us introduce D3, $\Dtb$ Chan-Paton embeddings $\gamma_{\theta,3}$,
$\gamma_{\theta,{\bar 3}}$ with $n_j$, $m_j$ eigenvalues $e^{2\pi i\,
j/N}$. Consider the  ${\bf 3\bar{3}+\bar{3}3}$ sector. In the NS sector
the complex tachyon, $\lambda|0\rangle$ survives the GSO, whereas the
massless states $\Psi_{-\frac12}|0\rangle$ do not. The orbifold projection
on the Chan-Paton factors for the tachyons is
\beqa
\lambda = \gamma_{\theta,3}\ \lambda\ \gamma_{\theta,\bar{3}}^{-1} &\quad
& \lambda = \gamma_{\theta,\bar{3}}\ \lambda \ \gamma_{\theta,3}^{-1}
\label{projtachyon}
\eeqa
leading to complex tachyons in the representation $\sum_{i=0}^{N-1} \;
[(n_i,{\ov m}_i)+(m_i,{\ov n}_i)]$. In the R sector, we obtain states
$\lambda|s_1,s_2,s_3,s_4\rangle$, with $s_i=\pm \frac 12$ and $\sum_i
s_i={\rm even}$. We get {\em right}-handed spacetime fermions $s_4=\frac
12$
with
the projection
\beqa
\begin{array}{lll}
\lambda_{3\bar{3}} = e^{2\pi i a_{\alpha}/N}\ \gamma_{\theta,3}\
\lambda_{3\bar{3}}\ \gamma_{\theta,\bar{3}}^{-1} & \quad &
\lambda_{\bar{3}3} = e^{2\pi i a_{\alpha}/N}\ \gamma_{\theta,\bar{3}}\
\lambda^{(a)}_{\bar{3}3}\ \gamma_{\theta,3}^{-1}
\end{array}
\eeqa
They transform in the representation $\sum_{i=0}^{N-1}\; [(n_i,{\ov
m}_{i+a_\alpha})+(m_i,{\ov n}_{i+a_\alpha})]$. 
The  ${\bf 3\bar{7_r}+\bar 7_r3}$ and ${\bf \bar{3}7_r+7_r\bar{3}}$
sectors are analogous to the ${\bf 37_i+7_i3}$ sector, with a few
modifications due only to the opposite GSO projections. Consider
introducing $\Dsb_3$-branes, with $\gamma_{\theta, 7_3}$ having $v_j$
eigenvalues $e^{2\pi i \, j/N}$, and let us center on the case $b_3={\rm 
even}$. Scalars arise from the NS sector, which has fermion zero modes
along the DN directions, $Y_1$, $Y_2$. Massless states have the form $
\lambda |s_1,s_2\rangle$, with $s_1=-s_2=\pm\frac 12$. The orbifold
projection for $|\frac 12,\frac 12\rangle$ gives
\beqa
\lambda_{3\ov{7}_3} = e^{-2\pi i(b_1-b_2)/2}\ \gamma_{\theta,3}\
\lambda_{3\ov{7}_3}\ \gamma_{\theta,\ov{7}_3}^{-1} & \quad &
\lambda_{\ov{7}_33} = e^{- 2\pi i(b_1-b_2)/2}\
\gamma_{\theta,\ov{7}_3}\ \lambda_{\ov{7}_33}\ \gamma_{\theta,3}^{-1}
\eeqa
We obtain complex scalars in the representation $\sum_{i=0}^{N-1}
[(n_i,{\ov v}_{i+\frac 12 (b_1-b_2)}) + (v_i,{\ov n}_{i+\frac
12(b_1-b_2)})]$.

Spacetime fermions arise from the R sector, which has fermion zero modes
along the DD and NN directions, $Y_3$, $Y_4$. Massless states are $\lambda
|s_3;s_4\rangle$, with $s_3=-s_4=\pm\s2$. The orbifold projection for
$s_4=\frac 12$ gives 
\beqa
\lambda_{3\ov{7}_3} = e^{2\pi ib_3/2}\ \gamma_{\theta,3}\
\lambda_{3\ov{7}_3}\ \gamma_{\theta,\ov{7}_3}^{-1} & \quad &
\lambda_{\ov{7}_33} = e^{2\pi ib_3/2}\ \gamma_{\theta,\ov{7}_3}\
\lambda_{\ov{7}_33}\ \gamma_{\theta,3}^{-1}
\eeqa 
We obtain {\em right}-handed fermions in the representation 
$\sum_{i=0}^{N-1}\; [(n_i,{\ov v}_{i-\frac 12b_3})+(v_i,{\ov n}_{i-\frac
12 b_3})]$. The spectrum in the ${\ov 3}7_3$, $7_3{\ov 3}$ is computed
analogously.

The analysis of gauge anomalies in the presence of antibranes can be
performed in analogy with that in sections 2.2 and 2.3 \cite{au,aiq1}. The
non-abelian gauge anomalies are equivalent to twisted RR tadpole
cancellation conditions, which have the form (\ref{tadpogen}) with the
replacement $\Tr \gamma_{\theta^k,p} \to\Tr\gamma_{\theta^k,p}-
\Tr\gamma_{\theta^k, {\bar p}}$. This reflects the opposite RR charges of 
branes and antibranes. Concerning $U(1)$ anomalies, one can show the
existence of two non-anomalous $U(1)$ per twist with fixed planes,
associated to combinations (\ref{charge}) for branes and antibranes,
respectively.

\section{T-Duality of the Orientifold  Models}

It is well known that $T$-duality changes Dirichlet and Neumann boundary
conditions, therefore $T$-duality with respect to the three complex planes
should map our models, where the standard model is embedded in D3-branes, 
to models where the standard model arises from D5, D7 or D9 branes. Since 
some standard-like models from D7 and D9 branes have recently been build
\cite{aiq1,aiq2}, it is worth seeing the explicit correspondence between
those models and our present constructions. In particular we will show
that, in a suitable T-dual version, the realistic sectors in the models in
\cite{aiq1,aiq2} correspond to local structures of D3- and D7-branes at
$\IC^3/\IZ_3$ singular points, of the type studied in section 3. Hence,
our results `explain' the natural appearance of certain features (like
hypercharge or $\BL$ $U(1)$ gauge factors, or three generations) in the
models in \cite{aiq1,aiq2}.

\bigskip
\noindent{\bf Dual of the Left-Right Model of \cite{aiq1}}
\bigskip

Let us start with the simplest model in \cite{aiq1,aiq2}, a left-right 
symmetric model constructed from a type IIB orientifold with D7$_3$-branes
at $Y_3=0$, and D3, $\Dtb$-branes at some of the fixed points. Since the 
D7-branes sit at the orientifold plane $O7$, the twist matrices
$\gamma_{\theta,7}$ and $\gamma_{\theta,3}$ have to satisfy the tadpole
condition
\beq
\Tr\gamma_7+3\left(\Tr \gamma_3-\Tr\gamma_{\bar 3}\right)\ =\ -4
\eeq
Taking  $\gamma_7=  \left(\tilde\gamma_7,\tilde\gamma_7^*\right)$ with 
\beq
\tilde\gamma_7\ =\ \diag\left(\alpha \id_3, \alpha^2 \id_2, \id_2; \id_2,
\alpha \id_7\right)
\label{twistaiq}
\eeq
and Wilson line in the first complex plane $\tilde\gamma_W\ =\ \diag
\left(\alpha \id_7,\id_9\right)$, the gauge group is $U(3)\times
U(2)_L\times U(2)_R \times\left[SO(4)\times U(7)\right]$. To satisfy the
tadpole constraint, we just need  to add D3-branes at the three points
$(-1,m,0)$ with $\gamma_3=\diag (\alpha,\alpha^2)$ (since 
$\Tr\gamma_7\gamma_W^2= -1$). As shown in \cite{aiq1,aiq2}, this is a
three-family model. Additional $\Dtb$-branes must be introduced to cancel
the untwisted tadpole, but we will not discuss them in detail here.

To find the dual of this model in the approach of the present paper
we have to dualize with respect to the first two complex planes so
that the D7$_3$-branes become D3-branes and viceversa. Also under
$T$-duality Wilson lines become displacement of the branes, although the
explicit mapping is rather involved. The best strategy in the present case
is to construct the model explicitly. 

Since the Wilson lines above only affect the 14 entries leading to the LR
model, these correspond to the D3-branes that sit outside the origin in
the T-dual version. Therefore we consider 18 D3-branes at the origin, with
twist matrix $\gamma_{\theta,3}\ = \ (\tilde\gamma_3,\tilde\gamma_3^*)$
and
\beqa
\tilde\gamma_3\ =\ \diag\left(\id_2, \alpha \id_7 \right)
\eeqa
This should correspond to the `hidden' part of the original model (last
entries in (\ref{twistaiq}), and indeed it leads to an $SO(4)\times
U(7)$ gauge group. Since the origin is an $O3$-plane, the tadpole
condition is (\ref{tadpoin}). Since $\Tr \gamma_3 =-3$ we can cancel
tadpoles introducing six D7-branes at $Y_3=0$ with
$\gamma_{\theta,7}=\diag(\alpha \id_3, \alpha^2 \id_3)$.

After introducing these D7-branes, we have to worry about tadpole
cancellation at the other fixed points in the $Y_1,Y_2$ planes, given by
(\ref{tadpoout}). At the points $(\pm 1, 0,0)$ (mapped into each other
under the orientifold action $(-1)^F\Omega R_1 R_2 R_3$) we
can put seven D3-branes on each, with twist matrix 
\beq
\gamma_{\theta,3}\ =\ \diag\left(\id_3, \alpha \id_2, \alpha^2
\id_2\right)
\eeq
giving rise to a three-generation $SU(3)\times SU(2)_L\times SU(2)_R
\times U(1)_{\BL}$ LR model of the type studied in section 3.4. 

Notice that we have introduced $18$ D3-branes at the origin and
$7+7=14$ at the $(\pm 1, 0, 0 )$ points, already saturating the total
number of allowed D3-branes ($32$). Therefore, twisted tadpoles at the
remaining fixed points should cancel without additional D3-branes. This
can be achieved by introducing a Wilson line in the $Y_2$ direction (hence
not affecting the points $(m, 0, 0 )$) , such that $\Tr\gamma_7\gamma_W
=\Tr\gamma_7\gamma_W^2=0$. The right choice is
\beq
\gamma_W\ =\ \diag \left(1,\alpha,\alpha^2; 1, \alpha^2, \alpha\right)
\eeq
Therefore we do not have to introduce D3-branes at any of the six fixed
points $(m,\pm 1, 0)$. Notice also that this Wilson line breaks the
D7-brane gauge group $U(3)$ to $U(1)^3$, as in the  original model. The
appearance of this Wilson line in the dual model reflects the fact that in
the original model the six D3-branes were distributed among three
different fixed points.

The last consistency requirement is to have an equal number of D7- and
$\Dsb$-branes. To satisfy it, we put three anti-D7-branes with
$\gamma_{\bar 7}=(1,\alpha,\alpha^2)$ at the two mirror locations
$Y_3=\pm 1$. Since this matrix is traceless there is no need
to add more D3-branes to cancel tadpoles. These anti-branes can move
freely in the $Y_3$ direction. Moreover, it is possible to turn on
continuous Wilson lines along their world-volume, i.e. the directions
$Y_1$, $Y_2$. This corresponds to possibility of moving the $\Dtb$-branes
in the original model in the three complex dimensions of the bulk.
 They are naturally dual to the anti
D3-branes of the original model.

\bigskip

\noindent{\bf A Variant Left-Right Orientifold Model and its Dual}
\bigskip

Let us consider a variant left-right orientifold model slightly
different
from the one in section 4.
We satisfy the tadpole equations (\ref{tadpoin}) at the origin
by having 12 D7-branes and 4 D3-branes with:
\beq
\gamma_7\ =\ \left(\alpha\id_6, \alpha^2\id_6\right)\ , \qquad
\gamma_3\ =\ \left( \alpha, \alpha^2, \alpha, \alpha^2\right)
\eeq
In the second plane, we add a Wilson line:
\beq
\gamma_W\ =\ \left( 1, \alpha, \alpha^2, \id_3, 1, \alpha, \alpha^2,
\id_3\right)
\eeq
Since $\Tr\gamma_7\gamma_W = \Tr\gamma_7\gamma_W^2 = -3$ we can put a
left-right model at each of the points $(0,\pm 1, 0)$
 in terms of 7 D3-branes
with 
\beq
\gamma_3\ =\ \left( \id_3, \alpha \id_2, \alpha^2 \id_2\right)
\eeq
satisfying the tadpole condition (\ref{tadpoout}).
In order to minimize the number of D3-branes needed to cancel tadpoles
(so we do  not exceed 32) it is convenient to add a second Wilson line
now in the third plane:
\beq
\gamma_W'\ =\ \left(\id_3, 1, \alpha, \alpha^2, \id_3, 1, \alpha,
 \alpha^2\right)
\eeq
Since $\Tr\gamma_7\gamma_W' = \Tr\gamma_7\gamma_W'^2 = -3$
we can cancel tadpoles at the points $(0, 0, \pm 1)$ by putting just 
one single D3-brane on each with $\gamma_3=1$. The advantage of adding
the second Wilson line is that at the points $(0, \pm 1, \pm 1)$ there
is no need to add any D3-brane because
$\Tr\gamma_7\gamma_W^m\gamma_W'^n=0$
for $m,n =1,2$. Therefore we have at the plane $Y_1=0$ 12 D7-branes
and a total of $4+7+7+1+1=20$ D3-branes.

At each of  the planes $Y_1=\pm 1$ we have to put 6 $\Dsb$-branes
and 6 D3-branes distributed on the fixed points $(\pm 1, m, n)$.
We achieve tadpole cancellation by having:
\beq
\gamma_{\ov 7}\ =\ \left( \alpha \id_3, \alpha^2 \id_3\right)\ ,
\qquad
\gamma_W\ =\ \left(1, \alpha, \alpha^2, 1, \alpha,
\alpha^2\right)
\eeq
So we have $\Tr\gamma_{\ov 7}=-3$ and can cancel tadpoles at 
the points $(\pm 1, 0, m)$ adding 2 D3-branes at each point with 
$\gamma_3=(\alpha,\alpha^2)$. This  saturates the number of D3-branes
($20+2\times 6=32$) which is OK since at the points $(\pm 1, \pm 1,
m)$
there is no need to add D3-branes since $\Tr\gamma_{\ov 7}\gamma_W=
\Tr\gamma_{\ov 7}\gamma_W^2=0$.

We can easily see that the gauge group coming from the D3-branes is 
$[U(3)\times U(2)_L\times U(2)_R]\times U(2)\times U(1)^7$ whereas
from the D7-branes is simply $U(1)^{12}$. The full spectrum may be
computed following the same lines as the left-right model of section
4.

In order to find the $T$ dual of this model, in the formalism of
 \cite{aiq1}\
 we know that at the plane
$Y_1=0$ we should have 20 D7 branes and 12 D3 branes. To obtain the
same gauge group as the model above we choose:
\beq
\tilde\gamma_7\ =\ \left(\alpha\id_3, \alpha^2\id_2, \id_2,
\alpha\id_2, \alpha^2\right)\ , \qquad \gamma_W\ =\ \left(
\alpha\id_7, \id_2, \alpha\right).
\eeq
Since $\Tr\gamma_7=-4$ and
$\Tr\gamma_7\gamma_W=\Tr\gamma_7\gamma_W^2=-1$
we cancel tadpoles by having 2 D3-branes at each of the 6 points $(0,
\pm 1, m)$ with $\gamma_3=(\alpha, \alpha^2)$.

At each of  the planes $Y_1=\pm 1$ we put 6 D7-branes with:
\beq
\tilde\gamma_7\ =\ \alpha\id_3\ ,\qquad \tilde\gamma_W\ =\
 \left(1, \alpha, \alpha^2\right)
\eeq
Since $\Tr\gamma_7=-3$ and
$\Tr\gamma_7\gamma_W=\Tr\gamma_7\gamma_W^2=0$
we cancel tadpoles by adding 2 anti D3 branes at each of the
points $(\pm 1, 0, n)$.

We can then see that there is an explicit D3/D7 duality among the two
models
indicating that they are $T$ dual of each other.
In both cases anti-branes are trapped, as in the orientifold models
of section 4.2 providing good examples of gravity mediated
supersymmetry breaking. One of the positive points about this
left-right  model is that the extra gauge symmetries are 
essentially abelian, and most of the $U(1)$ symmetries get broken by
the Green-Schwarz mechanism.
\vfill\eject

\end{document}